\documentclass[journal]{IEEEtran}
\usepackage{cuted}
\usepackage{verbatim}
\usepackage{graphicx}
\usepackage{amsmath, amsfonts, amssymb}
\usepackage{subfigure}
\usepackage{overpic,url}
\usepackage{rotating}
\usepackage{color}
\usepackage{enumerate}
\usepackage{amsthm}

\newtheorem{theorem}{Theorem}[section]
\newtheorem{lemma}[theorem]{Lemma}
\newtheorem{definition}[theorem]{Definition}
\newtheorem{corollary}[theorem]{Corollary}
\newtheorem{remark}{Remark}
\newtheorem{example}{Example}

\newcommand{\parenth}[1] {\left(#1\right)}
\newcommand{\braces}[1] {\left\{#1\right\}}
\newcommand{\brackets}[1] {\left[#1\right]}
\newcommand{\alphabet}[1] { {\mathsf #1}}

\newcommand{\reals}{\mathbb{R}}

\newcommand{\indicatorvbl}[1] {1_{\braces{#1}}}

\newcommand{\deq}{\equiv}
\newcommand{\twovec}[2]{\brackets{\begin{array}{c} #1 \\ #2\end{array}}}

\DeclareMathOperator*{\argmax}{arg\,max}

\DeclareMathOperator*{\arginf}{arg\,inf}

\renewcommand{\P}{\mathbb{P}}
\newcommand{\Q}{\mathbb{Q}}
\newcommand{\E}{ {\mathbb E}}
\newcommand{\prob}[1] {\P\parenth{#1}}

\newcommand{\probSimplex}[1]{ \mathcal{P}\parenth{\alphabet{#1}}}

\newcommand{\cF}{ {\cal F}}

\newcommand{\cB} {{\cal B}} 

\newcommand{\cV}{ {\alphabet{V}}}

\newcommand{\borel}[1]{ {\cal B} \parenth{\alphabet{#1}}}
\newcommand{\spaceV}{\parenth{\cV,\borelV}}
\newcommand{\borelV}{\borel{V}}

\newcommand{\cS}{ \alphabet{S}}
\newcommand{\cU}{ \alphabet{U}}

\newcommand{\equivInDistribution}{\overset{d}{=}}
\newcommand{\OSPU}{\Phi}
\newcommand{\NLF}{\Lambda} 

\newcommand{\regionDistStateCost}{\mathcal{R}}
\newcommand{\regionDistStateCostRandomized}{\widetilde{\regionDistStateCost}}
\newcommand{\randomizedPolicies}{\widetilde{\Pi}}

\newcommand{\totalcost}[2]{J_{n,#1}^{#2}}
\newcommand{\totalcostpi}{\totalcost{\pi}{\alpha}}
\newcommand{\totalcostpiprime}{\totalcost{\pi'}{\alpha}}
\newcommand{\totalcostpistar}{\totalcost{\pi^*}{\alpha}}
\newcommand{\totalcostbpi}{\totalcost{\bpi}{\alpha}}

\newcommand{\MarkovChainIFS}{\psi}

\newcommand{\IRC}{~reversibly feasible dynamics~}

\newcommand{\kldist}[2] {D \parenth{#1\|#2}}
\newcommand{\optimalInputDMC}{P^*_\State(\statecostfn,\PDMC,\statecostval)}

\newcommand{\optimalInputDMCpi}{P^*_\State(\statecostfn,\PDMC,\statecostval_\pi)}
\newcommand{\capacityCostFnVal}[1]{C\parenth{\statecostfn,\PDMC,#1}}
\newcommand{\capacityCostFn}{\capacityCostFnVal{\statecostval}}
\newcommand{\capacityCostFnpi}{\capacityCostFnVal{\statecostval_\pi}}

\newcommand{\rateDistortionFn}{R_n\parenth{\dist,P_{\Src^n},\distval}}
\newcommand{\rateDistortionFnPi}{R_n\parenth{\dist,P_{\Src^n},\distval_\pi}}

\newcommand{\optimalChannelRD}{P^*_{\Est^n|\Src^n}\parenth{\dist,P_{\Src^n},\distval}}
\newcommand{\optimalChannelRDPi}{P^*_{\Est^n|\Src^n}\parenth{\dist,P_{\Src^n},\distval_\pi}}

\newcommand{\PEstnGivenSrcnPolicyprime}{P^{\pi'}_{\Est^n|\Src^n}}
\newcommand{\PEstnGivenSrcnPolicy}{P^{\pi}_{\Est^n|\Src^n}}
\newcommand{\PEstnGivenSrcn}{P_{\Est^n|\Src^n}}
\newcommand{\POutnGivenSrcnPolicy}{P^{\pi}_{\Out^n|\Src^n}}

\newcommand{\bPEstnGivenSrcnPolicy}{P^{\bpi}_{\Est^n|\Src^n}}
\newcommand{\bPOutnGivenSrcnPolicy}{P^{\bpi}_{\Out^n|\Src^n}}

\newcommand{\PDMC}{P_{\Out|\State}}
\newcommand{\PDMCstate}{P_{\Out|\State=\state}}

\newcommand{\normalizationConstantExpFam}{\zeta}

\newcommand{\Src}{W}
\newcommand{\src}{w}
\newcommand{\cSrc}{\alphabet{\Src}}
\newcommand{\Psrc}{Q_{\Src}}

\newcommand{\kernelSrc}[2]{Q_{\Src}\parenth{#1|#2}}

\newcommand{\State}{X}
\newcommand{\state}{x}
\newcommand{\cState}{\alphabet{\State}}
\newcommand{\tState}{\tilde{\State}}
\newcommand{\tstate}{\tilde{\state}}

\newcommand{\cY} {\alphabet{Y}}

\newcommand{\Out}{Y}
\newcommand{\out}{y}
\newcommand{\cOut}{\alphabet{\Out}}

\newcommand{\bpi}{\bar{\pi}}

\newcommand{\cEnc}{\alphabet{E}}
\newcommand{\ctEnc}{\tilde{\alphabet{E}}}

\newcommand{\enc}{e}
\newcommand{\benc}{\bar{\enc}}
\newcommand{\encst}{\enc^*}

\newcommand{\tenc}{{\tilde{\enc}}}
\newcommand{\bencst}{{\benc^*}}
\newcommand{\encSC}{\bar{\enc}}

\newcommand{\cDec}{\alphabet{D}}
\newcommand{\dec}{d}
\newcommand{\decst}{\dec^*}

\newcommand{\bdecst}{{\bdec^*}}
\newcommand{\decSC}{\bar{\dec}} 
\newcommand{\bdec}{\bar{\dec}} 

\newcommand{\est}{z}
\newcommand{\Est}{Z}
\newcommand{\cEst}{\alphabet{\Est}}

\newcommand{\bdist}{\bar{\dist}}
\newcommand{\dist}{\rho}
\newcommand{\distval}{D}

\newcommand{\tdist}{\tilde{\dist}}
\newcommand{\statecostfn}{\eta}

\newcommand{\bstatecostfn}{\bar{\statecostfn}}

\newcommand{\statecostval}{L}
\newcommand{\bcost}{\bar{g}}
\newcommand{\valuefn}{V}

\newcommand{\Belief}{B}
\newcommand{\belief}{b}
\newcommand{\tEnc}{\tilde{E}}

\newcommand{\QNLFdenom}{P_{\NLF}}

\newcommand{\PnextYgivenBeliefEncEqualsZWithnMinusOne}{\QNLFdenom\parenth{\cdot|\est_{n-1},\tenc_{n}}}
\newcommand{\PnextYgivenBeliefEncEqualsBWithnMinusOne}{\QNLFdenom\parenth{\cdot|\belief_{n-1|n-1},\tenc_{n}}}
\newcommand{\PnextYEqualsyGivenBeliefEncEqualsBWithnMinusOne}{\QNLFdenom\parenth{dy|\belief_{n-1|n-1},\tenc_{n}}}

\newcommand{\Error}{E}

\newcommand{\PControlledMC}{Q_S}

\newcommand{\bitm}{\begin{itemize}}
\newcommand{\eitm}{\end{itemize}}
\newcommand{\benum}{\begin{enumerate}}
\newcommand{\eenum}{\end{enumerate}}
\newcommand{\beqa}{\begin{eqnarray}}
\newcommand{\eeqa}{\end{eqnarray}}
\newcommand{\beqas}{\begin{eqnarray*}}
\newcommand{\eeqas}{\end{eqnarray*}}
\newcommand{\baln}{\begin{align}}
\newcommand{\ealn}{\end{align}}
\newcommand{\balns}{\begin{align*}}
\newcommand{\ealns}{\end{align*}}

\newcommand{\tSrc}{\tilde{\Src}}

\newcommand{\GaussNoise}{V}
\newcommand{\NoiseVariance}{\sigma_v^2}
\newcommand{\kldistTermAnminusOne}{\kldist{\belief_{n-1|n-1}}{z_{n-1}}}
\newcommand{\arginfTermNMinusOne}{\arginf_{\belief_{n-1|n-1} \ll \est_{n-1} \ll \OSPU(\est_{n-2})}}

\title{Information-Theoretic Viewpoints on Optimal Causal Coding-Decoding Problems}

\author{Siva~K.~Gorantla and Todd~P.~Coleman \\
ECE Department \\
Coordinated Science Laboratory \\
University of Illinois \\
Urbana, IL \\
\{sgorant2,colemant\}@illinois.edu
}

\begin{document}
\maketitle
\begin{abstract}
In this paper we consider an interacting two-agent sequential decision-making problem consisting of a  Markov source process, a causal encoder with feedback, and a causal decoder.  Motivated by a desire to foster links between control and information theory, we augment the standard formulation by considering general alphabets and a cost function operating on current and previous symbols. Using dynamic programming, we provide a structural result whereby an optimal scheme exists that operates on appropriate sufficient statistics.  We emphasize an example where the decoder alphabet lies in a space of beliefs on the source alphabet, and the additive cost function is a log likelihood ratio pertaining to sequential information gain.  We also consider the inverse optimal control problem, where a fixed encoder/decoder pair satisfying statistical conditions is shown to be optimal for some cost function, using probabilistic matching. We provide examples of the applicability of this framework to communication with feedback, hidden Markov models and the nonlinear filter, decentralized control, brain-machine interfaces, and queuing theory.
\end{abstract}
\section{\bf Introduction} \label{sec:intro}\noindent
Many current and future societal problems involve designing and understanding networks of sequential decision-making entities cooperating in an uncertain environment.  Some of these entities may be physical/biological agents, whereas others might be computerized systems.  For example, cyber-physical systems feature interacting networks of physical processes that are noisily sensed and actuated by computational algorithms. The mammalian brain comprises another example, where the cooperative goals of sensing, perception, learning, and eliciting behavior are achieved via
coupled neural systems that interact via signaling across a noisy biological medium.

From an engineering system designer vantage point, obtaining optimal coordination strategies for a network of interacting decision-makers is in general computationally intractable \cite{Papadimitriou86}.  For a class of small networks (e.g. comprising a specific interaction structure between an encoder and a decoder), and an asymptotic performance objective, fundamental limits of performance can be addressed using the information theoretic concepts of communication and rate-distortion \cite{CoverThomas06}.   Identifying optimal strategies for sequential decision-making under uncertainty for a single agent, on the flipside, is traditionally addressed with control theoretic-principles of Markov decision theory \cite{bertsekas1978stochastic}.

From a scientific vantage point, the joint statistical dynamics between interacting decision-makers can provide insight into the cost or utility they are cooperatively optimizing.  For small networks (e.g. an encoder and decoder) with a limited statistical dynamics interaction structure, this has been addressed with the information-theoretic principle of source-channel probabilistic matching \cite{GastparRimoldiVertterli03}.
Inverse optimal control theory theory identifies cost functions for which a fixed strategy of one decision-maker is optimal \cite{kalman1964linear} and has been used in neural \cite{kording2004loss,todorov2004optimality} and cognitive science \cite{baker2007goal} applications.

It appears evident that understanding this class of problems for more general objectives and interaction structures can utilize insights from both information and control theory, but the differences in their philosophical starting points is striking:\\
{\bf Information Theory} problems, traditionally specify large but {\it fixed} time horizon $n$ for which some decisions are not made until this terminal point.   Even in problems where neither an observation nor a decision variable lies in a time-horizon dependent set (e.g. reproducing a source over a noisy channel with a fidelity criterion), Shannon's `separation theorem' \cite{Shannon48} shows that for very large $n$, it is sufficient to first decompose the problem into sub-problems, each of which contains some observations or decision variables with time horizon-dependent alphabet structure (e.g. of size $2^{nR}$) and a performance objective pertaining to constrained extremizing of mutual information.  As such, traditional information theoretic problem formulations have the following starting point:
\begin{enumerate}
\item[(a)] {\it time horizon-dependent alphabets}
\item[(b)] {\it some decisions made at final stage of long time horizons}
\item[(c)] {\it performance objective: extremize mutual information}
\end{enumerate}
{\bf Control Theory}
Markov decision theory problems typically involve observations of state variables' whose future statistics are impacted by their current values and the current `decision variable' that is under causal control of a decision-maker.  The alphabet size of observations and decision variables are typically {\it unrelated} to the time horizon $n$ of the problem.   Moreover, at each time step, a decision must be made based upon causal information up to that time.  Lastly, the performance objective is to minimize an expected sum of costs, each of which operates on current state, observation, and decision variables.  Structural results are typically desirable in such settings because they develop conditions relating the existence of explicit, non-random strategies that operate on sufficient statistics.  Succinctly, we can state this as follows:
\begin{enumerate}
\item[(a)] {\it time horizon-independent alphabets}
\item[(b)] {\it decisions made sequentially based on causal information}
\item[(c)] {\it performance objective: sum of costs operating on current observations and decision variables}
\end{enumerate}
So these two philosophies have striking differences.  Consider the class of `causal coding/decoding' problem that further demonstrates this:
\begin{figure}[hbtp]
\centering
\begin{overpic}[width=\columnwidth]{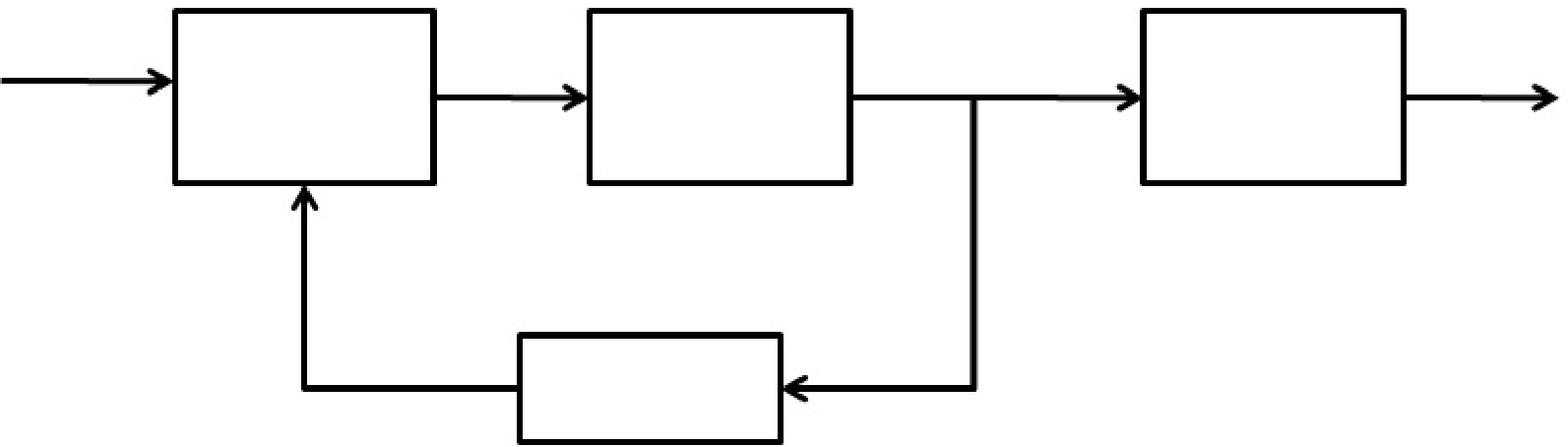}
\put(13,23){\bf causal}
\put(13,19){\bf encoder}
\put(74,23){\bf causal}
\put(74,19){\bf decoder}
\put(39,23){\bf noisy}
\put(39,20){\bf channel}
\put(37,3){\bf delay}
\put(91,23){\Large $\Est_i$}
\put(61,23.5){\Large $\Out_i$}
\put(30,23.5){\Large $\State_i$}
\put(1,25){\Large $\Src_i$}
\put(20,10){\Large $Y^{i-1}$}
\end{overpic}
\caption{Basic problem setup: an optimal causal coding/decoding problem.}
\label{fig:causalEncDec}
\end{figure}
At each time step $i$, the causal encoder's decision variable is the input $\State_i \in \cState$ to a noisy channel that is a causal function of source inputs $(W_1,\ldots,W_i)$ and the noisy channel outputs $Y_1,\ldots,Y_{i-1}$: $\State_i = e_i(W^i,Y^{i-1})$  The causal decoder's decision variable is a `source estimate' $\Est_i \in \cEst$ that is a causal function of channel outputs $(Y_1,\ldots,Y_i)$: $\Est_i=d_i(Y^i)$.
They jointly design their strategies $\pi=(e,d)$ to minimize a function $J_{n,\pi}$ pertaining to an expected sum of costs:
   \beqa
    J_{n,\pi} = \E_{e,d} \brackets{\sum_{i=1}^n g(\Src_i,\Est_i)} \label{eqn:intro:WalrandVaraiyaObjective}
    \eeqa
    Some aspects of the problem appear to make it amenable to a control theoretic analysis:  (a) the source alphabet $\cSrc$ is unrelated to the time horizon $n$ and (b) the sequential decision-making and additive costs , (c) the performance objective \eqref{eqn:intro:WalrandVaraiyaObjective} operates additively on observations/decision variables in the vicinity of each time $i$ as compared to only at the final time horizon $n$.  The presence of the noisy channel in the loop possibly make it amenable to an information-theoretic analysis: mutual information could plausibly provide tight bounds on attainable costs.  On the flipside, neither agent's observations at any time point are a nested version of the other's and so they have a `non-classical' information structure \cite{witsenhausen71separation} - making this a `hard' control problem.
    Analogously, the `hard' delay constraint pertaining to causal decoding and typical `hard decision' assumption of $\cSrc$,$\cEst$ being in discrete, time-horizon independent alphabets typically render information-theoretic techniques irrelevant to the understanding of these `real-time' problems \cite{witsenhausen1978informational,teneketzis1980communication}.

    In this paper we consider a causal coding/decoding problem where $\Src$ is a Markov source process.  We consider additive cost functions
operating of the form $g(\src_i,\state_i,\est_{i-1},\est_i)$. We do not impose assumptions (e.g. finiteness)
on alphabets of the variables.  Our motivation for this more general framework is an example (Section~\ref{sec:intro:example}) motivated by feedback communication where the source alphabet is continuous, the  decoder alphabet lies in a space of {\it beliefs} on the source alphabet, and the additive cost function is a log likelihood ratio pertaining to {\it sequential information gain}.  Using dynamic programming, we provide a structural result whereby an optimal scheme exists that operates on appropriate sufficient statistics.  We also consider the inverse optimal control problem,
where a fixed encoder/decoder pair satisfying a sufficient statistical condition is shown to be optimal for some cost function, using probabilistic matching. We provide examples of the applicability of this framework to communication with feedback, hidden Markov models and the nonlinear filter,
decentralized control, brain-machine interfaces, and queuing theory.

\subsection{\bf Example: Communication over a Noisy Channel with Feedback and the Sequential Information Gain Cost}\label{sec:intro:example}
    We now consider the traditional feedback communication model and how its assumptions - along with traditional `real-time' problem assumptions - can be modified so that fundamental limits are unchanged but the frameworks align.
Consider the traditional information-theoretic communication model with feedback, consisting of an encoder, a decoder, and a fixed block length $n$.  The encoder has a message $W \in \cSrc=\{1,\ldots,2^{nR}\}$. It specifies $n$ inputs to the channel, $X_1,\ldots,X_n$.  The channel is memoryless and non-anticipative where $P_{Y|X}(y|x)$ is the statistics of the output given the input.  At each time step $i$, the encoder selects the message $W$ and the previous channel outputs $Y_1,\ldots,Y_{i-1}$ at time $i$, to specify the next channel input $X_i$. The decoder, at time $n$, having acquired channel outputs $Y_1,\ldots,Y_n$, specifies a single decision, $\hat{W}_n \in \cSrc$.  The question asked in information theory is, how large can $R$ be such that for sufficiently large $n$, there exist encoders and decoders for which
$\P(\hat{W}_n \neq W) \to 0$?   To demonstrate the existence of such encoders and decoders, a {\it random coding} argument \cite{CoverThomas06} and the laws of large numbers are typically invoked.

Recently, a development by Shayevitz \& Feder \cite{ShayevitzFeder07,ShayevitzFeder08,shayevitz2009optimal}, has re-visited a philosophically different way to frame the feedback communication model - dating back to the 1960s \cite{horstein1963stu,schalkwijk1966csaOne,schalkwijk1966csaTwo}- that has a more dynamical systems and control theoretic flavor.
\begin{figure}[hbtp]
\centering
\begin{overpic}[width=\columnwidth]{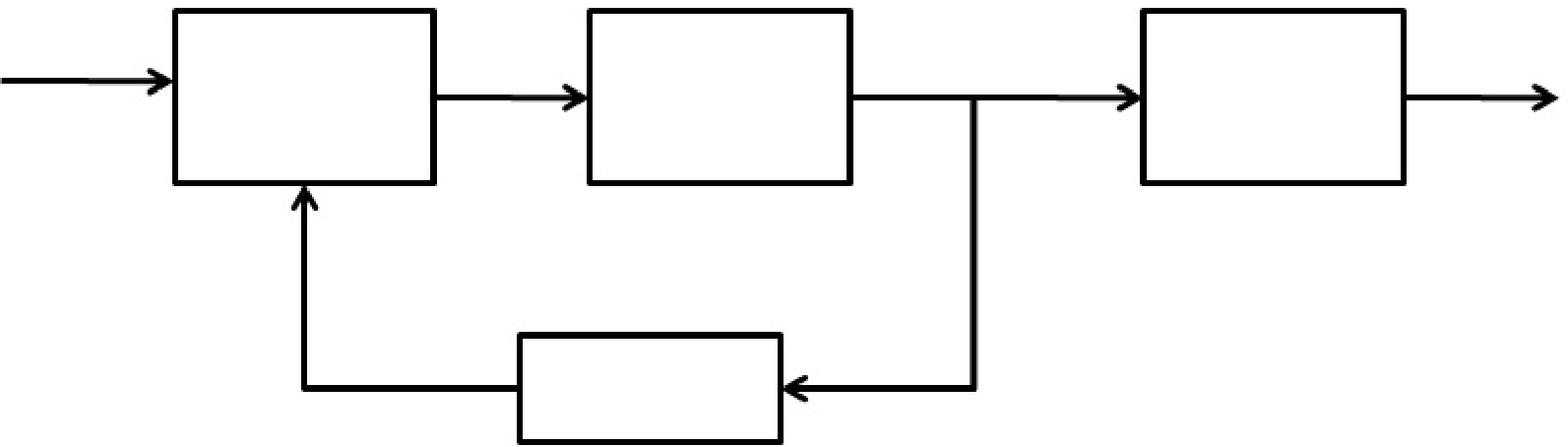}
\put(13,23){\bf causal}
\put(13,19){\bf encoder}
\put(73,23){\bf causal}
\put(74,19){\bf decoder}
\put(39,23){\bf noisy}
\put(39,20){\bf channel}
\put(37,3){\bf delay}
\put(91,24){\Large $\Belief_{i|i}$}
\put(61,23.5){\Large $\Out_i$}
\put(30,23.5){\Large $\State_i$}
\put(1,25){\Large $\Src$}
\put(20,10){\Large $Y^{i-1}$}
\end{overpic}
\caption{Communication of a message point $\Src$ with causal feedback over a memoryless channel. }
\label{fig:channelCodingFB}
\end{figure}
Consider the following changes to the standard information theoretic formulation that more closely resembles a causal coding/decoding problem, shown in Figure~\ref{fig:channelCodingFB}:
\begin{itemize}
  \item \textit{Message Point:} $\Src_i=\Src$ is equally likely over interval $\cSrc = [0,1]$.
   \item \textit{Decoder:} At each $i$ (not only at time $n$), the decoder specifies $Z_i=\Belief_{i|i}$, the posterior belief
   about $\Src$ given $Y_1,\ldots,Y_i$: $\Belief_{i|i}(A) \triangleq \prob{W_i \in A | Y^i}$.
  \item \textit{Achievability:} As shown in Fig~\ref{fig:discretizedPMFs}, with a set of uniform quantizers $\parenth{q_{iR}: [0,1] \to 2^{iR}, \;i \geq 1}$, a rate $R$ is achievable if $\Belief_{i|i}\parenth{\braces{w: q_{iR}(w)=q_{iR}(W)}} \to 1$.
\end{itemize}
\begin{figure}[hbtp]
\centering
\begin{overpic}[width=0.3\columnwidth]{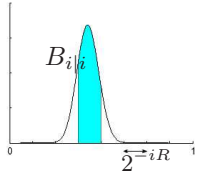}
\put(60,-4){\small $2^{-iR}$}
\put(22,49){$\Belief_{i|i}$}
\end{overpic}
\caption{Representation of the posterior belief $B_i$ in terms of its density.}
\label{fig:discretizedPMFs}
\end{figure}
 Note the importance of $\cSrc$ being a {\it continuous interval} and $\cEst$ being the space of beliefs on $\cSrc$, $\probSimplex{\cSrc}$, in order for this `real-time' flavored problem to relate to traditional information-theoretic notions of achievability. The fundamental limits under both formulations are equivalent \cite{shayevitz2009optimal}, where achieving capacity subject to channel input cost $\eta(\state)$ constraints pertains  to maximizing the mutual information $I(\Src;Y^n)$ \cite{Shannon48}.  A time-invariant `posterior matching' encoding scheme in Figure~\ref{fig:channelCodingFB}'s framework achieves capacity on general memoryless channels  \cite{shayevitz2009optimal}.  Moreover, it is an optimal solution to a stochastic control problem \cite{ColemanISIT09} whose cost function at each time step is related to the {\it sequential information gain} $I(W;Y_i|Y^{i-1})$:
\beqas
 I(W;Y^n)=  \sum_{i=1}^n I(W;Y_i|Y^{i-1})=\sum_{i=1}^n \E \brackets{\log \frac{d\Belief_{i|i}}{d\Belief_{i-1|i-1}}(\Src) }
 \eeqas
 Note that the sequential information gain term represents the reduction in $\Src$'s uncertainty from the previous posterior belief $\Belief_{i-1|i-1}$ to the current, and so each term in the sum operates on $\Src$, $\Belief_{i|i}$, {\bf and} $\Belief_{i-1|i-1}$.
 This alludes to a generalization of causal coding/decoding problems with a cost function $g(\src_i,\state_i,\est_{i-1},\est_i)$, which in this case
 could plausibly be
 \beqa
 g(\src_i,\state_i,\est_{i-1},\est_i) = -\log \frac{d\est_{i}}{d\est_{i-1}}(\src_i) + \alpha \statecostfn(\state_i),  \label{eqn:intro:informationGainCost}
\eeqa
where $\Est_i \in \cEst= \probSimplex{\cSrc}$ is a {\it decision variable} that can be any belief about the message.
In this manuscript, we plan to build on this example and formulate general problems that capture this generalization and further elucidate an interplay between information theory and control theory within the context of both designing optimal strategies and performing inverse optimal control to characterize cost functions for which fixed strategies are optimal.


\subsection{\bf Related Work}\label{sec:intro:relatedWork}
The interplay between information and control theory has been established when treating a `message point' as a real-valued point within the context of control over noisy channels \cite{tatikonda2000control,tatikonda2004stochastic,tatikonda2004control,elia2004bode,martins2008feedback,liu2009convergence,ardestanizadeh2010control}
and feedback information theory \cite{horstein1963stu,schalkwijk1966csaOne,schalkwijk1966csaTwo,ShayevitzFeder07,shayevitz2009optimal,shayevitz2009fixedPoints}.

Cost functions pertaining to log likelihood ratios and making decision variables pertaining to beliefs have been used within the context of sequential prediction \cite{merhav2002universal,cesa2006prediction} and signal compression/classification \cite{aiyer2005lloyd}, relating thermodynamics and information theory with inference on hidden Markov models \cite{mitter2004variational,mitter2005information}, linearly solvable Markov decision problems \cite{todorov2009efficient}, and feedback information theory \cite{tatikonda2009capacity,ColemanISIT09,bae2010posterior,AnandKumarCDC10}.

Causal coding-decoding problems akin to Figure~\ref{fig:causalEncDec} have been studied extensively when $\Src$ is a Markov process, as it typically enables the use of dynamic programming to demonstrate the existence of optimal strategies where agents use posterior beliefs as state variables. In the case of all-discrete alphabets, this was demonstrated in \cite{WalrandVaraiya83},\cite{mahajan2008design}.  In the case of all-real alphabets, $\Src$ a Gauss-Markov process, and $\PDMC$ an additive Gaussian channel, this was demonstrated in \cite{bansal1988sdc}\cite[Ch. 6]{tatikonda2000control} where additionally an explicit optimal scheme consisting of  `innovations-encoding'  and `minimum-mean-squared error decoding' strategy was constructed.  
\cite{AnandKumarCDC10} considered the case of $\cSrc$ discrete and the objective to maximize mutual information, but $\Est_i$ was not a decision variable.  The case of general alphabets $\cSrc$ was considered in \cite{borkar2001optimal,yuksel2008causal}, but the purpose was quantization and thus $\cEst$ was discrete and the cost function balanced squared error distortion and quantizer output entropy rate. Note that the information gain scenario in Section~\ref{sec:intro:example} does not fall within any of the aforementioned works.

Control-theoretic approaches to inverse optimal control involving a single-agent system have been developed classically
for the case of a known policy  \cite{kalman1964linear,casti1980general} where a control-Lyapunov function acts as an optimal value function
and imposes constraints on candidate cost functions.  Inverse reinforcement learning additionally requires inferring the single-agent's policy based on experimental data \cite{ng2000algorithms,abbeel2004apprenticeship,dvijothaminverse2010}, and it has been applied to solve extremely challenging engineering problems  \cite{abbeel2007application} and within the context of neural \cite{kording2004loss,todorov2004optimality} and cognitive \cite{baker2007goal} science.

Information-theoretic approaches to inverse optimal control for a two-agent (encoder/decoder) system relate costs to likelihood ratios through the variational equations for the rate-distortion and capacity-cost function \cite{csiszar1974extremum,csiszar1982information}, but the problem formulations do not consider cost functions of the form $g(\src_i,\state_i,\est_{i-1},\est_i)$ and either consider encoder/decoder interactions \cite{GastparPhDThesis2002,GastparRimoldiVertterli03} that do not have dynamics akin to the feedback loop and random process input in  Figure~\ref{fig:causalEncDec} or have very specific statistical assumptions (e.g. the Gauss-Markov source and additive Gaussian channel \cite{bansal1988sdc}\cite[Ch. 6]{tatikonda2000control}).

\subsection{\bf Paper Outline and Main Results}\label{sec:intro:mainResults}
We now outline the paper, where in each section we provide bullet points about how it differs from other formulations and its main results.

{\bf Section~\ref{sec:defns}}  provides mathematical notation and definitions that will be used throughout the manuscript.

{\bf Section~\ref{sec:problemSetup}} provides the problem setup. We emphasize the following properties that make it differ from traditional approaches:
\bitm
\item the Markov process source has a general alphabet $\cSrc$
\item the traditional cost function $g(\src_i,\est_i)$ is replaced by
     \beqa
      g(\src_i,\state_i,\est_{i-1},\est_i) = \dist(\src_i,\est_{i-1},\est_i) + \alpha \statecostfn(\state_i) \label{eqn:intro:generalcostfn}
      \eeqa
\item decision variables lie in arbitrary spaces $\cState$ and $\cEst$
\eitm

{\bf Section~\ref{sec:MainResults}} considers a fixed cost function \eqref{eqn:intro:generalcostfn} and finding optimal coordination strategies $(\enc,\dec)$.  Results include:
 \bitm
 \item a structural result demonstrating the existence of optimal coordination strategies operating on sufficient statistics, capturing traditional results \cite{WalrandVaraiya83} as a special case.
 \eitm

 {\bf Section~\ref{sec:directedInfo}} considers the {\it sequential information gain} cost function \eqref{eqn:intro:informationGainCost} with $\cEst = \probSimplex{\cSrc}$ and finding optimal coordination strategies. Results include:
 \bitm
 \item an optimal coordination strategy always specifies $\Est_i=\Belief_{i|i}$
 \item a characterization of the problem as cost-penalized maximization of mutual information $I(\Src^n;\Out^n)$
 \eitm
  The first result uses dynamic programming and the second law of thermodynamics for Markov chains \cite{chigansky2009intrinsic}.  It synergizes with work in \cite{mitter2004variational} but differs in how this is cast in the causal coding/decoding framework and the information gain cost \eqref{eqn:intro:informationGainCost}.

  {\bf Section~\ref{sec:SCmatching}} considers the inverse optimal control scenario with cost functions of the form \eqref{eqn:intro:generalcostfn}.
   Results demonstrate that a `stationary Markov' coordination strategy $(\benc,\bdec)$ is inverse optimal when the channel outputs $Y_1,\ldots,Y_n$ are statistically independent.  The technique constructs the induced $\dist$ and $\statecostfn$  from the variational equations to the rate-distortion and capacity-cost functions and can be interpreted as a `source-channel matching' \cite{GastparPhDThesis2002,GastparRimoldiVertterli03} generalization
 applicable to the causal coding/decoding problem with cost function \eqref{eqn:intro:generalcostfn}.  It is also shown how in some situations, this sufficient condition reduces to time-reversibility of a Markov chain, thus further demonstrating a relationship between thermodynamics and information theory that has been developed in \cite{propp1985thermodynamic,mitter2005information}.

 {\bf Section~\ref{sec:examples}}, provides example problems for which the aforementioned results apply, and shows how:
  \bitm
  \item under a particular constraint, the hidden Markov model and nonlinear filter \cite{elliott1995hidden} are an optimal coordination strategy for the information gain cost \eqref{eqn:intro:informationGainCost} with $\cSrc=\cState$
  \item the posterior matching scheme  \cite{shayevitz2009optimal} is an optimal coordination strategy for
        the information gain cost \eqref{eqn:intro:informationGainCost} and source model $\Src_i=\Src_{i-1}$ with $\cSrc=[0,1]$
  \item the structural results aid the design of optimal and `user-friendly' coordination strategies for brain-machine interfaces  \cite{OmarEtAkBCIIJHJCIsubmittedNov09}
  \item inverse control optimal Markov coordination policies with cost \eqref{eqn:intro:generalcostfn} exist for:
  \bitm
    \item Gauss-Markov source, AGN channel pair
    \item Markov counting-function source, $Z$ channel pair
    \item Markov counting-function source, `inverted $E$' channel pair
  \eitm
 \eitm
 The first example is related to the variational characterization of the optimality of the nonlinear filter \cite{mitter2004variational},
  but is different due to the information gain cost \eqref{eqn:intro:informationGainCost}.
 The second example generalizes the result of \cite{ColemanISIT09} because here, $\Est_i$ is a decision variable.
  In the inverse optimal control examples, the Gaussian channel case pertains to the decentralized control problems in
    \cite[Ch 6]{tatikonda2000control},\cite{bansal1988sdc} with quadratic state cost and squared error distortion; the $Z$ channel case pertains to the $\cdot/M/1$ queue for timing channels \cite{AnantharamVerdu96,SundaresanVerdu06}. the `inverted $E$' channel pertains
     to Blackwell's trapdoor communication channel \cite{Blackwell,AhlswedeKaspiTrapdoor,cuff06}.

{\bf Section~\ref{sec:discussionConclusion}} provides a discussion and conclusion, followed by references and an appendix of proofs.


\section{\bf Definitions and Notations}\label{sec:defns}
{\bf Probabilistic Notation}
\begin{itemize}
\item For a sequence $a_1,a_2,\ldots$, denote $a_i^j$ as $(a_i,\ldots,a_j)$ and $a^j \triangleq a_1^j$.
\item
Denote the probability space with sample space $\Omega$, sigma-algebra $\cF$, and probability measure $\P$ as $(\Omega, \cF, \P)$.
\item For a given $(\Omega,\cF)$ and a Borel space $(\cV,\borelV)$, denote any measurable function $X: \Omega \to \cV$ as a random object.  If
$\cV=\reals$, then $X$ is termed a random variable.
\item Upper-case letters $V$ represent random objects and lowercase letters  $v \in \cV$ represent their realizations.
\item For two probability measures $\P$ and $\Q$ on $(\Omega,\cF)$, we say that $\P$ is absolutely continuous with respect to $\Q$ (denoted by $\P \ll \Q$) if $\Q(A) = 0$ implies $\P(A) = 0$ for all $A \in \cF$.
If $\P \ll \Q$, denote the Radon-Nikodym derivative as any random variable $\frac{d\P}{d\Q}: \Omega \to \reals$ that satisfies
\[ \P(A) = \int_{\omega \in A} \frac{d\P}{d\Q}(\omega) \Q(d \omega),\;\;\; A \in \cF.\]
\item Denote $\probSimplex{V}$ as the space of probability measures on $\spaceV$.  For any
random object $V: \Omega \to \cV$, denote
\[ P_V(A) \triangleq \prob{V \in A} \triangleq \prob{\braces{\omega: V(\omega) \in A}}, \quad A \in \borelV.\]
\item Denote the conditional probability distribution of one random object $V$ given that another $U$ takes on $u$ as
\[ P_{V|U=u}(A) \triangleq \prob{V \in A | U=u}, \;\;\; A \in \borelV.\]
\end{itemize}
{\bf Markov Chains Notation:}
\begin{itemize}
\item A random process $V=(V_i: i \geq 1)$ is a Markov chain if
\beqa
P_{V_{i+1}|V^i=v^i}(A) = P_{V_{i+1}|V_i=v_i}(A), \;\;\; A \in \borelV.
\eeqa
It is {\it time-homogenous} if $P_{V_{i+1}|V_i=v_i}(A)= Q(A|v_i)$.
\item A Markov chain is {\it time-reversible} if the forward and reverse time processes are statistically indistinguishable:
    \begin{eqnarray}
\vspace{-0.05in}
(V_j: 1 \leq j \leq n) &\equivInDistribution&
(V_{n-j+1}: 1 \leq j \leq n) \label{eqn:defn:reversibleMC}
   \end{eqnarray}
   where $\equivInDistribution$ denotes equivalence in distribution.
\end{itemize}
{\bf Information Theoretic Notation:}
\begin{itemize}
\item Given two probability measures $P,Q \in \probSimplex{V}$, define the {\it Kullback-Leibler divergence} as
\begin{align}
\kldist{P}{Q} \deq \begin{cases}
\int_{\cV} \log \frac{dP}{dQ}(v) P_V(dv), & \text{if } P \ll Q\\
+ \infty, & \text{otherwise }
\end{cases}\label{eqn:defn:divergence}
\end{align}
\item Given two sets of conditional distributions $(P_{V|U=u},P'_{V|U=u}\in \probSimplex{V}: u \in \mathcal{U})$ and a
distribution $P_U \in \probSimplex{U}$, define the {\it conditional divergence} as
\beqa
\!\!\kldist{P_{V|U}}{P'_{V|U}|P_U} \!\!\triangleq \!\!\int_{\mathcal{U}} \!\! \kldist{P_{V|U=u}}{P'_{V|U=u}} P_U(du)
\eeqa
\item Consider a set of conditional distributions $(P_{V|U=u},\in \probSimplex{V}: u \in \mathcal{U})$ and a distribution $P_U \in \probSimplex{U}$. This induces a marginal distribution $P_V \in \probSimplex{V}$. The mutual information is given by
\beqa
I(P_{V|U},P_U) \triangleq I(V;U)  \triangleq \kldist{P_{V|U}}{P_{V}|P_U}. \label{eqn:defn:mutualInformation}
\eeqa
$U$ and $V$ are  independent if and only if $I(V;U)=0$.
\item The conditional mutual information is given by
\beqa
I(W;Y_2|Y_1) = \kldist{P_{W|Y_1,Y_2}}{P_{W|Y_1}|P_{Y_1,Y_2}}.\label{eqn:defn:conditionalMutualInformation}
\eeqa
\item The chain rule for mutual information is given by
\beqa
I(W;Y^n) &=& \sum_{i=1}^n I(W;Y_i|Y^{i-1}).\label{eqn:defn:chainRuleMutualInformation} \\
\Rightarrow  I(\Src^n;\Out^n)
&=& \sum_{i=1}^n I(\Src^n;\Out_i|\Out^{i-1}) \nonumber
\eeqa
\item Consider a memoryless channel $\PDMC=(Q_{Y|\State=\state} \in \probSimplex{\cOut}: \state \in \cState)$, a cost function $\statecostfn:\cState \to \reals_+$, and an upper bound $\statecostval \in \reals_+$.  Define the capacity-cost function as $\capacityCostFn$ \cite{mceliece2002theory} and its maximizing distribution $\optimalInputDMC$ as:
\begin{eqnarray}
\!\!\!\!\!\!\!\!\!\!\!\!\optimalInputDMC &\triangleq& \!\!\!\!\!\!\!\!\!\argmax_{P_\State \in \probSimplex{\State} s.t. \E[\statecostfn(\State)] \leq \statecostval} \!\!\!\!\!\!I(P_\State,\PDMC) \label{eqn:defn:OptimalInputcapacityCostFn} \\
\!\!\!\!\capacityCostFn &\triangleq& I \parenth{\optimalInputDMC,\PDMC}.\label{eqn:defn:capacityCostFn}
\end{eqnarray}
\end{itemize}
\section{\bf Problem Setup}\label{sec:problemSetup}
Throughout this discussion, we consider $4$ random processes $\Src,\State,\Out,\Est$ associated
with Borel metric spaces  $\cSrc,\cState,\cOut,\cEst$ that are coupled according to Figure~\ref{fig:causalEncDec}.
The natural time ordering for the causal construction of the four random objects through time is given by:
\[ \ldots,\Est_{i-1},\underbrace{\Src_i, \State_i, Y_i,\Est_i}_{\text{$i$th epoch}}, \Src_{i+1},\ldots \]

{\bf The input process:}\\
W is a time-homogenous Markov process such that for any $A \in \cF_{\cSrc}$:
\beqa
P_{\Src_{i+1}|\Src^{i}=\src^i,\State^i=\state^i,Y^i=y^i}(A)&=&
\!\!P_{\Src_{i+1}|\Src_i=\src_i}(A) \label{eqn:defn:tSource:a} \\
&\deq& \kernelSrc{A}{\src_i}  \label{eqn:defn:tSource}
\eeqa

{\bf The causal encoder:}\\
The {\it causal encoder} at time $i$ has causal information about the source, $\Src^i$, and causal feedback about the channel outputs, $Y^{i-1}$, to specify the next channel input, $\State_i$,
\beqa
 \state_i = \enc_i(\src^i,y^{i-1}). \label{eqn:defn:policy}
\eeqa
We define the aspect of $\enc_i \in \cEnc_i$ that maps $\Src_i$ to $\State_i$ as $\tenc_i \in \ctEnc$ where $\ctEnc$ is a space of Borel-measurable functions $f: \cSrc \to \cState$:
\beqa
\tenc_{i}(\src^{i-1},y^{i-1})(\cdot)= e_{i}\parenth{\twovec{\cdot}{w^{i-1}},y^{i-1}} \deq \tenc_{i}(\cdot) . \label{eqn:defn:tenc}
\eeqa
and we define $\cEnc_i$ to be the space of Borel-measurable functions $f: \cSrc^i \times \cOut^{i-1} \to \cState$ such that $\tenc_i \in \ctEnc$ for all $w^{i-1}$ and $y^{i-1}$.\\
{\bf The memoryless non-anticipative channel:}\\
$\State_i \in \cState$ is passed through a time-homogenous, non-anticipative, memoryless channel to produce $Y_i \in \cY$;
for any $A \in \cF_{\cY}$:
\begin{eqnarray}
P_{\Out_i|\Out^{i-1}=\out^{i-1}, \State^n=\state^n, \Src^n=\src^n}(A) =
P_{\Out|\State}(A|\state_i).
\label{eqn:defn:DMC}
\end{eqnarray}

{\bf The causal decoder:}\\
Lastly, the causal decoder at time $i$ uses causal channel outputs, $Y^i$ to specify $\Est_i \in \cEst$.
Define $\cDec_i$ as a space of Borel-measurable functions $f: \cOut^i \to \cEst$ and $\cDec = \cDec_1 \times \ldots \times \cDec_n$.  Then the causal decoder $\dec \in \cDec$ is a sequence of functions $\dec = \parenth{\dec_i:\;\; 1\leq i \leq n}$:
\beqa
\est_i = \dec_i(\out^i) \label{eqn:defn:decoder}
\eeqa

{\bf Belief update:}\\
In the above discussion on the causal decoder, we deliberately consider $\cEst$ to be general, not necessarily equal to $\cSrc$.  Indeed, as we shall see, in some cases we set $\cEst = \probSimplex{\cSrc}$ so that
the outputs of the causal decoder represent {\it beliefs} about the source at time i.   Define the beliefs $\Belief_{i|j} \in \probSimplex{\Src}$ about the source at time $i$ given the decoder's observations up until time $j \leq i$ as, for any $A \in \borel{\Src}$:
\begin{subequations} \label{eqn:defn:beliefs}
\beqa
\Belief_{i|j}(A) &\triangleq& \prob{\Src_i \in A | Y^j},
\label{eqn:defn:beliefs:a}\\
\belief_{i|j}(A) &\triangleq& P_{\Src_i|Y^j=y^j}(A),\; \label{eqn:defn:beliefs:b}
\eeqa
\end{subequations}
The beliefs can be interpreted as state variables that can be updated sequentially given new observations.  The {\it nonlinear filter} $\NLF: \probSimplex{\Src} \times \cOut \times \ctEnc \to \probSimplex{\Src}$
and {\it one step prediction update} $\OSPU: \probSimplex{\Src} \to \probSimplex{\Src}$ rules are given by \cite{elliott1995hidden}:
\begin{eqnarray}
\!\!\!\NLF(\belief,y,\tenc)(d\src) &=&
\frac{d\PDMC\parenth{\cdot|\tenc(\src)}}{d\QNLFdenom(\cdot|\belief,\tenc)}(y) \times \OSPU(\belief)(d\src)
                                                         \label{eqn:defn:nonlinearfilter} \\
\!\!\!\OSPU(\belief)(d\src) &\triangleq& \int_{\src' \in \cSrc} \kernelSrc{d\src}{\src'} \belief(d\src') \label{eqn:defn:onestepupdaterule} \\
\!\!\!\!\!\!\QNLFdenom(dy|\belief,\tenc) &\triangleq& \int_{\src' \in \cSrc}\PDMC(dy|\tenc(\src')) \OSPU(\belief)(d\src')    \label{eqn:defn:NLFdemoninator}
\end{eqnarray}
\eqref{eqn:defn:nonlinearfilter} can be interpreted
as a standard manifestation of Bayes' rule: the numerator is simply the {\it likelihood}, the denominator is a {\it normalization constant}, and the coefficient $\OSPU(\belief)$ is simply the {\it prior}.
The aforementioned two equations specify how the beliefs are sequentially updated:
\begin{lemma}[\cite{WalrandVaraiya83},\cite{elliott1995hidden}] \label{lemma:recursiveBeliefUpdate}
For any $i$ and encoder policy $e_i$ with associated $\tenc_i$ given by \eqref{eqn:defn:tenc}, the following holds:
\begin{subequations}
\begin{eqnarray}
\belief_{i|i-1} &=&  \OSPU \parenth{\belief_{i-1|i-1}} \label{eqn:lemma:recursiveBeliefUpdate:b} \\
\belief_{i|i} &=& \NLF\parenth{\belief_{i-1|i-1}, y_i, \tenc_i} \label{eqn:lemma:recursiveBeliefUpdate:a}
\end{eqnarray}
\end{subequations}
\end{lemma}
\noindent In Section~\ref{sec:MainResults}, we  demonstrate using a structural result how the beliefs arise as sufficient statistics in our main problem.  In Section~\ref{sec:directedInfo}, we  demonstrate how they additionally serve as optimal decision variables with information gain cost \eqref{eqn:intro:informationGainCost}.

{\bf Additive cost function:}\\
Denote a coordination strategy, also termed policy, as $\pi=(e_1,\ldots,e_{n},d_1,\ldots,d_{n})$ and the set of all feasible policies as $\Pi=\braces{\pi: e_i \in \cEnc_i, \; d_i \in \cDec_i}$
The causal encoder and decoder $e$ and $d$ are cooperating to achieve a common goal. The performance of their cooperation is measured in terms of an expected sum of costs over time horizon $n$ with the following structure:
\beqa
  \totalcostpi = \E_{\pi} \brackets{\sum_{i=1}^{n} \dist(\Src_i,\Est_{i-1},\Est_i) +  \alpha \statecostfn(\State_i)}
 \label{eqn:defn:JSCC:expectedCost}
\eeqa
The above expectation is taken with respect to an initial distribution $P_{\Src_0,\Est_{0}}$ where $\Est_{0}$ is assumed known to the encoder and decoder.
 We assume that the functions  $\dist$ and $\statecostfn$ along with constant $\alpha$ have the following structure:
\begin{itemize}
\item $\dist: \cSrc \times \cEst \times \cEst \to \reals_+$ is a `distortion-like' source cost, that relates the distortion between the source at time $i$ and the outputs in the vicinity of time $i$.
\item $\statecostfn: \cState \to \reals_+$ is a `power-like'  channel input cost that penalizes channel inputs that deviate significantly from nominal desired values
\item $\alpha \in \reals^+$ balances the relative importance of the two costs.
\end{itemize}
\begin{definition}
We say that a sequential encoder-decoder  pair $\pi^* \in \Pi$ is (globally) optimal if
\begin{eqnarray}
\totalcostpistar \leq \totalcostpi \;\; \text{ for all } \pi \in \Pi.  \label{eqn:defn:optimalestimatorPolicyPair}
\end{eqnarray}
\end{definition}
\section{\bf Main Structural Results}\label{sec:MainResults}
\begin{figure}[hbtp]
\centering
\begin{overpic}[width=\columnwidth]{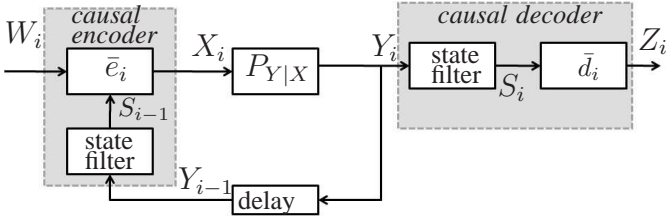}
\put(10,29){\it causal}
\put(10,26){\it encoder}
\put(15,21){\large $\benc_i$}
\put(17,15){$S_{i-1}$}
\put(12,10){state}
\put(12,7){filter}
\put(0,26){\large $\Src_i$}
\put(28,24){\large $\State_i$}
\put(36,21){\large $\PDMC$}
\put(26,4){\large $Y_{i-1}$}
\put(35,1){delay}
\put(55,24){\large $Y_i$}
\put(65,29){\it causal decoder}
\put(64,23){state}
\put(64,20){filter}
\put(86,21){$\bdec_i$}
\put(74,18){\large $S_i$}
\put(95,25){\large $\Est_i$}
\end{overpic}
\caption{Structural Result and Sufficient Statistics}
\label{fig:SeparationPrincipleGeneral}
\end{figure}

In this section, we prove that - under mild technical assumptions - for a general class of cost functions $(\dist,\statecostfn,\alpha)$ inducing an average cost specified in \eqref{eqn:defn:JSCC:expectedCost}, an optimal belief-based policy-estimator pair exists with the structure as shown in Fig \ref{fig:SeparationPrincipleGeneral}.

We first consider the basic solution approach to the problem by first demonstrating an example with two time steps.  The essence of the idea is as follows: 
\begin{subequations} \label{eqn:decentralizedControl}
\begin{eqnarray}
&& \!\!\!\!\!\!\!\!\!\!\!\!\min_{e_1,d_1,e_2,d_2} \sum_{i=1}^2 \E \brackets{\dist(\Src_i,\Est_{i-1},\Est_i) + \alpha \statecostfn(\State_i)}
  \nonumber \\ 
&=& \E \brackets{\underbrace{\min_{e_1} \E\brackets{\alpha \statecostfn(\State_1)|\tilde{E}_{1} }}_{\text{controller at stage } 0}} \nonumber \\
&+& \E \brackets{
\underbrace{\min_{e_2,d_1} \E \brackets{\dist(\Src_1,\Est_0,\Est_1) + \alpha \statecostfn(\State_2)\Big|\Est_{0},Y^1,\tilde{E}_{2},Z_1}}_{\text{controller at stage } 1}} \nonumber \\
&+& \E \brackets{
\underbrace{\min_{d_2} \E \brackets{\dist(\Src_2,\Est_1,\Est_2)\Big|\Est_{1},Y^2,Z_2}}_{\text{controller at stage } 2}}
\end{eqnarray}
\end{subequations}
Note that by grouping $(d_i,\enc_{i+1})$ in this manner, in stage $i$, $d_i: \cOut^{i} \to \cEst$ and $e_{i+1}: \cSrc^{i+1} \times \cOut^i \to \cState$ have access to a common piece of information, $\out^i$ (and thus also $\est_{i}$).  Note that $\tenc_{i+1}(\src^{i},y^{i})(\cdot) \deq \tenc_{i+1}(\cdot): \cSrc \to \cState$ is a mapping as given by \eqref{eqn:defn:tenc}, whose alphabet, $\ctEnc$, does not grow with $i$.  We secondly consider the belief $\belief_{i|i}$, which is a function of $(y^i,\enc_1,\ldots,\enc_i)$ and whose alphabet, $\probSimplex{\cSrc}$, does not grow with $i$.
The `control' action taken by the encoder (decoder) are given by $\tilde{E}_{i+1}$ ($\Est_i$) respectively.

We next demonstrate that the conditional expectations in \eqref{eqn:decentralizedControl} can be described in terms of these and other variables whose alphabets do not grow with $i$:
\begin{lemma}\label{lemma:structuralresult:fixedalphabetdynamics}
For a fixed policy $\pi=(\enc,\dec)$, define $b'=\belief_{i|i} \in \probSimplex{\cSrc}$ as in \eqref{eqn:defn:beliefs},
and $\tenc_{i+1}(\src^{i},y^{i})(\cdot) \triangleq \tenc_{i+1}(\cdot) \in \ctEnc$ as in \eqref{eqn:defn:tenc}.
Define the state space $\cS = \cEst \times \probSimplex{\Src}$ and control space $\cU = \tilde{\alphabet{E}} \times \cEst$
with $s_i \in \cS, u_i \in \cU$ given by
\begin{eqnarray}
s_i = (\est_{i-1},\belief_{i|i}),  \qquad u=(\tenc_{i+1},\est_i).
\label{eqn:defn:MDP:statesControls}
\end{eqnarray}
Then
\begin{subequations} \label{eqn:defn:iteratedExpectationCosts}
\beqa
&&\E\brackets{\dist(\Src_i,\Est_{i-1},\Est_i)  | \Est_{i-1},\Out^i, \Est_i} = \bdist(S_i,\Est_i)
    \label{eqn:defn:iteratedExpectationCosts:bdist}   \\
&&\E\brackets{ \statecostfn(\State_{i+1}) |  \Est_{i-1},\Out^i, \tilde{E}_{i+1}} = \bstatecostfn(S_i,\tEnc_{i+1})
\label{eqn:defn:iteratedExpectationCosts:bstatecostfn}
\eeqa
\end{subequations}
\end{lemma}
To emphasize, this demonstrates state ($s$) and control ($u$) variables whose alphabets do not grow with $i$, for which `distortion' and `cost' like functions solely operate on. The definitions of $\bdist$ and $\bstatecostfn$ along with the lemma's proof can be found in Appendix~\ref{appendix:proof:lemma:structuralresult:fixedalphabetdynamics}.
We now demonstrate that these state and control variables comprise a controlled Markov chain:
\begin{lemma}\label{lemma:controlledMC}
The state $s_i = (\est_{i-1},\belief_{i|i})$ and control $u=(\tenc_{i+1},\est_i)$ variables comprise a controlled Markov chain:
\begin{align}
\text{(a)}&\;\;
\totalcostpi=\E_\pi \brackets{\sum_{i=0}^{n} \bcost_i(S_i,U_i) }\label{eqn:decentralizedControl:example:b} \\
&\;\; \bcost_i(s,u) \triangleq
\begin{cases}
\alpha \bstatecostfn(s_i,\tenc_{i+1}) & i=0 \\
\bdist(s_i,z_i) + \alpha \bstatecostfn(s_i,\tenc_{i+1}) & 1\leq i \leq n-1 \\
\bdist(s_i,z_i) & i=n
\end{cases} \label{eqn:decentralizedControl:defn:tcost}\\
\text{(b)}&\;\;P_{S_{i+1}|S^i,U^i}(ds_{i+1}|s^i,u^i) = P_{S_{i+1}|S_i,U_i}(ds_{i+1}|s_i,u_i) \nonumber \\
&\;\;\;\;\;\;\;\;\;\;\;\;\;\;\;\;\;\;\;\;\;\;\;\;\;\;\;\;\;\;\;\;\;\;\;\;\;\;\; \triangleq \PControlledMC(ds_{i+1}|s_i,u_i) \nonumber
\end{align}
\end{lemma}
The proof of (a) follows directly from the law of iterated expectation and the definition \eqref{eqn:defn:iteratedExpectationCosts}.
The proof of (b) can be found in Appendix~\ref{appendix:proofLemmaControlledMC}.  Now define
the cost-to-go function at stage $n-k$ as $\valuefn_{n-k}: \cS \to \reals$.  Then for
$\valuefn_{n+1}(s) \deq 0$ and $k=1,\ldots,n$ define:
\begin{eqnarray}
\!\valuefn_{n-k}(s) \!=\!  \inf_{u \in \cU} \brackets{\bcost_{n-k}(s,u) \!+\!
\int_{s'} \!\!\valuefn_{n-k+1}(s') \PControlledMC(ds'|s,u)
}\label{eqn:decentralizedControl:DPrecursion}
\end{eqnarray}
This allows for us to state our main theorem of this section:
\begin{theorem} \label{thm:mainTheoremGeneralProblem}
If for each $s \in \cS$, the infimum in \eqref{eqn:decentralizedControl:DPrecursion} is attained and the functions $(\valuefn_k: k=0,\ldots,n)$ are universally measurable, then  there exists an optimal encoder/decoder policy $(\enc^*,\dec^*)$ pair of the form
\begin{subequations}
\begin{eqnarray}
\encst_{i+1}(\src^{i+1},y^i) &\equiv& \bencst_{i+1}(\src_{i+1},\est_{i-1},\belief_{i|i}) \label{eqn:thm:mainThmGeneralProblem:a}\\
\decst_i(y^i) &\equiv& \bdecst_i(\est_{i-1},\belief_{i|i}) \label{eqn:thm:mainThmGeneralProblem:b}
\end{eqnarray} \label{eqn:thm:mainThmGeneralProblem}
\end{subequations}
\end{theorem}
\vspace{-0.3in}
\begin{IEEEproof}
Using standard dynamic programming arguments,  \cite[Chapter 8]{bertsekas1978stochastic} we have that $\totalcostpistar \geq \E[\valuefn_0(S_0)]$.
Next, $\totalcostpistar = \E[\valuefn_0(S_0)]$ and it can be implemented by a policy of the form \eqref{eqn:thm:mainThmGeneralProblem} by a policy
that attains the infimum of \eqref{eqn:decentralizedControl:DPrecursion} for each $s$ \cite[Prop 8.6]{bertsekas1978stochastic}.
\end{IEEEproof}
The structural result in graphical form is shown in Figure~\ref{fig:SeparationPrincipleGeneral}.
Note that within the causal encoder,
the first process is a filter that computes sufficient statistics.  From here, these sufficient statistics are
given to another encoder, $\benc_i$, that uses them, along with the current source value, $\Src_i$, to specify the next channel
input $\State_i$.  Analogously, the causal decoder is comprised of first the same recursive filter that computes sufficient
statistics, followed by another decoder, $\bdec_i$, that computes $\Est_i$.

We now note that `universal measurability' \cite{bertsekas1978stochastic} is usually satisfied:
\begin{remark}
Standard technical assumptions guarantee universal measurability and that the infimum is attained; one example is as follows:
(a) $\cSrc$, $\cState$, $\cOut$, and $\cEst$ are compact Borel metric spaces, (b) $\dist$ and $\statecostfn$ are lower semi-continuous, (c) $\PDMC(dy|x)$ and $\kernelSrc{dw}{w'}$ are continuous stochastic kernels, and (d) $\ctEnc$ is an equicontinuous space of functions.
\end{remark}
We also note our result generalizes the classical result of Walrand and Varaiya \cite{WalrandVaraiya83}:
\begin{remark}
This result instantiates the result in \cite{WalrandVaraiya83} which assumes all alphabets are finite, $\statecostfn \deq 0$, and
$\dist(\src_i,\est_{i-1},\est_i) \deq \dist(\src_i,\est_i)$:  (i) because of the finite alphabets and costs, the infimum is attained in \eqref{eqn:decentralizedControl:DPrecursion}; (ii) \eqref{eqn:defn:MDP:statesControls} can be collapsed to $s_i = \belief_{i|i}$ because of the absence of $\est_{i-1}$ in the function $\dist$. Secondly, our proof technique differs from \cite[Sec. IV]{WalrandVaraiya83}
in that we replace the three-step proof technique of (\cite[Thm 1, Lemma 1, Thm 2]{WalrandVaraiya83}) - which includes two DP arguments (\cite[Thm 1,Thm 2]{WalrandVaraiya83}) - with a single DP argument.
\end{remark}
However, our emphasis is not solely on allowing general alphabets or using the cost function of a particular form - both of these have in essence been accomplished using state augmentation and dynamic programming over general spaces.  Rather, our emphasis is to carefully augment standard formulations to uncover an interplay information theory and control theory problems, as we shall see in the next section.
\section{\bf The Sequential Information Gain Cost} \label{sec:directedInfo}
In this section, we specifically consider a class of problems that are not covered in traditional causal coding/decoding frameworks
\cite{WalrandVaraiya83,mahajan2008design},\cite{bansal1988sdc},\cite[Ch. 6]{tatikonda2000control}.

Traditional problems consider cost functions of the form $\dist(\src_i,\est_i)$ and assume that either all alphabets are finite \cite{WalrandVaraiya83,mahajan2008design}, or $\cSrc=\cEst=\reals$ \cite{bansal1988sdc},\cite[Ch. 6]{tatikonda2000control}.
Motivated by the feedback communication example in Section~\ref{sec:intro} A, we now assume that $\cEst = \probSimplex{\cSrc}$, the space of possible beliefs on the source.  Secondly, we construct $\dist(\src_i,\est_{i-1},\est_i)$ to be a log-likelihood ratio that is suggestive of an `information gain'-like quantity.

The following Lemma describes the relationship between $I(\Src^n;Y^n)$ and $I(\Src^n \to Y^n)$ for our problem setup \eqref{eqn:defn:tSource}-\eqref{eqn:defn:decoder}. Because there is no feedback loop from $Y$ to the generative process of $\Src$, these two quantities are equivalent:
\begin{lemma}\label{lemma:problemSetup:mutualInfoSimplification}
For any `sufficient statistic operating' encoder $\pi \in \Pi$ satisfying \eqref{eqn:thm:mainThmGeneralProblem:a}, i.e. $\state_{i+1}=\benc_{i+1}(\src_{i+1},\est_{i-1},\belief_{i|i})$, the following holds:
\[ I(\Src^n;Y^n) = I(\Src^n \to Y^n) = \sum_{i=1}^n I(\Src_i;Y_i|Y^{i-1}).\]
\end{lemma}
The proof is in Appendix~\ref{proof:lemma:problemSetup:mutualInfoSimplification}.  Note that from the structural result in Theorem~\ref{thm:mainTheoremGeneralProblem}, there is no loss in performance for restricting attention to encoders of the form \eqref{eqn:thm:mainThmGeneralProblem:a}.  Under such encoders, note that the mutual information can be expressed as an accumulation of {\it sequential information gains},
\begin{subequations}
\begin{eqnarray}
I(\Src^n;Y^n) &=&  \sum_{i=1}^n I(\Src_i;Y_i|Y^{i-1}) \label{eqn:maximizationDirInfo:expansionMutualInfo:a} \\
              &=&  \sum_{i=1}^n \E\brackets{\kldist{B_{i|i}}{B_{i|i-1}}} \label{eqn:maximizationDirInfo:expansionMutualInfo:b} \\
              &=&  \sum_{i=1}^n \E\brackets{\log\frac{dB_{i|i}}{d\OSPU(B_{i-1|i-1})}(\Src_i)} \label{eqn:maximizationDirInfo:expansionMutualInfo:c}
\end{eqnarray} \label{eqn:maximizationDirInfo:expansionMutualInfo}
\end{subequations}
where \eqref{eqn:maximizationDirInfo:expansionMutualInfo:a} follows from Lemma~\ref{lemma:problemSetup:mutualInfoSimplification};
\eqref{eqn:maximizationDirInfo:expansionMutualInfo:b} follows from \eqref{eqn:defn:conditionalMutualInformation} and \eqref{eqn:defn:beliefs}; and
\eqref{eqn:maximizationDirInfo:expansionMutualInfo:c} follows from \eqref{eqn:defn:divergence} and \eqref{eqn:defn:onestepupdaterule}.

One may consider finding encoder policies $\enc$ in order to maximize $I(\Src^n;Y^n)$, using a state space approach over the space of beliefs. \cite{ColemanISIT09} formulated a stochastic control problem where $\Belief_{i-1|i-1}$ is a state variable and the only decision variable is the causal encoder's strategy - the decoder did not specify a decision variable $\Est_i$.  There, it was shown that when $\Src$ is uniformly distributed on $\cSrc=[0,1]$ and $(\Src_i=\Src:\;i \geq 1)$, the causal encoder given by the posterior matching scheme by Shayevitz and Feder \cite{shayevitz2009optimal} is an optimal solution to a control problem where costs are related to conditional mutual informations \eqref{eqn:maximizationDirInfo:expansionMutualInfo:b}.  Anand and Kumar have recently considered a related problem
where \cite{AnandKumarCDC10} where $(\Src_i\;i \geq 1)$ is a general Markov process over a {\it finite} alphabet, and the cost function is a conditional mutual information.  There, also, however, the decoder did not specify a decision variable $\Est_i$.

In this setting, we do not treat $\Belief_{i|i}$ as a state variable; rather, we first consider a problem in the framework of causal coding/decoding, where the decoder's decision variable $\Est_i$ can be {\it any} possible belief: $\cEst = \probSimplex{\cSrc}$.  In order to reward larger information gains, we define an appropriate cost pertaining to the negative logarithm of the Radon-Nikodym derivative evaluated at $\src_i$ that is inspired by the expansion of mutual information given in \eqref{eqn:maximizationDirInfo:expansionMutualInfo:c}:
  \begin{eqnarray}
  \dist(\src_i,\est_{i-1},\est_i) = \begin{cases}
                                         -\log \frac{d\est_i}{d \OSPU(\est_{i-1})}(\src_i)  & \text{if } \est_i \ll \OSPU(\est_{i-1})\\
                                        \infty, & \text{otherwise}
                                     \end{cases} \label{eqn:maximizationDirInfo:CostFn:costfn}
  \end{eqnarray}
The reason we assign $\dist=\infty$ when $\est_i \ll \OSPU(\est_{i-1})$ is because under any reasonable belief-setting strategy,
if the belief about $\Src_i$ given $\Out^{i-1}$ - the one-step prediction update \eqref{eqn:defn:onestepupdaterule} given by $\OSPU(\est_{i-1})$ -
assigns zero probability mass to $A \in \cB(\cSrc)$, then the belief about $\Src_i$ given $\Out^i$ - which is given by $\Est_i$ should also.

We emphasize here that the beliefs on the source are themselves decision variables, which are what the causal decoder must specify.  This viewpoint has been used within the sequential prediction literature \cite{merhav2002universal} and statistical signal processing \cite{aiyer2005lloyd} but appears to not have been used as frequently in the literature that attempts to draw synergies between information theory and control.

Define $Z_{0}(A) =  \prob{\Src_0 \in A}$, the distribution on $\Src_0$.  We now state the following useful Lemma that decomposes the cost into the state and distortion parts, that act on different aspects of the control input:
\begin{lemma} \label{lemma:loglikelihoodratiocost}
Under the information gain criterion \eqref{eqn:maximizationDirInfo:CostFn:costfn}, for a state variable
$s_i = (\est,\belief)$ and control variable $u_i=(\tenc,\est')$,
\begin{eqnarray}
\!\!&&\!\!\!\!\!\!\!\!\bstatecostfn(s_i,\tenc) =  \int_{\src \in \cSrc} \statecostfn \parenth{\tenc\parenth{\src}} \OSPU(\belief)(d\src)  \label{eqn:lemma:loglikelihoodratiocost:b}\\
\!\!&&\!\!\!\!\bdist(s_i,\est') = \!\!\begin{cases}
                       \kldist{\belief}{\est'}\!\!-\!\!\kldist{\belief}{\OSPU(\est)}  & \!\!    \belief\!\ll\!\est'\!\ll \OSPU(\est) \\
                        \infty    & \text{otherwise}
                   \end{cases}  \label{eqn:lemma:loglikelihoodratiocost:c}
\end{eqnarray}
\end{lemma}
The proof can be found in Appendix~\ref{appendix:proof:lemma:loglikelihoodratiocost}.

With this, we state the main theorem of our section.  It says that when treating beliefs as decision variables, under the
information gain criterion \eqref{eqn:maximizationDirInfo:CostFn:costfn},
the optimal decision rule for the decoder is to select its  belief about $\Src_i$ to be $z_i = \belief_{i|i}$, and the optimal decision rule for the encoder is to maximize mutual information subject to a cost on channel inputs:
\begin{theorem}\label{thm:optimalDecoderMaximizationDirInfo}
Under cost criterion \eqref{eqn:lemma:loglikelihoodratiocost:c}, there exists an optimal encoder/decoder policy $(\enc^*,\dec^*)$ pair of the form
\begin{eqnarray}
\encst_{i+1}(\src^{i+1},y^i) &\equiv& \bencst_{i+1}(\src_{i+1},\belief_{i|i}) \label{eqn:lemma:structuralResultSimplifiedInformationCost:a}\\
\decst_i(y^i) &\equiv& \bdecst_i(\belief_{i|i})=\belief_{i|i} \label{eqn:lemma:structuralResultSimplifiedInformationCost:b}
\end{eqnarray}
where $\belief_{i|i}= \NLF(\belief_{i-1|i-1},y_i,\bencst_i(\cdot,\belief_{i-1|i-1}))$ and the optimal cost is given by
\begin{eqnarray}
\totalcostpistar &=&\min_{\enc \in \cEnc} -I(\Src^n; Y^n) + \alpha   \E_{\enc}  \brackets{\sum_{i=1}^{n}\statecostfn(\State_i)}.
\end{eqnarray}
\end{theorem}

\begin{figure}[hbtp]
\centering
\begin{overpic}[width=\columnwidth]{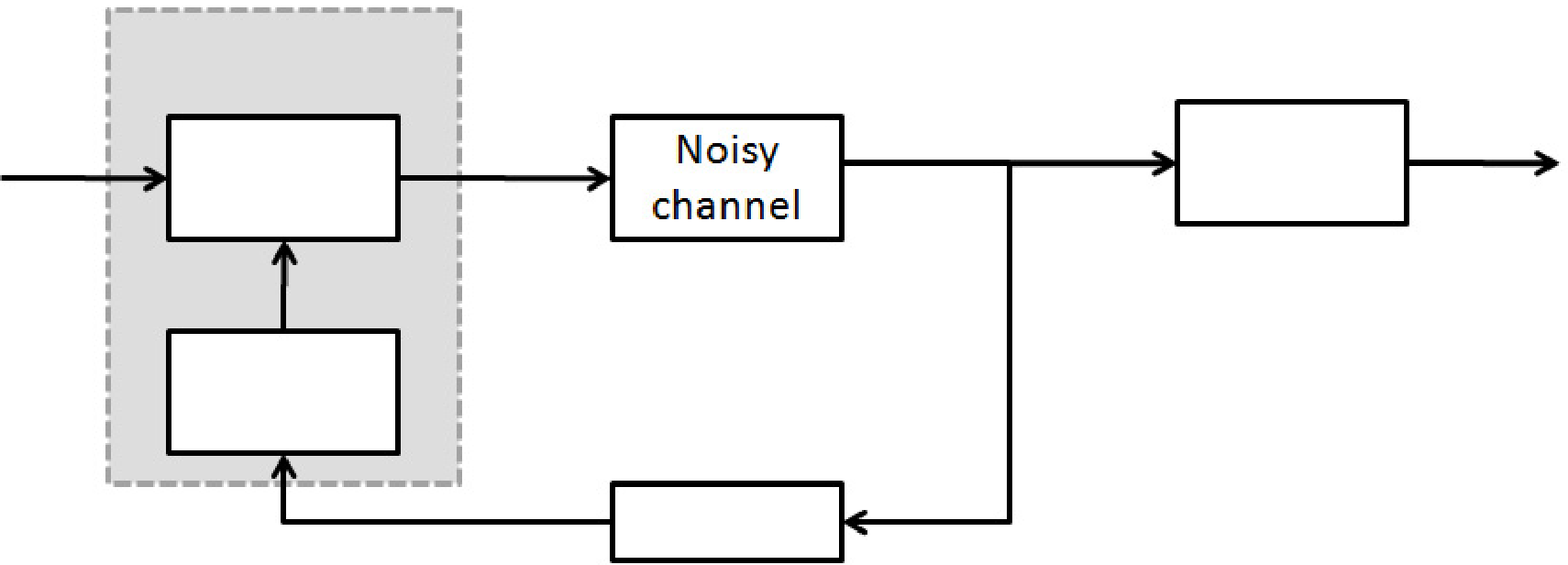}
\put(12,32){\it causal}
\put(12,29){\it encoder}
\put(11,11){\small nonlinear}
\put(13,8.5){\small filter}
\put(16,23){\large $\bencst_{i}$}
\put(1,26){\large $\Src_i$}
\put(31,26){\large $\State_i$}
\put(57,27){\large $\Out_i$}
\put(90,27){\large $\Belief_{i|i}$}
\put(18,17){\small $\Belief_{i-1|i-1}$}
\put(75,26){\small nonlinear}
\put(79,23){\small filter}
\put(42,2){delay}
\end{overpic}
\caption{Simplified structural result with $\cEst = \probSimplex{\cSrc}$ and sequential information gain cost \eqref{eqn:lemma:loglikelihoodratiocost:c}.}
\label{fig:informationGainCostStructuralResult}
\end{figure}
The proof can be found in Appendix~\ref{sec:proof:lemma:maximizationDirInfo}.

\begin{remark}
The proof of Theorem~\ref{thm:optimalDecoderMaximizationDirInfo} (Appendix~\ref{sec:proof:lemma:maximizationDirInfo}) uses dynamic programming, the second law of thermodynamics for Markov chains, and exploits how the divergence acts as a Lyapunov function for the stability of the nonlinear filter.  This further demonstrates an interesting relationship between information theory and thermodynamics \cite{mitter2004variational,mitter2005information}.  This idea of using beliefs as decision variables where the posterior belief is optimal has been used in sequential prediction \cite{merhav2002universal} and in variational approaches to nonlinear estimation \cite{mitter2004variational}, but within the constext of causal coding decoding problems, this is to the best of our knowledge, new.
\end{remark}

We will demonstrate in the examples section how this relates to the hidden Markov model and the nonlinear filter as well as the Posterior Matching Scheme  \cite{shayevitz2009optimal} for communication of a message point over a noisy channel with feedback.
\section{\bf Inverse Optimal Control with Stationary Markov Coordination Strategies} \label{sec:SCmatching}
In the last section, we demonstrated that for a specific ``information-gain'' related cost function \eqref{eqn:maximizationDirInfo:CostFn:costfn}, there existed an optimal encoder policy of the form $\State_i = \benc^*_i(\Src_i,\Est_{i-1})$ and decoder policy of the form $\Est_i = \Belief_{i|i} = \NLF\parenth{\Belief_{i-1|i-1},Y_i,\benc^*_i(\cdot,\Belief_{i-1|i-1})} = \bdec(Z_{i-1},Y_i)$.

\begin{figure}[hbtp]
\centering
\begin{overpic}[width=\columnwidth]{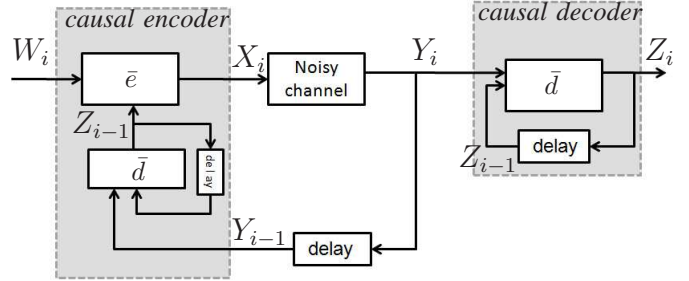}
\put(8,38){\it causal encoder}
\put(17,29){$\encSC$}
\put(9,22){\large $\Est_{i-1}$}
\put(18,15){$\decSC$}
\put(0,33){\large $\Src_i$}
\put(33,32){\large $\State_i$}
\put(33,6){\large $Y_{i-1}$}
\put(60,33){\large $Y_i$}
\put(70,39){\it causal decoder}
\put(80,28){$\decSC$}
\put(67,17){\large $\Est_{i-1}$}
\put(95,33){\large $\Est_i$}
\end{overpic}
\caption{A Stationary Markov coordination strategy}
\label{fig:SCmatching:StationaryMarkovPolicies}
\end{figure}

In light of this, we now consider a general Markov process $\Src \in \cSrc$ and $\Est \in \cEst$ where $\cEst$ need not be $\probSimplex{\cSrc}$ and fix
the coordination strategy $\bpi$ to be {\it stationary Markov} (SM), meaning that for fixed functions
$\encSC: \cSrc \times \cEst \to \cState$ and $\decSC: \cEst \times \cOut \to \cEst$, the following holds:
\begin{subequations} \label{eqn:SCmatching:SMpolicies}
\begin{eqnarray}
\state_i &=& \encSC(\src_i,\est_{i-1}) \label{eqn:SCmatching:SMpolicies:enc} \\
\est_i &=& \decSC(\est_{i-1},y_i) \label{eqn:SCmatching:SMpolicies:dec}.
\end{eqnarray}
\end{subequations}
See Figure~\ref{fig:SCmatching:StationaryMarkovPolicies}.
\eqref{eqn:SCmatching:SMpolicies:dec} is sometimes termed a decoder of `finite-memory' \cite{WalrandVaraiya83},\cite{mahajan2008sequential}.
Since the encoder and decoder both utilize $\est_{i-1}$, this can be also interpreted as a collection of `equi-memory' encoders and decoders \cite[Definition 6.3.2]{tatikonda2000control}.
For a fixed SM coordination strategy $\bpi$, we compare it against arbitrary policies of the form $\pi=(e_1,\ldots,e_{n},d_1,\ldots,d_{n})$
where $e_i: \cSrc^i \times \cOut^{i-1} \to \cState$ is given in \eqref{eqn:defn:policy}, and
$d_i: \cOut^i \to \cEst$ is given in \eqref{eqn:defn:decoder}.  Here we identify the structure of cost functions $\dist(\src_i,\est_{i-1},\est_i)$ under which $\bpi$ is globally optimal.
\begin{definition} \label{defn:inverseControlOptimality}
A coordination strategy $\pi$ is inverse-control optimal for
a source-channel pair $(P_{\Src^n},\PDMC)$ if $\totalcostpi \leq \totalcostpiprime$ for all $\pi' \in \Pi$ for some $\alpha \geq 0$,
 $\dist: \cSrc \times \cEst \times \cEst \to \reals$ and $\statecostfn: \cState \to \reals_+$ in \eqref{eqn:defn:JSCC:expectedCost}.
\end{definition}

To develop high-level conditions under which $\pi$ is indeed inverse-control optimal, we first develop some preliminary machinery.
Fix a specific $\dist$ and $\statecostfn$ function.  For any coordination strategy $\pi$, define $\PEstnGivenSrcnPolicy$ as the conditional distribution induced statistical law under $\pi$ and also define:
\beqa
\distval_\pi &\triangleq& \frac{1}{n}\E_{\pi} \brackets{\sum_{i=1}^{n} \dist(\Src_i,\Est_{i-1},\Est_i)}  \label{eqn:SCmatching:defn:D} \\
\statecostval_\pi &\triangleq&  \frac{1}{n} \E_{\pi} \brackets{\sum_{i=1}^{n} \statecostfn(\State_i)} \label{eqn:SCmatching:defn:a}
\eeqa
Define the rate-distortion function for $P_{\Src^n}$ and $\rho$ as \cite{CoverThomas06}
\beqa
\!\!\!\!\!\!\!\!\rateDistortionFn \!\!\!\!\!\!&\triangleq\!\!\!\!\!\!\!\!\!\!\!\!\!\!\!\!\!\!\!\!\!\!\!\!\!\!\!\!\!\!\!\!&
\!\!\!\!\!\!\!\!\min_{\PEstnGivenSrcn: \E\brackets{ \frac{1}{n}\sum_{i=1}^{n}\dist(\Src_i,\Est_{i-1},\Est_i)} \leq \distval }
\frac{1}{n}I(P_{\Src^n},\PEstnGivenSrcn) \label{eqn:defn:RateDistFn}
\eeqa
and denote $\optimalChannelRD$ as the minimizer in \eqref{eqn:defn:RateDistFn}.
We now state the following standard lemma:
\begin{lemma}\label{lemma:RlessthanC}
Fix a $\pi$,$P_{\Src^n}$, $\PDMC$, $\dist$, and $\statecostfn$.  Then
\begin{subequations} \label{eqn:RlessthanC}
\beqa
\rateDistortionFnPi &\leq& \frac{1}{n} I\parenth{P_{\Src^n},\PEstnGivenSrcnPolicy} \label{eqn:RlessthanC:aa} \\
&\leq& \frac{1}{n}I\parenth{P_{\Src^n},\POutnGivenSrcnPolicy} \label{eqn:RlessthanC:bb} \\
&\leq& \frac{1}{n} \sum_{i=1}^n I(P^\pi_{\State_i},\PDMC)\label{eqn:RlessthanC:cc} \\
&\leq& \capacityCostFnpi  \label{eqn:RlessthanC:dd}
\eeqa
\end{subequations}
where equality holds if and only if:
\bitm
\item (a) $\PEstnGivenSrcnPolicy = \optimalChannelRDPi$
\item (b) $I\parenth{P_{\Src^n},\PEstnGivenSrcnPolicy} =I\parenth{P_{\Src^n},\POutnGivenSrcnPolicy}$
\item (c) $I(Y_i;Y^{i-1})=0$ for each $i$
\item (d) $P^\pi_{\State_i} = \optimalInputDMCpi$ for each $i$
\eitm
\end{lemma}
The proof is standard \cite{CoverThomas06} but for the sake of completeness, we include it in Appendix~\ref{sec:proof:lemma:RlessthanC}.
 This leads to an intermediate sufficient condition for inverse control optimality that
applies for {\it any} $\pi \in \Pi$ (e.g. $\pi$ need not be stationary-Markov):
\begin{lemma}\label{lemma:SCmatching:RDequalsCapacitysufficientForGlobalOptimality}
If a policy $\pi \in \Pi$ results in \eqref{eqn:RlessthanC} holding with equality, then it is inverse control optimal.
\end{lemma}
\begin{IEEEproof}
Note that $\totalcostpi=\distval_\pi+\alpha \statecostval_\pi = \left<(1,\alpha),(\distval_\pi,\statecostval_\pi)\right>$.
Define $\regionDistStateCost=\{ (\distval_{\pi'},\statecostval_{\pi'}: \pi'): \pi' \in \Pi \}$.
Define $\randomizedPolicies$ to be the set of randomized policies in $\Pi$.
Note that any $\pi' \in \randomizedPolicies$ still induces a conditional distribution $\PEstnGivenSrcnPolicyprime$ and thus an induced
$\distval_{\pi'}$ and $\statecostval_{\pi'}$ so that we may define
$\regionDistStateCostRandomized=\{ (\distval_{\pi'},\statecostval_{\pi'}: \pi'): \pi' \in \randomizedPolicies\}$.
Clearly, $\regionDistStateCost \subset \regionDistStateCostRandomized$, and secondly, $\regionDistStateCostRandomized$ is convex.
Next, note that if \eqref{eqn:RlessthanC} holds with equality for some $\pi \in \Pi$, then
$\distval_{\pi} \leq \distval_{\pi'}$ for for any $\pi' \in \randomizedPolicies$ for which $\statecostval_{\pi'} \leq \statecostval_\pi$
from the definition of $R_n(\dist,P_{\Src^n},\distval_\pi)$ in \eqref{eqn:defn:RateDistFn} and $\capacityCostFnpi$ in \eqref{eqn:defn:OptimalInputcapacityCostFn}
(See also \cite[Lemma 1]{GastparRimoldiVertterli03}).
Thus $(\distval_\pi,\statecostval_\pi)$ is a boundary point of $\regionDistStateCostRandomized$.
Therefore there exists a supporting hyperplane parametrized by $\alpha \geq 0$ that intersects $(\distval_{\pi},\statecostval_{\pi})$:
\[ \totalcostpi=\left<(1,\alpha),(\distval_{\pi},\statecostval_{\pi})\right> \leq \left<(1,\alpha),(\distval_{\pi'},\statecostval_{\pi'})\right>=
\totalcostpiprime.\]
for all $\pi' \in \randomizedPolicies \supset \Pi$  \cite{bertsekas1999nonlinear}, where $\langle \cdot, \cdot \rangle$ denotes inner product.
\end{IEEEproof}
We now consider SM policies for which condition (c) in Lemma~\ref{lemma:RlessthanC}
holds, and demonstrate a stationary Markov relationships between $(\Src^n,\Est^n)$ random variables:
\begin{lemma} \label{lemma:SC:markovDecoder}
If an SM coordination strategy $\bpi$ \eqref{eqn:SCmatching:SMpolicies} induces the channel outputs $(Y_i: 1 \leq i \leq n)$ being i.i.d., then
\begin{subequations}
\begin{eqnarray} \label{eqn:lemma:SC:markovDecoder}
 P^{\bpi}_{\Est_i|\Est^{i-1}=\est^{i-1},\Src^n = \src^n} (d\est_i) \deq
 Q^{\bpi}_{\Est'|\Est,\Src'}(d\est_i|\est_{i-1},\src_i)
 \label{eqn:lemma:SC:markovDecoder1}\\
 P^{\bpi}_{\Est_i|\Est^{i-1}=\est^{i-1}}(d\est_i) \deq
 Q^{\bpi}_{\Est'|\Est}(d\est_i|\est_{i-1}) 
 \label{eqn:lemma:SC:markovDecoder2}
\end{eqnarray}
\end{subequations}
\end{lemma}
The proof of this can be found in Appendix~\ref{appendix:proof:lemma:SC:markovDecoder} and exploits the equivalence between
a random process being a time-homogeneous Markov chain and it being represented as an iterated function system \cite{ykifer}.
With this, we can now state the main theorem of this section:
\begin{theorem}\label{thm:SCmatching}
If under a SM policy $\bpi$, $(Y_i: 1 \leq i \leq n)$ are i.i.d. and
$I\parenth{P_{\Src^n},\bPEstnGivenSrcnPolicy} =I\parenth{P_{\Src^n},\bPOutnGivenSrcnPolicy}$,
then $\bpi$ is inverse control optimal with $\dist$ and $\statecostfn$ given by:
\begin{subequations}
\begin{eqnarray}
&& \!\!\!\!\!\!\!\!\!\!\!\! \statecostfn(\state) \propto_+  \kldist{\PDMCstate}{P^{\bpi}_{\Out}} 
\label{eqn:thm:SCmatching:a}\\
&& \!\!\!\!\!\!\!\!\!\!\!\!\!\!\!\!\!\!\!\!\!\!\!\!\!\!\!
\dist(\src_i,\est_{i-1},\est_i) \propto_+ -  \log
\frac{dQ^{\bpi}_{\Est'|\Est,\Src'}(\cdot|\est_{i-1},\src_i)}{dQ^{\bpi}_{\Est'|\Est}(\cdot|\est_{i-1})}
\parenth{\est_i} 
\label{eqn:thm:SCmatching:b}
\end{eqnarray}
\end{subequations}
where $\propto_+$ denotes proportional to with a positive constant.
\end{theorem}
\begin{IEEEproof}
Note that it suffices to show that  \eqref{eqn:RlessthanC} holds with equality and then invoke Lemma~\ref{lemma:SCmatching:RDequalsCapacitysufficientForGlobalOptimality}.
First note that from the theorem definition, clearly conditions (b) in Lemma~\ref{lemma:RlessthanC} holds with equality.
Since $(Y_i: i=1,\ldots,n)$ are i.i.d. and since the channel is memoryless, it follows that the $(\State_i: i=1,\ldots,n)$ are identically distributed
and so condition (c) in Lemma~\ref{lemma:RlessthanC} holds with equality.
Thus the two remaining conditions are to show that conditions (a) and (d) in Lemma~\ref{lemma:RlessthanC} hold with equality.

The variational equations for an optimal solution to \eqref{eqn:defn:RateDistFn} state that a necessary and sufficient condition
for $\PEstnGivenSrcnPolicy = \optimalChannelRDPi$ is the following relationship \cite{csiszar1974extremum}:
\beqa
\frac{dP_{\Est^n|\Src^n=\src^n}}{dP_{\Est^n}}(\est^n)= \normalizationConstantExpFam(\src^n) e^{-\alpha_2 \parenth{\sum_{i=1}^{n} \dist(\src_i,\est_{i-1},\est_i)} } \label{eqn:proof:SCmainthm:f}
\eeqa
For our case, note that
\beqa
\!\!\!\!\!\!\log\!\!\frac{dP^{\bpi}_{\Est^n|\Src^n=\src^n}}{dP^{\bpi}_{\Est^n}}(\est^n) \!\!\!\!\!\!&=&
\!\!\!\!\!\!\!\sum_{i=1}^n \log  \frac{dP^{\bpi}_{\Est_i|\Est^{i-1}=\est^{i-1},\Src^n=\src^n}}{dP^{\bpi}_{\Est_i|\Est^{i-1}=\est^{i-1}}}(\est_i) \nonumber\\
\!\!\!\!\!\!\!\!\!\!\!\!&=&\!\!\!\!\!\!
\sum_{i=1}^n \log\frac{d Q^{\bpi}_{\Est'|\Est,\Src'}(\cdot|\est_{i-1},\src_i)}{d Q^{\bpi}_{\Est'|\Est}(\cdot|\est_{i-1})}(\est_i)
\label{eqn:proof:SCmainthm:g}
\eeqa
where \eqref{eqn:proof:SCmainthm:g} follows from Lemma~\ref{lemma:SC:markovDecoder}.
Thus we see that with $\dist$  given by \eqref{eqn:thm:SCmatching:b},
from \eqref{eqn:proof:SCmainthm:f} we see that condition (a) of Lemma~\ref{eqn:RlessthanC} holds with equality.

Lastly, condition (d) of Lemma~\ref{eqn:RlessthanC} holds with equality if and only if each $P^{\bpi}_{\State_i} \sim \optimalInputDMCpi$.  Variational arguments
\cite[Lemma 3]{GastparRimoldiVertterli03},\cite[p. 147]{csiszar1982information} demonstrate that this criterion is equivalent to
\eqref{eqn:thm:SCmatching:a}.
\end{IEEEproof}

\begin{corollary} \label{cor:SCmatching:decoderInvertible}
if the function $\decSC(\est_{i-1},\cdot) \triangleq \decSC_{\est_{i-1}}(\cdot)$ in \eqref{eqn:SCmatching:SMpolicies:dec} is invertible, then
condition \eqref{eqn:thm:SCmatching:b} in Theorem~\ref{thm:SCmatching} becomes
\[ \dist(\src_i,\est_{i-1},\est_i) \propto_+  \log
\frac{dP_{Y|\State=\encSC(\src_i,\est_{i-1})}}{dP^{\bpi}_Y}\parenth{\decSC^{-1}_{\est_{i-1}}(\est_i)}
\]
\end{corollary}

We now first relate this to `source-channel' matching and how it is in some sense it is also `natural' within the inverse control
framework to have a distortion function of the form $\dist(\src_i,\est_{i-1},\est_i)$:
\begin{remark}
The problem setup leading up to Theorem~\ref{thm:SCmatching} is philosophically inspired by  the
`source channel matching' work \cite{GastparRimoldiVertterli03,GastparPhDThesis2002} -
but here, we are relating this to a causal coding-decoding problem with {\bf causal encoder feedback},
 and {\bf time-invariant additive costs}.
These two properties appear to make the distortion function $\dist(\src_i,\est_{i-1},\est_i)$ - as compared to $\dist(\src_i,\est_i)$ - crucially
important: note the time-invariant statistical relationships in Lemma~\ref{lemma:SC:markovDecoder} and how they
relate to $\dist(\src_i,\est_{i-1},\est_i)$ in \eqref{eqn:proof:SCmainthm:g} and Corollary~\ref{cor:SCmatching:decoderInvertible}
pertaining to condition (a) in Lemma~\ref{lemma:RlessthanC}.  With this more general $\dist(\src_i,\est_{i-1},\est_i)$  framework,
we can characterize time-invariant cost functions for problems where neither $\Src$ nor $\Est$ are stationary (see the
linear quadratic Gaussian decentralized control  and M/M/1 queue examples in Section~\ref{sec:examples}).
\end{remark}

Next, we demonstrate how time-reversibility of an appropriately defined Markov chain can serve for Theorem~\ref{thm:SCmatching} - and thus inverse control optimality -
to hold for SM coordination strategies.

\subsection{Time-Reversibility of Markov Chains and Inverse Optimal Control} \label{sec:timereversibility}
Time reversibility plays an important role in disciplines concerning dynamical systems, e.g. in physics (conservation laws); statistical mechanics (in terms of equilibrium states); stochastic processes (e.g. queuing networks \cite{kelly1979reversibility,Gallager96Discrete}  and
convergence rates of Markov chains \cite[ch 20]{meyn2009markov}); and biology (e.g. trans paths in ion channels \cite{alvarez2006equivalence}).
However, its use in acting as a sufficient condition to saturate fundamental information-theoretic limits
appears to be somewhat limited.  One special noteworthy exception is how Mitter and colleagues have related Markov chain reversibility to
rate of entropy production in non-equilibrium thermodynamics \cite{propp1985thermodynamic},\cite[Remark 2.1]{mitter2005information}.

In queuing systems, the celebrated Burke's theorem \cite{Gallager96Discrete,hsuBurke1976behavior} uses Markov chain time reversibility to show that, in a certain stochastic dynamical system - an M/M/1 queue in steady-state - {\it the state of the system (queue) at  time $i$ is {\it statistically independent} of all outputs (departures) before time $i$}.
This observation has been used in proving achievability theorems using for queuing timing channels \cite{AnantharamVerdu96,SundaresanVerdu06,ColemanITWVolos09},
and for implementing recursive schemes that maximize mutual information according to the converse to the channel coding theorem with feedback \cite{horstein1963stu,schalkwijk1966csaOne,schalkwijk1966csaTwo,shayevitz2009optimal}.
Here we demonstrate how time reversibility of Markov chains provides a sufficient condition for inverse optimal control with SM coordination strategies.

We first note that from \eqref{eqn:defn:tSource}, $\Src$ is a time-homogenous Markov chain and so it can be represented as an iterated function system \cite{ykifer}:
\beqa
 \Src_i =\MarkovChainIFS(\tSrc_i,\Src_{i-1}) \deq \MarkovChainIFS_{\tSrc_i}(\Src_{i-1}), \;\;\;i \geq 1
\eeqa
where $\tSrc_i$ are i.i.d.  To ensure, \eqref{eqn:defn:tSource:a}, we assume
\begin{eqnarray}
I(\tSrc_i;\State^{i-1},Y^{i-1})=0. \label{eqn:assumption:timeReversibility:tsrc}
\end{eqnarray}

We next suppose the structure of the SM coordination strategy is such that the following assumption holds
\begin{definition}\label{assumption:timeReversibility}
We say that the SM coordination strategy $\bpi=(\benc,\bdec)$ elicits `\IRC' if
$I\parenth{P_{\Src^n},\PEstnGivenSrcnPolicy} =I\parenth{P_{\Src^n},\POutnGivenSrcnPolicy}$ and the statistical
dynamics can be described as
\beqa
\State_i &=& f(\tState_{i-1},\tSrc_i) \deq f_{\tState_{i-1}}(\tSrc_i) \label{eqn:SCmatching:timeReversibility:c}\\
\tState_i &=& g(\State_i,\Out_i) \deq g_{\State_i}(\Out_i) \label{eqn:SCmatching:timeReversibility:d}
\eeqa
where $f_{\tState_{i-1}}: \cSrc \to \cState$ and $g_{\State_i}: \cOut \to \cState$ are $\mathbb{P}$-a.s. invertible functions for $i=1,\ldots,n$.
\end{definition}
Note that $\tstate$ in condition \eqref{eqn:SCmatching:timeReversibility:d} is the update to the state {\it after} the output of the channel is taken into consideration and {\it before} the source $\src$ is updated to the state.

\begin{figure}[hbtp]
\centering
\begin{overpic}[width=0.7\columnwidth]{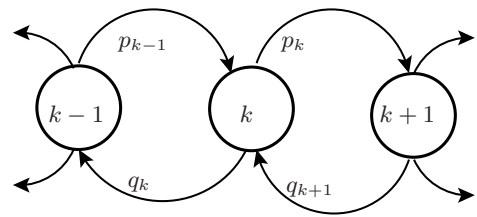}
\put(23,36){\small $p_{k-1}$}
\put(25,6){\small $q_k$}
\put(59,5){\small $q_{k+1}$}
\put(58,36){\small $p_k$}
\put(49,20){\small $k$}
\put(8,20){\small $k-1$}
\put(79,20){\small $k+1$}
\end{overpic}
\vspace{-0.1in}
\caption{A birth-death Markov chain $\State$.}
\label{fig:birthDeathChain}
\end{figure}

We now show an example that is  related to feedback communication with posterior matching \cite{shayevitz2009optimal}
\begin{example}\label{example:inverse-reversible-compatible:a}
Let $\cSrc = \cState = [0,1]$ and $\cEst = \probSimplex{\cSrc}$.  Then the `posterior matching' scheme \cite{shayevitz2009optimal} given by
\begin{subequations}\label{eqn:inverse-reversible-compatible:a}
\beqa
\tSrc_0 &\sim& \text{unif}[0,1], \;\;\; \tSrc_i \deq 1,  i \geq 1 \label{eqn:inverse-reversible-compatible:a:a}\\
\Src_0 &=& \tSrc_0 \;\;\; \Src_i=\Src_{i-1}, \; i \geq 1 \label{eqn:inverse-reversible-compatible:a:b}\\
\Est_i &=& \Belief_{i|i}  \label{eqn:inverse-reversible-compatible:a:c} \\
\State_0 &=& 0, \; \State_i = \tState_{i-1}, i \geq 1 \label{eqn:inverse-reversible-compatible:a:d} \\
\!\!\!\!\tState_i &=& \Est_{i-1}([0,\Src_i]) =  F_{X|Y}(\State_i|\Out_i) \label{eqn:inverse-reversible-compatible:a:e}
\eeqa
\end{subequations}
elicits~\IRC.
Note that this clearly is a SM coordination policy because for the decoder $\Belief_{i|i}$ is given by the nonlinear filter,
and for the encoder, this follows from the first equality in \eqref{eqn:inverse-reversible-compatible:a:e}.
To verify that the last equality in \eqref{eqn:inverse-reversible-compatible:a:e} holds, see \cite[Corollary 6]{shayevitz2009optimal}.
\end{example}
Next, we consider a SM coordination strategy that results in a {\it birth-death} Markov chain \cite{kelly1979reversibility,Gallager96Discrete}
where $\State$ can increase or decrease by at most $1$ from time $i$ to time $i+1$ (see Figure~\ref{fig:birthDeathChain}):
\begin{example}\label{example:inverse-reversible-compatible:b}
Let $\tilde{\cSrc}=\cSrc=\cState=\cOut=\cEst=\mathbb{F}$ for some field.  Then the following SM coordination strategy
\begin{subequations}\label{eqn:inverse-reversible-compatible:b}
\beqa
\Src_i &=& \Src_{i-1} + \tSrc_i \\
\Est_i &=& \Est_{i-1} + \Out_i \\
\tState_i &=& \State_i - \Out_i \\
\State_i &=& \Src_i - \Est_{i-1} = \tState_{i-1} + \tSrc_{i}
\eeqa
\end{subequations}
elicits~\IRC.  This follows from inspection.
\end{example}
See Figure~\ref{fig:birthDeathChain}.
\begin{lemma}\label{lemma:SCmatching:Xbirthdeathchain}
Consider an SM coordination strategy with dynamics given by \eqref{eqn:inverse-reversible-compatible:b}
where $\prob{\tSrc_i \in \{0,1\}}=\prob{Y_i \in \{0,1\}}$.  If $\State$ is a time-reversible Markov chain, then $(\State,\tState)$ is jointly a time-reversible Markov chain,
$Y_i$ are i.i.d., and $\bpi$ is inverse-control optimal.
\end{lemma}
The proof that $(\State,\tState)$ is jointly stationary and that $Y_i$ are i.i.d. is a generalization  \cite{martin2009batch} of the discrete-time Burke's theorem \cite{hsuBurke1976behavior}
from queuing theory.  From there, the Lemma follows by simply invoking Definition~\ref{defn:inverseControlOptimality} and Theorem~\ref{thm:SCmatching}.
As of now, time-reversibility is only wed to inverse control optimality in the algebraic setup of Example~\ref{example:inverse-reversible-compatible:b}.
This alludes to there being a more general statement under which time-reversibility implies inverse optimality:
\begin{lemma}\label{lemma:timeReversibility}
If an SM coordination strategy $\bpi$ elicits~\IRC and
$(\State,\tState)$ is jointly a time-reversible Markov chain, then $\bpi$ is inverse-control optimal.
\end{lemma}

\begin{IEEEproof}
We now develop a generalization to the discrete-time proof of Burke's theorem \cite{hsuBurke1976behavior,martin2009batch} from queuing theory.
Note that from Assumption~\ref{assumption:timeReversibility} that
\begin{eqnarray}
\tSrc_i &=& f_{\tState_{i-1}}^{-1}(\State_i) \label{eqn:SCmatching:timeReversibility:e}\\
\Out_i &=& g_{\State_i}^{-1} (\tState_i) \label{eqn:SCmatching:timeReversibility:f}
\end{eqnarray}
Now note from the time-reversibility assumption, we have that
\beqa
\!\!\parenth{\State_1,\tState_1,\ldots,\tState_{i-1},\State_i}
\!\!\equivInDistribution\!\!
\parenth{\State_{2i-1},\tState_{2i-2},\ldots,\tState_{i},\State_i}
\label{eqn:SCmatching:defn:rever:equivInDist}
\eeqa
Re-arranging terms, we have
\beqa
\parenth{\State_i,\tState^{i-1},\State^{i-1}} &\equivInDistribution& \parenth{\State_i,\tState_i^{2i-2},\State_{i+1}^{2i-1}} \label{eqn:SCmatching:timeRev:proof:a}
\\
\Rightarrow I(\State_i;\tState^{i-1},\State^{i-1}) &=& I(\State_i;\tState_i^{2i-2},\State_{i+1}^{2i-1}) \label{eqn:SCmatching:timeRev:proof:b} \\
\Leftrightarrow I(\State_i;\Out^{i-1},\State^{i-1}) &=& I(\State_i;\tState_i^{2i-2},\tSrc_{i+1}^{2i-1}) \label{eqn:SCmatching:timeRev:proof:c}
\eeqa
where \eqref{eqn:SCmatching:timeRev:proof:c} follows from invariance of mutual information to bijective transformations and \eqref{eqn:SCmatching:timeReversibility:c}-\eqref{eqn:SCmatching:timeReversibility:d},
\eqref{eqn:SCmatching:timeReversibility:e}-\eqref{eqn:SCmatching:timeReversibility:f}; Analogously, from \eqref{eqn:SCmatching:timeRev:proof:a},
\beqa
I(\State_i;\State^{i-1}|\tState^{i-1}) &=& I(\State_i;\State_{i+1}^{2i-1}|\tState_i^{2i-2}) \label{eqn:SCmatching:timeRev:proof:d} \\
\Leftrightarrow I(\State_i;\State^{i-1}|\Out^{i-1}) &=& I(\State_i;\tSrc_{i+1}^{2i-1}|\tState_i^{2i-2}) \label{eqn:SCmatching:timeRev:proof:e}
\eeqa
Therefore
\begin{eqnarray}
I(\State_i;Y^{i-1}) &=& I(\State_i;\tSrc_{i+1}^{2i-1}) \label{eqn:SCmatching:timeReversibility:h} \\
                    &=& 0 \label{eqn:SCmatching:timeReversibility:i}
\end{eqnarray}
where \eqref{eqn:SCmatching:timeReversibility:h} follows from the chain rule of mutual information \eqref{eqn:defn:chainRuleMutualInformation} and subtracting \eqref{eqn:SCmatching:timeRev:proof:e} from
\eqref{eqn:SCmatching:timeRev:proof:c}; and
\eqref{eqn:SCmatching:timeReversibility:i} follows from \eqref{eqn:assumption:timeReversibility:tsrc}.
Because of the nature of the memoryless channel $\PDMC$ in \eqref{eqn:defn:DMC}, it follows that $I(Y_i;Y^{i-1})=0$ for all $i$. Moreover,
because Markov chain reversibility implies stationarity, it follows that $(Y_i: \;1\leq i\leq n)$ are i.i.d.
Thus we can invoke Theorem~\ref{thm:SCmatching}.
\end{IEEEproof}

\subsection{Information Gain Cost and Inverse Optimal Control}\label{sec:SCmatching:infogain}
In the beginning of this section, we motivated the definition of stationary Markov coordination strategies,
of the type $\state_i=\encSC_i(\src_i,\est_{i-1})$ and $\est_i=\decSC_i(\est_{i-1},\out_i)$ by noting from Section~\ref{sec:directedInfo} that
such an optimal decoder exists when $\cEst = \probSimplex{\cSrc}$ and the cost function is of the ``information-gain'' related structure \eqref{eqn:maximizationDirInfo:CostFn:costfn}:
\beqa
\State_i = \benc^*_i(\Src_i,\Belief_{i-1|i-1}) = \encSC_i(\Src_i,\Est_{i-1})\nonumber\\
Z_i = \Belief_{i|i}= \NLF\parenth{\Belief_{i-1|i-1},Y_i,\benc^*_i(\cdot,\Belief_{i-1|i-1})}= \decSC_i(Z_{i-1},Y_i)\nonumber
\eeqa
We now demonstrate that the information gain cost function in Section~\ref{sec:directedInfo} can be seen to be a consequence of our inverse
optimal control framework for any coordination strategy for $(Y_i: 1\leq i \leq n)$ are i.i.d. and $\bdec$ is the nonlinear filter:
\begin{lemma} \label{lemma:SCmatching:infogain}
Let $\cEst = \probSimplex{\cSrc}$.  If a SM coordination strategy $\bpi$ contains a nonlinear filter decoder
$\est_i=\decSC(\est_{i-1},\out_i) = \NLF(\est_{i-1},\out_i,\encSC(\cdot,\out_{i-1}))$ and $(Y_i: 1 \leq n)$ are i.i.d., then
$\bpi$ is inverse control optimal with information gain distortion  $\dist(\src_i,\est_{i-1},\est_i)=-\log \frac{d \est_i}{d \OSPU(\est_{i-1})}(\src_i)$
and state cost function $\statecostfn(\state)=\kldist{\PDMCstate}{P^{\bpi}_{Y}}$.  The optimal cost is given by
\beqa
\totalcostbpi = (\alpha-1) n \capacityCostFnpi.
\eeqa
\end{lemma}
\begin{IEEEproof}
First note that under this policy $\bpi$, $\Est_i = \Belief_{i|i}$.  Now note that clearly
\begin{subequations} \label{eqn:proof:SCmatching:infogain:posteriors}
\beqa
\!\!\!\!\!\!\!\!&& \!\!\!\!\!\!\!\! \prob{\Src_i \in A | Y^i} = Z_i(A) =\prob{\Src_i \in A | \Est_{i-1},\Est_i},  \\
\!\!\!\!\!\!\!\!&& \!\!\!\!\!\!\!\! \prob{\Src_i \in A | Y^{i-1}} =\OSPU(Z_{i-1})(A)= \prob{\Src_i \in A | \Est_{i-1}}.
\eeqa
\end{subequations}
As such, we have that,
\beqa
\frac{d \est_i}{d \OSPU(\est_{i-1})}(\src_i) &=& \frac{dP^{\bpi}_{\Src_i|Y^i}}{dP^{\bpi}_{\Src_i|Y^{i-1}}}(\src_i) \nonumber \\
                                             &=& \frac{dP^{\bpi}_{\Src_i|\Est_{i-1}=\est_{i-1},\Est_i=\est_i}}{dP^{\bpi}_{\Src_i|\Est_{i-1}=\est_{i-1}}}(\src_i) \label{eqn:lemma:SCmatching:infogain:a}\\
                                             &=& \frac{dP^{\bpi}_{\Est_i|\Est_{i-1}=\est_{i-1}, \Src_i=\src_i}}{dP^{\bpi}_{\Est_i|\Est_{i-1}=\est_{i-1}}}(\est_i)
                                             \label{eqn:lemma:SCmatching:infogain:b}\\
                                             &=& \frac{dQ^{\bpi}_{\Est'|\Est,\Src}(\cdot|\est_{i-1},\src_i)}{dQ^{\bpi}_{\Est'|\Est}(\cdot|\est_{i-1})}(\est_i)
\parenth{\est_i}\label{eqn:lemma:SCmatching:infogain:c}
\eeqa
where \eqref{eqn:lemma:SCmatching:infogain:a} follows from \eqref{eqn:proof:SCmatching:infogain:posteriors};
\eqref{eqn:lemma:SCmatching:infogain:b} follows from a simple application of Bayes' rule:
$\frac{\prob{A|B,C}}{\prob{A|B}}=\frac{\prob{C|A,B}}{\prob{C|B}}$; and \eqref{eqn:lemma:SCmatching:infogain:c} follows from Lemma~\ref{lemma:SC:markovDecoder}.  Also, since $Z_i = \Belief_{i|i}$, it follows that $I(\Src^n;Y^n) = I(\Src^n;\Est^n)$.  Thus Theorem~\ref{thm:SCmatching} applies and so $\bpi$ is
inverse control optimal.  To characterize the final cost, note that for the associated $\alpha$,
\beqa
\totalcostbpi &=& -I(\Src^n;\Out^n) + \alpha  \E_{\encSC} \brackets{  \sum_{i=1}^{n} \statecostfn(\State_i)} \label{eqn:lemma:SCmatching:infogain:dd} \\
&=& -n \capacityCostFn + \alpha  \E_{\encSC} \brackets{  \sum_{i=1}^{n} \statecostfn(\State_i)} \label{eqn:lemma:SCmatching:infogain:d} \\
           &=& -n \capacityCostFn + \alpha  \parenth{\sum_{i=1}^{n} I(\State_i;Y_i)} \label{eqn:lemma:SCmatching:infogain:e}\\
           &=& (\alpha-1)n \capacityCostFn \label{eqn:lemma:SCmatching:infogain:f}
\eeqa
where \eqref{eqn:lemma:SCmatching:infogain:dd} follows from Theorem~\ref{thm:optimalDecoderMaximizationDirInfo}; \eqref{eqn:lemma:SCmatching:infogain:d} follows from the fact that Theorem~\ref{thm:SCmatching} applies which means that \eqref{eqn:RlessthanC} holds with equality;
 \eqref{eqn:lemma:SCmatching:infogain:e} follows from the definition of mutual information and that
$\statecostfn(\state)=\kldist{\PDMCstate}{P^{\bpi}_{Y}}$; and \eqref{eqn:lemma:SCmatching:infogain:f} follows from the fact that  \eqref{eqn:RlessthanC} holds with equality.
\end{IEEEproof}

Traditionally, inverse optimal control is performed through  finding a control-Lyapunov function \eqref{eqn:SCmatching:SMpolicies},
which involves performing a sequential decomposition of the problem and finding a consistent value function \eqref{eqn:decentralizedControl:DPrecursion}.
When $\cEst=\probSimplex{\cSrc}$, this can be done using only the decision variables and stationary-Markov coordination strategies as in Section~\ref{sec:directedInfo}:
$\State_i = \benc_i(\Src_i,\Est_{i-1})$ and $\Est_i = \Belief_{i|i} = \NLF\parenth{\Belief_{i-1|i-1},Y_i,\benc_i(\cdot,\Belief_{i-1|i-1})} = \bdec(Z_{i-1},Y_i)$.
This means that using a control-Lyapunov approach, first a sequential decomposition resting  upon the structural result work in Section~\ref{sec:MainResults}
would be needed, with the additional effort of showing that coordination strategies of the structural result form \eqref{eqn:thm:mainThmGeneralProblem} can be
reduced to stationary Markov strategies of the form \eqref{eqn:SCmatching:SMpolicies}.
However, our inverse control optimality sufficient conditions apply for a general $\cEst$ (which need not be $\probSimplex{\cEst}$) and do not involve a sequential decomposition.
As such, the approach developed in this section - when applicable - appears to require`less effort' than typically required in  arriving at an inverse optimal control result.
\section{\bf Examples}\label{sec:examples}
In this section, we provide examples of the Theorems and Lemmas from previous sections.

\subsection{\bf Likelihood Ratio Cost and Information Gain: HMMs and the Nonlinear Filter} \label{sec:directedInfo:NonlinearFilter}
We now demonstrate that the information gain cost framework of Section~\ref{sec:directedInfo}  demonstrates the causal coding/decoding and information-theoretic optimality of the nonlinear filter in a specific sense.   Related work on using variational principles to characterize the nonlinear filter was reported in  \cite{mitter2004variational}. However, demonstrating that the nonlinear filter is acting as an optimal controller with respect to this information gain cost function, is - to the best of our knowledge - new.  We start by considering the following assumptions:
\bitm
\item [(i)] the source and channel inputs have the same alphabets: $\cSrc = \cState$
\item [(ii)] the causal encoder alphabet $\cEnc_i = \{ e_i: \cSrc^i \times \cOut^{i-1} \to \cState \}=\{=\}$ where $=$ is the identity function: $\state_i = \src_i$.
\eitm
Under these conditions, the only feasible encoder simply specifies $\src_i$ as the channel inputs, and thus this becomes a hidden Markov model.
 \begin{figure}[hbtp]
\centering
\begin{overpic}[width=\columnwidth]{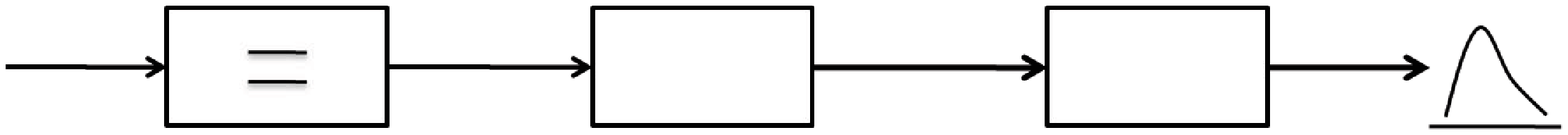}
\put(2,7){\large $\Src_i$}
\put(29,7){\large $\State_i$}
\put(39,5){\large $\PDMC$}
\put(59,7){\large $Y_i$}
\put(67.5,6){\small nonlinear}
\put(71,3){\small filter}
\put(93,9){\large $\Est_i$}
\end{overpic}
\caption{The information gain cost when the encoder set consists of only the identity function.  This becomes a hidden Markov model where the nonlinear filter is an optimal solution.}
\label{fig:nonlinearfilter}
\end{figure}

 As such, we can consider maximizing the mutual information from $\Src^n$ to $Y^n$ over all possible causal decoder policies.  As such, the optimal design of $\enc_i$ disappears and the focus becomes optimal design of $\{\dec_i \}$.  We now show that, assuming $\Est_0(A) = \prob{\Src_0 \in A}$, the optimal policy for the decoder is given by the true posterior - which can be computed recursively using the nonlinear filter:
\begin{lemma}
Under assumptions (i) and (ii) above, and cost functions $\statecostfn \deq 0$ and $\dist$ given by \eqref{eqn:maximizationDirInfo:CostFn:costfn}
\[  \dist(\src_i,\est_{i-1},\est_i) = \begin{cases}
                                         -\log \frac{d\est_i}{d \OSPU(\est_{i-1})}(\src_i)  & \text{if } \est_i \ll \OSPU(\est_{i-1})\\
                                        \infty, & \text{otherwise}
                                     \end{cases},
\]
the policy $\bpi$ consisting of the identity function encoder and nonlinear filter decoder $\Est_i= \Belief_{i|i} = \NLF(\Est_{i-1},Y_{i},=)$, is globally optimal where $J^\alpha_{n,\pi^*} = -I(\Src^n;Y^n)$.
\end{lemma}
\begin{IEEEproof}
Because $\cEnc_i$ is a singleton consisting of the identity function, and because $\statecostfn(\state)=0$, this follows directly from Theorem~\ref{thm:optimalDecoderMaximizationDirInfo}.
\end{IEEEproof}

\subsection{\bf Likelihood Ratio Cost and Information Gain: Feedback Communication of a Message Point}\label{sec:directedInfo:commFB}
Given that the natural mathematical framework to handle feedback is control theory, we consider the problem of communication over noisy channels with feedback from the dynamical systems perspective, and make use of recent sequential approaches to communication.  This viewpoint has been made largely possible by a recent development in the information theory literature - the posterior matching (PM) scheme \cite{shayevitz2009optimal} - which generalizes other `message-point' style feedback communication schemes \cite{schalkwijk1966csaOne,schalkwijk1966csaTwo,horstein1963stu}: rather than $nR$ bits, a message point on the interval $\left[0,1\right]$ is considered.  The notion of ``decoding $nR$ bits'' now becomes equivalent to determining the message point within an interval of length $2^{-nR}$ at the receiver (see Section~\ref{sec:intro:example}).

The implementational details and fundamental limits are completely in line with traditional communication paradigms (see \cite{shayevitz2009optimal}) but there are subtle, yet striking differences. Because the message point is a point on the $\left[0,1\right]$ line, there is no pre-specified block length; the system operates to sequentially give the user the information that is ``still missing" at the receiver.  Moreover, at each time step, the decoder specifies an output $\Est_i \in \probSimplex{\cSrc}$, which is a belief about the message point. We now demonstrate how this notion of communication, and the problem of finding the optimal encoder with feedback, can be captured with our framework.  Moreover, we will demonstrate that the PM scheme is an optimal solution to the problem.

\begin{figure}[hbtp]
\centering
\begin{overpic}[width=1.02\columnwidth]{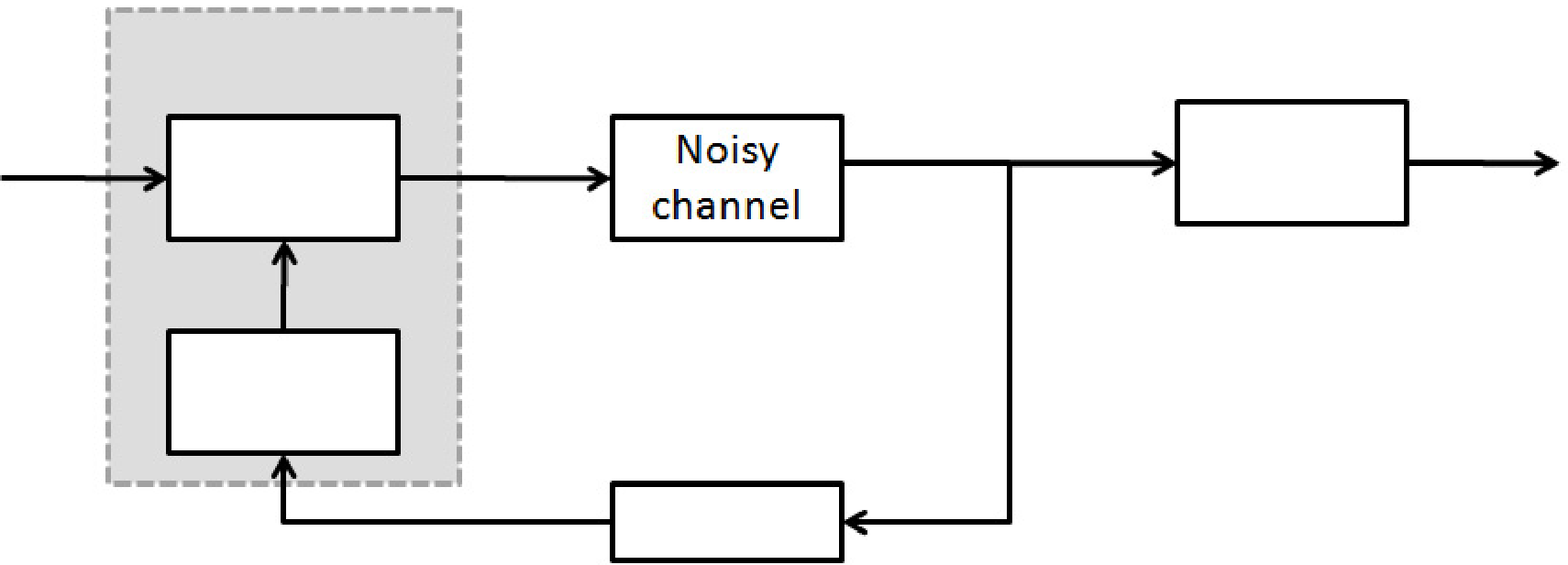}
\put(12,32){\it causal}
\put(12,29){\it encoder}
\put(11,11){\small nonlinear}
\put(13,8.5){\small filter}
\put(16,23){\large $\bencst$}
\put(1,26){\large $\Src$}
\put(31,26){\large $\State_i$}
\put(57,27){\large $\Out_i$}
\put(90,27){\large $\Belief_{i|i}$}
\put(18,17){\small $\Belief_{i-1|i-1}$}
\put(74.5,26){\small nonlinear}
\put(79,23){\small filter}
\put(42,2){delay}
\end{overpic}
\caption{Posterior matching scheme by Shayevitz \& Feder, interpreted as a time-invariant manifestation of the simplified structural result in Figure~\ref{fig:informationGainCostStructuralResult}.
}
\label{fig:pmscheme}
\end{figure}

Let $\cSrc = [0,1]$ and $\cEst = \probSimplex{\Src}$.  Further, let the source process be the `repetition' Markov process $(\Src_i=\Src: i \geq 1)$ with $\Src$ uniformly distributed over $[0,1]$.
If we assume that there is an expected cost constraint
$\frac{1}{n} \sum_{i=1}^n \E\brackets{\statecostfn(\State_i)} \leq \statecostval$, then we may formulate a communication problem
of communicating a message point over a memoryless channel with causal feedback.  First note that the mutual information between the message point and observations is given by
\beqas
\frac{1}{n} I(\Src;Y^n) &=& \frac{1}{n} \sum_{i=1}^n I(W;Y_i|Y^{i-1}) \\
                           &=& \sum_{i=1}^n \E \brackets{\log \frac{d\Belief_{i|i}}{d\Belief_{i|i-1}}(\Src) }
\eeqas
Shannon's converse to the channel coding theorem with feedback tells us that in order to achieve capacity, this aforementioned quantity must asymptotically be maximized.  This allows for us to consider the following maximization problem
\[ \max_{\pi \in \Pi} I(\Src;Y^n) + \alpha \E_\pi\brackets{\sum_{i=1}^n \statecostfn(\State_i)}, \]
where $\alpha$ serves as a Lagrange multiplier such that under an optimal policy, the average state cost is upper bounded by $\statecostval$.
We note that this can be captured in a causal coding/decoding framework by considering the sequential information gain
distortion function \eqref{eqn:maximizationDirInfo:CostFn:costfn}.  From Lemma~\ref{lemma:SCmatching:infogain}, we note that a sufficient condition for optimality to this control problem is for
\begin{itemize}
\item $I(Y_i;Y^{i-1})=0$  for all $i$
\item $\State_i \sim \optimalInputDMC$, given in  \eqref{eqn:defn:OptimalInputcapacityCostFn}, for all $i$
\end{itemize}
Let $\cState = \reals$ and denote $F_{\State}(\cdot)$ as the cumulative distribution function of the optimal input distribution $\optimalInputDMC$.  The Posterior matching (PM) scheme \cite{shayevitz2009optimal} simultaneously enables the two properties to hold for each $i$ and is given by:
\begin{subequations} \label{eqn:pmschemeoptimal}
\beqa
\State_i &=& F_{\State}^{-1} \parenth{ F_{\Src|Y^{i-1}}(\Src|Y^{i-1})} \label{eqn:pmschemeoptimal:a}\\
  &=& F_{\State}^{-1} \parenth{ \Belief_{i-1|i-1}([0,\Src])} \label{eqn:pmschemeoptimal:b}\\
  &=& \benc \parenth{\Src, Z_{i-1}} \label{eqn:pmschemeoptimal:c}
\eeqa
\end{subequations}
where \eqref{eqn:pmschemeoptimal:c} follows from
 Theorem~\ref{thm:optimalDecoderMaximizationDirInfo} and because $\Src_i=\Src$.  Note that the $F_{\Src|Y^{i-1}}$ operation constructs a uniform-$[0,1]$ random variable that is independent of the past channel outputs, and the $F_{\State}^{-1}$ shaping operation enables each input to be drawn according to the optimal channel input distribution $\optimalInputDMC$.
Note that from \eqref{eqn:pmschemeoptimal:c}, the PM scheme can be interpreted as `minimal' from our structural result in the causal coding/decoding framework in Section~\ref{sec:directedInfo}.  Moreover, the causal encoder is {\it time-invariant}, and so likewise for the decoder
acting as the nonlinear filter; thus, this means the PM scheme also can be interpreted as an instance of the inverse optimal control framework via Lemma~\ref{lemma:SCmatching:infogain}.  See Figure~\ref{fig:pmscheme} and its relationship with Figure~\ref{fig:informationGainCostStructuralResult}.
Also note that in some cases, the time-reversibility condition for inverse optimal control in Section~\ref{sec:timereversibility} is applicable:
from Example\ref{example:inverse-reversible-compatible:a}, the PM scheme \eqref{eqn:pmschemeoptimal} elicits~\IRC. Consider the additive Gaussian noise channel.  Under the PM scheme, $(\State_i,\Out_i: i \geq 1)$ are jointly Gaussian (see \eqref{eqn:inverse-reversible-compatible:a:e} and \cite[Example 1]{ShayevitzFeder07}).  Note that since $\State_i = \tState_{i-1}$ in \eqref{eqn:inverse-reversible-compatible:a:d}, the joint reversibility sufficient condition is equivalent to reversibility of the Markov chain $\State$.  Since all stationary Gaussian processes are time-reversible, we see that in this scenario, the time-reversibility framework for our inverse optimal control framework is linked to the PM scheme.
Although our control problem only addresses the maximization of mutual information - which is a necessary condition for reliable communication by the converse to the channel coding theorem -  it can be shown that reliable communication, as defined in Section~\ref{sec:intro:example}, results as a consequence of the mutual information maximization control problem under mild technical conditions \cite{shayevitz2009optimal}.

\subsection{\bf Structural Result: Brain-Machine Interfaces}\label{sec:directedInfo:BMI}
A brain-machine interface (BMI) is a system that elicits a direct communication pathway between a human and an external device.  In many cases, it is the objective of the human to control an external device merely by imagination, and the external device acquires neural signals, actuates some physical system, and perceptual feedback is given to the user to complete the loop.  We now demonstrate how our structural result can be applied to the design of brain-machine interfaces that have a `user-friendly' structure: displaying the minimal amount of useful perceptual feedback to the user, and designing an interaction strategy between the user and the external device.

Consider a brain-machine interface where a human has a desired high-level intent represented by the Markov process $(\Src_i: i \geq 1)$.  At each time step, the human imagines a control signal $\State_i$ which is statistically linked to neural activity $\Out_i$ that is observed by the external device. For example, the statistics of $\Out_i$ are different when imagining a left-oriented movement $\State_i=0$ as compared to imagining a right-oriented movement $\State_i=1$ \cite{McCormickMaColemanICASSP2010}.  At each time step, the external device maps all its recorded observations $\Out^i$ to actuate some system, whose state is given by $\Est_i$. Equally as important, the user gets perceptual feedback from the external device and allows this, along with causal information about the high-level intent, $\Src^i$, to specify the subsequent imagined control signal $\State_i$.

Without loss of generality, because we do not know yet what perceptual feedback is the most relevant, we could consider a scenario where all information available to the decoder at any time $i$ is fed back to the subject.  Secondly, we may assume that we are planning to design the coordination strategy between the user and the interface: not only how the interface should take its observations and actuate the plant, but also what perceptual feedback should be specified back to the user {\it and} how the user should react to the perceptual feedback to specify the subsequent control signal $\State_i$.  In such a case, this problem boils down to our problem formulation in Section~\ref{sec:problemSetup}.  Note that because of the causal nature of the problem, real-time constraints with a human in the loop obviate the possibility of using `block-coding' like paradigms.  Secondly, such settings are more complicated than simply optimally representing intent with an arithmetic coding procedure as in \cite{wills2006dasher} - because of the inherent uncertainty also due to the noisy channel mapping intent to neural signals.

Almost all previous approaches to design BMIs failed to consider how the desired control signals change in response to sensory feedback.  For example, many previous schemes simply attempt to recursively estimate $\State_i$ from $\Out^i$ under the assumption that ($\State_i: i \geq 1$) is a Markov process.  However, as we know from our structural result, for an arbitrary objective with additive cost function, it is crucially important for the system to keep a running estimate, or belief, on $\Src_i$ given $Y^i$.  Moreover, it is critically important that the user and the system agree on an interaction protocol that specifies both {\it what} sensory feedback is provided to the user (e.g. the sufficient statistics) and {\it how} the user should react to this feedback in pursuit of high-level intent (e.g. the function $\benc_i$).

\begin{figure}[hbtp]
\centering
\begin{overpic}[width=1.02\columnwidth]{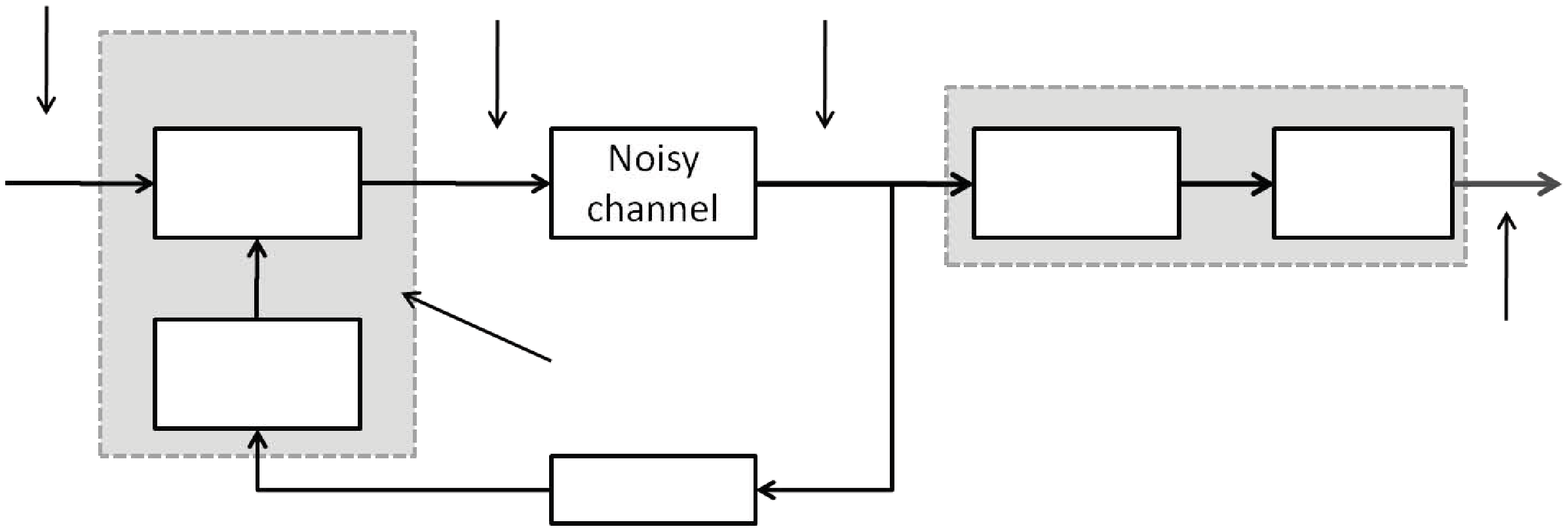}
\put(12,21){\large user}
\put(17,15){$S_{i-1}$}
\put(13,10){\small state}
\put(13,7){\small filter}
\put(1,23){\small $\Src_i$}
\put(-3.5,36.5){\small user's high}
\put(-3.5,33.5){\small level intent}
\put(29,23){\small $\State_i$}
\put(17,36.5){\small user's next imagined}
\put(24,33.5){\small command}
\put(27,4){\small $Y_{i-1}$}
\put(35,11){\small feedback}
\put(35,9){\small to the user}
\put(37,1){delay}
\put(50,22.5){\small $Y_i$}
\put(50,36.5){\small recorded}
\put(48,33.5){\small neural signal}
\put(64,22){\small state}
\put(64,19){\small filter}
\put(83,21){\small plant}
\put(74,18){ $S_i$}
\put(93,23){\small $\Est_i$}
\put(89,12){\small state of}
\put(89,9){\small system}
\end{overpic}
\caption{Structural result within the context of a brain-machine interface: in an optimal system, the user acts as part of the causal encoder.  The other part accumulates all causal observations and summarizes them into sufficient statistics acting as perceptual feedback to the user.}
\label{fig:SeparationPrincipleBMI}
\end{figure}

Our structural result says that first a state filter can construct sufficient statistics $S_i =(\Est_{i-1},\Belief_{i|i})$,
and then the external device can actuate the plant using $S_i$ and the user only needs to be fed back $S_{i-1}$ as perceptual feedback.  This information, along with the current high-level goal $\Est_i$, is all that is needed to specify an optimal causal encoder $\benc_i$.  See Figure~\ref{fig:SeparationPrincipleBMI}. $\;$\\

In \cite{OmarEtAkBCIIJHJCIsubmittedNov09}, we instantiate this idea in an EEG-based BMI in two steps. We assume the high-level intent can be mathematically represented as a Markov process $(W_i=W: i\geq 1)$ on $\cSrc=[0,1]$ for which $W$ is uniformly distributed over the $[0,1]$ line.  As such, this means we are assuming that the whole high-level intent is known to the user at all times.  To relate this to a variety of practical applications, the user interprets the message point as a countably infinite sequence of symbols $D=(D_1,D_2,\ldots)$ in an {\it ordered} countable set $\mathcal{D}$ with a known statistical model (typically modeled as a fixed-order Markov process).  Examples of the sequence $D$ include an infinite sequence of text characters or an infinite sequence of small path arcs pertaining to a smooth path of bounded curvature.  We use arithmetic coding \cite{CoverThomas06} to develop a one-to-one mapping between any such sequence $D$ and a point $\Src = \tau(D)$ uniformly distributed on the $[0,1]$ line.
We subsequently use an EEG system and specify a binary-input (left/right motor imagery) noisy channel with a spatial filter to extract beliefs $\Belief_{i|i}$ sequentially \cite{McCormickMaColemanICASSP2010}.  With this, we implement the Posterior Matching scheme for the binary symmetric channel \cite{horstein1963stu,shayevitz2009optimal}.  Here, what is nice for a human in the loop is that $\benc_i=\benc$, and secondly, for the BSC, it only requires a {\it functional} of the posterior $\Belief_{i-1|i-1}$ to be given to the encoder at time $i$:  the median (denoted as $m(B_{i-1|i-1})$) \cite{horstein1963stu,ShayevitzFeder07}:
 \beqa
  \State_i = \begin{cases}
                0, & \Src < m(B_{i-1|i-1})\\
                1, & \Src \geq m(B_{i-1|i-1})
 \end{cases}. \label{eqn:PM:BSC}
 \eeqa
Because of the one-to-one mapping $\tau$, at time $i$, this can be implemented by visually displaying the median path $\tau^{-1}(m(B_{i-1|i-1}))$ on the screen and instructing the user to obey the time-invariant PM scheme \eqref{eqn:PM:BSC} within the context of the median path.  This simply means performing a lexicographically comparison to $D$ (i.e. identify the first location where the sequences differ and perform a symbol-based comparison).  We have successfully implemented this to demonstrate reliable text spelling and two-dimensional smooth path specification.  Secondly, wedding with arithmetic coding with the PM scheme has the added benefit that a natural `propagation' of uncertainty ensues: the locations where $D$ and $\tau^{-1}(m(B_{i-1|i-1}))$ differ increase to later and later parts of their sequences; this leads to a natural real-time implementation plausibility.  Remote-control of an unmanned aerial vehicle using this paradigm has recently been shown in \cite{akce2010remote}.

We also comment how the PM scheme by Shayevitz and Feder \cite{shayevitz2009optimal} is particularly relevant here: formulating this problem as one where the encoder has one of $2^{nR}$ hypotheses would mean that the human agent attempting to elicit neural control of an external device would have to implement an a strategy that differentiates possible inputs based upon one of $2^{nR}$ hypotheses.  Even with visualization, this could be cumbersome.  Moreover, it is unclear how the design specification would change when $n=100$ as compared to when $n=101$.
Remarkably, using the posterior matching framework makes this problem truly solvable both theoretically and practically - by simply changing the starting point to be $\cSrc=[0,1]$ and $\cEst=\probSimplex{\cSrc}$ and defining an appropriate information gain cost criterion. These observations speak to the fragility at which information theoretic problems with the same fundamental limits can be formulated.

The structural result demonstrated in this paper now enables the opportunity to design many brain-machine interface paradigms for a variety of cost functions beyond the  the information gain paradigm and with assumption that $\Src_i=\Src_{i-1}$. The structural result has the potential more generally to enable an interesting intersection of desires on one platform: (i) guaranteed optimality from a decision-theoretic viewpoint; (ii)
elucidation of the minimal amount of perceptual feedback information required to optimally display to the user; and (iii) potential ease-of-use when (e.g. when $\benc_i=\benc$ and it has a simple operational interpretation).

\subsection{\bf Inverse Optimal Control: Gauss-Markov source and AGN channel}
Here we show that a stationary Markov coordination strategy consisting of a linear `estimation error' encoder and MMSE decoder is inverse-control optimal for a Gauss-Markov $\Psrc$ and a power-constrained additive Gaussian channel.  A variant of this problem for $\dist(\src_i,\est_{i-1},\est_i) \deq \dist(\src_i,\est_i)=(\src_i-\est_i)^2$ has been studied by \cite{bansal1988sdc},\cite{tatikonda2000control}.

Let $\cSrc=\cState=\cOut=\cEst=\mathbb{R}$. The source is a Gauss-Markov process with i.i.d. $\tSrc_i \sim \mathcal{N}\parenth{0,\sigma_m^2}$,
\begin{subequations} \label{eqn:gaussianEx}
  \begin{eqnarray}
   && \Src_{0} \sim \mathcal{N}\parenth{0,\frac{\sigma_m^2\NoiseVariance}{\statecostval+\NoiseVariance(1-\rho^2)}}, \;\;\;\; \label{eqn:gaussianEx:init} \\
  &&\Src_{i} = \rho \Src_{i-1} + \tilde{\Src}_i \;\;\; i \geq 1,  \label{eqn:gaussianEx:src} \\
  &&    I(\tSrc_i;\State^{i-1},\Out^{i-1})=0, \;\;\; i \geq 1 \label{eqn:gaussianEx:src:innovations}
  \end{eqnarray}

\end{subequations}
Note that we are not assuming that $\Src$ is stationary.  As such, this problem can be connected to problems in `control over noisy channels.  In such problems with quadratic cost and linear Gaussian dynamics, the essence of optimally solving the control over noisy channels problem is optimally solving this causal coding/decoding `active tracking' problem \cite{bansal1988sdc},\cite{tatikonda2000control}.

The channel additive with Gaussian noise (AGN):
  \begin{eqnarray}
  \Out_i &=& \State_i + \GaussNoise_i \;\;\; \GaussNoise_i \sim \mathcal{N}\parenth{0,\NoiseVariance}\label{eqn:gaussianEx:out}
  \end{eqnarray}
A typical objective in practice is to design an encoder and decoder than can minimize the mean-squared error in estimating the source process, i.e., minimize $ J(\enc^n,\dec^n) = \E\brackets{\sum_{i=1}^{n} \parenth{\Est_i - \Src_i}^2 + \alpha \State_i^2}$.  It is known \cite{bansal1988sdc},\cite{tatikonda2000control} that an optimal linear coordination strategy exists, pertaining to
``error'' encoding and MMSE estimation decoding:
\begin{subequations} \label{eqn:gaussianEx:optencdec}
\begin{eqnarray}
   \State_i &=& \beta_i \parenth{\Src_i - \E\brackets{\Src_i|\Out^{i-1}}} \label{eqn:gaussianEx:optencdec:a} \\
    \Est_i &=& \E\brackets{\Src_i|\Out^{i}}  \label{eqn:gaussianEx:optencdec:b}
\end{eqnarray}
\end{subequations}
where $\beta_i$ are time-varying normalizing constants  that result in $\State_i \sim \mathcal{N}(0,\statecostval)$ for all $i$, and the power-constraint $\statecostval$ depends on the value of $\alpha$.  

\begin{figure}[hbtp]
\centering
\resizebox{\columnwidth}{!}{
\begin{overpic}[width=1.05\columnwidth]{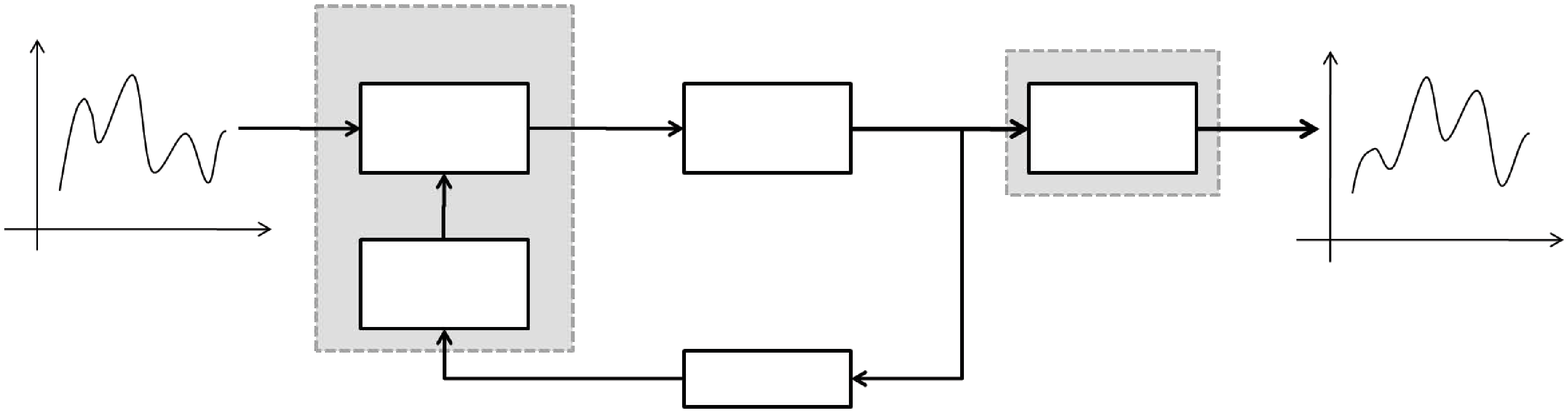}
\put(25,24){\small \it causal}
\put(25,21){\small \it encoder}
\put(26,16.5){\large $+$}
\put(25,12){\small $-$}
\put(30,12){\small $\Est_{i-1}$}
\put(23,8){\small MMSE}
\put(23.5,5.5){\small estim.}
\put(8,24){\large $\Src_i$}
\put(38,19){\small $\State_i$}
\put(45,18){\small AGN}
\put(43.5,15.5){\small channel}
\put(36,3){\small $Y_{i-1}$}
\put(45,1){ delay}
\put(59,19){\small $Y_i$}
\put(62,24){\small \it causal decoder}
\put(66,18){\small MMSE}
\put(66.5,15.5){\small estim.}
\put(90,23){\large $\Est_i$}
\end{overpic}
}
\caption{With $\Psrc$ Gauss-Markov and $\PDMC$ an AGN channel, ``error'' encoding and MMSE estimation decoding is inverse control optimal.
The induced cost function is squared error.}
\label{fig:GaussianFig}
\end{figure}

We now consider observing this problem from the lens of inverse optimal control for a distortion function of the form $\dist(\src_i,\est_{i-1},\est_i)$:
\begin{lemma} \label{lemma:gaussianEx}
For the problem setup in \eqref{eqn:gaussianEx},
define the following Stationary Markov coordination policy
      \begin{subequations} \label{eqn:gaussianEx:optencdec:ver2}
      \begin{eqnarray}
   \State_i &=& \beta \parenth{\Src_i - \rho \Est_{i-1}} \label{eqn:gaussianEx:optencdec:ver2:a}\\
   \Est_i &=& \rho \Est_{i-1} + \gamma Y_i \label{eqn:gaussianEx:optencdec:ver2:b}
      \end{eqnarray}
\end{subequations}
where $\beta = \sqrt{\frac{\statecostval}{C}},\gamma = \frac{\sqrt{\statecostval C}}{\statecostval+\NoiseVariance}$, and $C = \frac{\sigma_m^2}{1-\rho^2\frac{\NoiseVariance}{\statecostval+\NoiseVariance}}$.
\begin{itemize}
\item (a)
The policy pair in \eqref{eqn:gaussianEx:optencdec:ver2} is inverse control optimal
   \begin{subequations}
   \begin{eqnarray}
\!\!\!\!\!\!\!\!\!\!\!\!\!\!\!  \statecostfn(\state_i) \!\!\!\!\!\!&\propto_+&\!\!\! \state_i^2  \label{eqn:gaussianEx:parameterCostfn:state} \\
   \!\!\!\!\!\!\!\!\!\!\!\!\!\!\!  \dist(\src_i,\est_{i-1},\est_i) \!\!\!\!\!\!&\propto_+&  \!\!\!\!\!\!(\src_i-\est_i)^2 - \frac{\NoiseVariance}{\statecostval+\NoiseVariance}(\src_i-\est_{i-1})^2  \label{eqn:gaussianEx:parameterCostfn}
   \end{eqnarray}
   \end{subequations}
    \item (b) The total cost can be represented as a weighted MMSE cost given by $J^\alpha_{n,\pi}$:
\begin{eqnarray}
\!\!\!\!\!\!\!\!\!\!\!\!&&\E\brackets{\sum_{i=1}^{n} \rho(\Src_i,\Est_{i-1},\Est_i)} \nonumber \\
\!\!\!\!\!\!\!\!\!\!\!\!&=& \E \brackets{\sum_{i=1}^{n} \parenth{\Est_i-\Src_i}^2 + \parenth{\frac{1}{1-\frac{\NoiseVariance\rho^2}{\statecostval+\NoiseVariance}}}
 \parenth{\Est_{n} - \Src_{n}}^2} \nonumber
\end{eqnarray}
\end{itemize}
\end{lemma}
The proof is provided in Appendix \ref{appendix:gaussian}.
\begin{remark}
The policy-pair in \eqref{eqn:gaussianEx:optencdec} is optimal for a mean-square distortion cost (MMSE) problem for Gauss-Markov sources except that the last reconstruction has higher penalty. For $n \to \infty$, the cost for which \eqref{eqn:gaussianEx:optencdec} is optimal is exactly equivalent to a MMSE cost problem $\lim_{n \to \infty }\frac{1}{n}\E \brackets{\sum_{i=1}^{n} \parenth{\Est_i-\Src_i}^2 + \alpha \State_i^2  }$. Thus, asymptotically, we can recover the results of \cite{bansal1988sdc},\cite[Ch. 6]{tatikonda2000control} using inverse-optimal control and time-invariant cost functions.
\end{remark}

\subsection{\bf Inverse Optimal Control: the M/M/1 Queue}
Here we show that the $\cdot/M/1$ queue's dynamics can be interpreted as a stationary Markov coordination strategy
that is inverse control optimal for $\Psrc$ being a Poisson process.   It is well-known from Burke's theorem \cite{Gallager96Discrete,hsuBurke1976behavior} that for a Poisson process of rate $\lambda$
entering a $\cdot/M/1$ queue, in steady state the queue state at time $t$ is independent of
the output before time $t$.
We now demonstrate that this statement has implications not only for the capacity of queuing timing channels \cite{AnantharamVerdu96,SundaresanVerdu06,ColemanITWVolos09,gorantla2010reversible}, but also for inverse optimal control.

Divide time into units of interval $\Delta$ where $\Delta \ll 1$. The input to the queue $\Src_i$ represents the number of arrivals to the queue till time $i$. For a Poisson source, $(\Src_i:i \geq 1)$ is the discrete-time equivalent of the counting function representation of a Poisson process.
\beqa \label{eqn:estc:src}
\kernelSrc{\src_i}{ \src_{i-1}} = \begin{cases}
                                                \lambda \Delta, & \text{if }\src_i = \src_{i-1}+1 \\
                                                1-\lambda \Delta, & \text{if }\src_i = \src_{i-1} \\
                                                0, & \text{otherwise}
                                             \end{cases}
\eeqa
In other words, $\Src_i = \Src_{i-1} + \tSrc_i$ where $\tSrc_i$ are i.i.d. with $\prob{\tSrc_i=1}=\lambda \Delta$.
Assume the following model for the channel:
\begin{eqnarray}
\PDMC(1|\state)=\begin{cases}
             0 & \state=0 \\
             \mu \Delta & \state > 0
           \end{cases} \label{eqn:ZChann}
\end{eqnarray}
For a queuing system, note that this means that a departure $(\Out_i=1)$ can only occur when the number of customers in the queue is positive, and the likelihood of a departure in that scenario for a bin of length $\Delta$ is $\mu\Delta$.  Continuing on with the queuing analogy, note that we   represent $\Est$ as the counting function representation of the departure process as  $\Est_i = \sum_{k \leq i} \Out_k$ where $\Out_k \in \{0,1\}$. $\State_i$ is the queue size representing the number of customers in the queue at the i-th time instant: $\State_i = \Src_i-\Est_{i-1}$. Thus, the update equations for the state $\State_i$ and output of the queue $\Est_i$ are linear stationary Markov policies given by
\begin{subequations} \label{eqn:estc:policies}
\begin{eqnarray}
  \State_i &=& \Src_i - \Est_{i-1}  \\
  \Est_i &=& \Est_{i-1} + \Out_i
\end{eqnarray}
\end{subequations}
The departure at i-th time instant $\Out_i$ depends on the state by the following discrete memoryless `Z' channel model:
That is, there will be no departure if the queue is empty, and there will be departure with probability $\mu\Delta$ if the queue is not empty. The initial number of arrivals $\Src_0$ is drawn according to $\prob{\Src_0 = k} = \parenth{1-\frac{\lambda}{\mu}} \parenth{\frac{\lambda}{\mu}}^k, \; k \geq 0$ and the initial number of departures $\Est_{0}=0$.
\begin{figure}[hbtp]
\centering
\resizebox{\columnwidth}{!}{
\begin{overpic}[width=1.1\columnwidth]{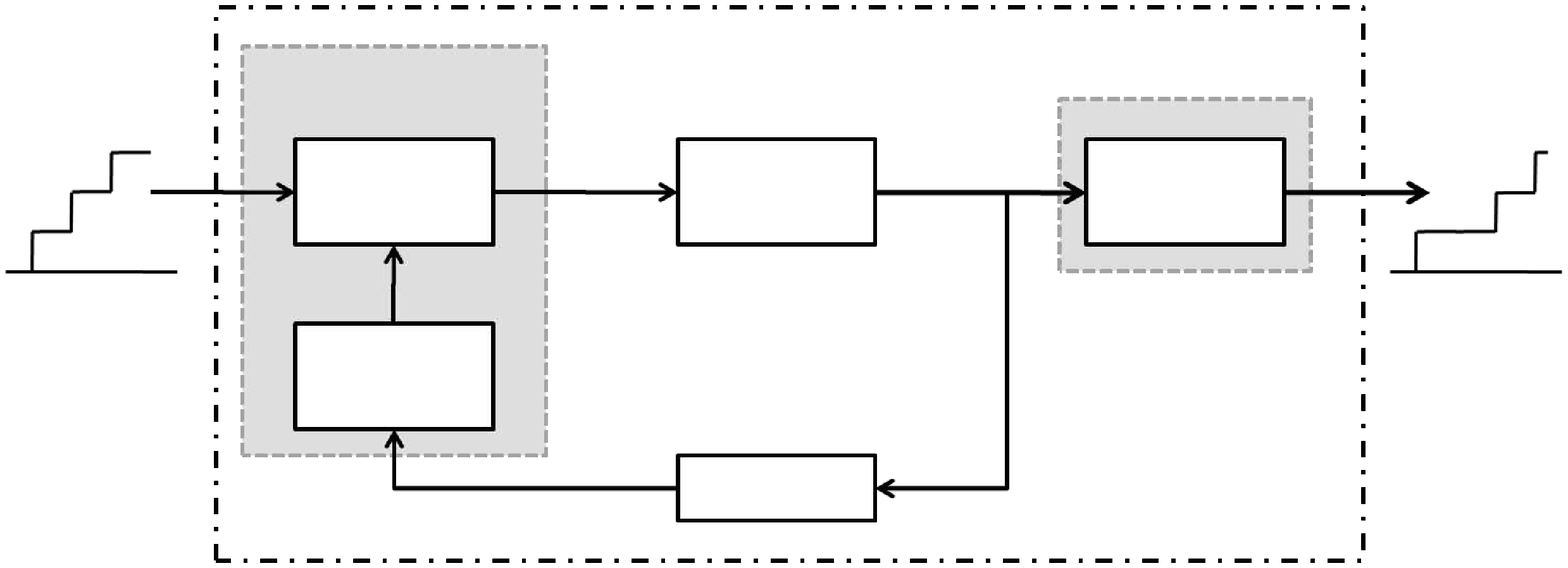}
\put(18,31){\small \it causal}
\put(18,28){\small \it encoder}
\put(23,21.5){\Large $+$}
\put(20,17){ \small $-$}
\put(27,17){$\Est_{i-1}$}
\put(19,10.5){\small accumul.}
\put(5,29){\large $\Src_i$}
\put(36.45,25){\large $\State_i$}
\put(48,24){$Z$}
\put(45,21.5){\small channel}
\put(36,6){\large $Y_{i-1}$}
\put(46,4){ delay}
\put(60,25){\large $Y_i$}
\put(65,31){\small \it causal decoder}
\put(70.5,23.5){\small accumul.}
\put(93,26){\large $\Est_i$}
\put(72,6){\large \bf $\cdot/M/1$}
\put(75,3){\large queue}
\end{overpic}
}
\caption{With $\Psrc$ a Poisson process and $\PDMC$ a Z channel, the $\cdot/M/1$ queue is inverse control optimal.}
\label{fig:SeparationPrincipleESTC}
\end{figure}
Note that the aggregate statistical dynamics of $P^{\bpi}_{\Est^n|\Src^n}$ in Figure \ref{fig:SeparationPrincipleESTC}
are precisely that of the discrete-time exponential server timing channel, also termed a $\cdot/M/1$ queue of rate $\mu$, which is a first-come, first-serve queuing system with i.i.d. service times geometrically distributed  of rate $\mu$ \cite{Gallager96Discrete}.  As $\Delta \to 0$, this becomes the continuous-time $\cdot/M/1$ queue.
\begin{figure}[htbp]
\begin{minipage}[b]{0.48\linewidth}
\centering
\begin{overpic}[width=1.2\columnwidth]{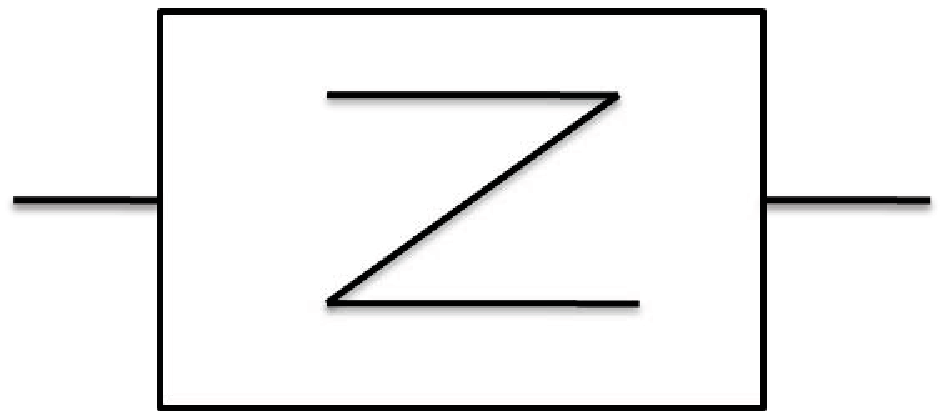}
\put(85,24){\large $\Out_i$}
\put(50,36){\large$1$}
\put(46,5){\large $\mu\Delta$}
\put(5,24){\large $\State_i$}
\put(20,30){\large $0$}
\put(18.5,10){\large $>0$}
\put(68,30){\large $0$}
\put(68,10){\large $1$}
\end{overpic}
\caption{$\PDMC$ for the $\cdot/M/1$ queue sampled at length-$\Delta$ intervals}
\label{fig:ESTC-ChannelLaw}
\end{minipage}
\hspace{0.5cm}
\begin{minipage}[b]{0.5\linewidth}
\centering
\resizebox{0.9\columnwidth}{!}{
\begin{overpic}[width=\columnwidth]{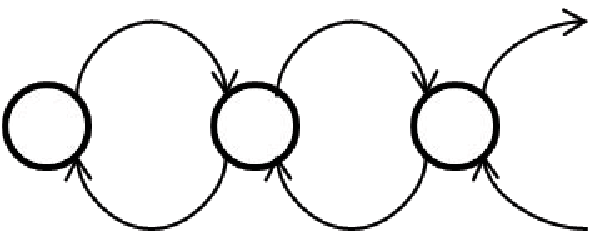}
\put(6,17){\large $0$}
\put(21,37){\large $\lambda\Delta$}
\put(21,-3.5){\large $\mu\Delta$}
\put(40,17){\large $1$}
\put(54,37){\large $\lambda\Delta$}
\put(54,-3.5){\large $\mu\Delta$}
\put(73,17){\large $2$}
\put(82,37){\large $\lambda\Delta$}
\put(82,-4){\large $\mu\Delta$}
\end{overpic}
}
\caption{birth-death chain for $\State$ in the $M/M/1$ queue}
\label{fig:ESTCstatetransitions}
\end{minipage}
\end{figure}
From standard queuing theory it follows that $\State$ is a birth-death Markov chain in steady-state with distribution
\begin{eqnarray}
  \pi_k &=& \parenth{1-\frac{\lambda}{\mu}} \parenth{\frac{\lambda}{\mu}}^k \;\;, k\geq 0.
\end{eqnarray}
Therefore Lemma~\ref{lemma:SCmatching:Xbirthdeathchain} holds and note that the fixed coordination strategy given by
\eqref{eqn:estc:policies} is inverse-control optimal for
\begin{eqnarray}
  \dist(\src_i,\est_{i-1},\est_i)
  \!\!\!\!\!&=&\!\!\!\!\! \begin{cases}
                                          -\log (1-\lambda\Delta), & \state_i=0,y_i=0; \\
                                          \log \frac{1-\mu\Delta}{1-\lambda\Delta}, & \state_i>0,y_i=0; \\
                                          \log \frac{\mu}{\lambda} , & \state_i>0,y_i=1; \\
                                          +\infty, & \text{otherwise}.
                                        \end{cases}\nonumber\\
\Rightarrow \lim_{\Delta \to 0}   \dist(\src_i,\est_{i-1},\est_i)
\!\!\!\!\!&=&\!\!\!\!\! \begin{cases}
     0, & y_i=0 \\
     \log \frac{\mu}{\lambda}, & \state_i > 0, y_i = 1 \\
     \infty & \text{otherwise}
    \end{cases} \label{eqn:examples:c}
\end{eqnarray}
Figure~\ref{fig:SeparationPrincipleESTC} is akin to \cite[Fig 4]{AnantharamVerdu96}, where it is shown that this insight and \eqref{eqn:examples:c} leads
to the derivation of the capacity of the exponential server timing channel.
\begin{remark}
Though the ESTC is time-varying, non-memoryless, and has non-linear dynamics from a inter-arrival time viewpoint \cite{AnantharamVerdu96}, when viewed appropriately, its internal structure consists of a time-invariant memoryless `Z' channel and a feedback loop comprising a linear SM coordination strategy $\bpi$.  Moreover, for a Poisson process input, $\bpi$ is inverse control optimal.  As such, the internal structure of the $\cdot/M/1$ queue can be interpreted as an optimal decentralized controller. Also, note how the internal structure is exactly synonymous to the Gaussian case \eqref{eqn:gaussianEx} (\cite{bansal1988sdc},\cite{tatikonda2000control}) in that the encoder and decoder are both linear dynamical systems.
\end{remark}

The result differs from the source-channel matching results in \cite[Sec 3]{GastparPhDThesis2002} for two reasons: i) the problem is approached through an inter-arrival viewpoint in \cite{GastparPhDThesis2002}, while we use counting function representation (inputs and outputs to the queue). ii) The dynamics of the ESTC are fixed and \cite{GastparPhDThesis2002} considers a possible encoder between the poisson process and the ESTC input, and a decoder between ESTC output and the reconstruction and show that the encoder and decoder should be identity mappings. In our case, the linear encoder and decoder policies are fixed and internal to the structure of the queue dynamics. As a consequence, the source-channel matching results has to be performed over a less complicated memoryless `Z' channel.

Other extensions to queuing timing channels fit within this framework as well: see for example the variety of queuing systems in \cite{martin2009batch} for which joint reversibility holds.  Similar results hold for other queuing timing channels, such as:
\begin{itemize}
  \item $\cdot/M/c$ queue: There are $c$ servers each with an i.i.d exponential service time. In this case, the queue dynamics-the linear encoder and the decoder will be the same (Fig \ref{fig:SeparationPrincipleESTC}). The structure of memoryless channel ($\PDMC$) will depend on $c$.
  \item  `The queue with feedback' \cite[p 204-205]{Gallager96Discrete}. Here, with probability $1-p_0$ departures from the queue instantaneously return to the input of the queue (independent of all other processes). The `effective' $Z$ channel changes $\mu\Delta$ to $p_0 \mu \Delta$ and all other arguments hold.
\end{itemize}

\subsection{\bf Inverse Optimal Control: Blackwell's Trapdoor Channel}
Here we show that the internal structure of Blackwell's trapdoor channel can be interpreted as a stationary Markov coordination strategy
that is inverse control optimal.

\begin{figure}[hbtp]
\centering
\resizebox{0.6\columnwidth}{!}{
\begin{overpic}[width=\columnwidth]{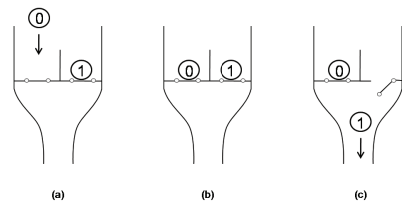}
\end{overpic}
}
\caption{Blackwell's Trapdoor Channel}
\label{fig:trapdoorchannelfig}
\end{figure}

Consider `the chemical (trapdoor) channel' \cite{Blackwell,AhlswedeKaspiTrapdoor,cuff06} as shown in Fig \ref{fig:trapdoorchannelfig}. Initially (Fig. \ref{fig:trapdoorchannelfig}a), a ball labeled either 0 (red) or 1 (blue) is present in one of the two slots. Then (Fig. \ref{fig:trapdoorchannelfig}b) a ball, either a 0 or 1, is placed in the empty slot, after which (Fig. \ref{fig:trapdoorchannelfig}c) one of the trapdoors opens at random with probability $(\frac{1}{2},\frac{1}{2})$. The ball lying above the open door then falls through. The door closes (as in Fig. \ref{fig:trapdoorchannelfig}a) and the process is repeated.

Let $\tSrc_i \in \{0,1\}$ and $\Out_i \in \{0,1\}$ represent the color of the ball that is input and output of the trapdoor respectively. Define the channel input $\State_i$ to pertain to the composition of balls before one of the doors is opened (Fig \ref{fig:trapdoorchannelfig}b). That is, $\State_i \in \{0,1,2\}$ where $\State_i=0$ represents two red balls (0,0), $\State_i=1$ represents a blue ball and a red ball (0,1)and $\State_i=2$ represent two blue balls (1,1). Thus, the dynamics are given by
\begin{eqnarray}
  \State_i &=& \State_{i-1}+\widetilde{\Src}_i-\Out_{i-1} \label{eqn:trapdoorpolicies:enc}
\end{eqnarray}
From a counting function viewpoint, let $\{\Src_i\}$ and $\{\Est_i\}$ be the counting processes representing the number of blue balls that were input and output from the system. Hence $\State_i$, as defined above tells about the composition of balls, or equivalently the number of blue balls that are `in' the system at time $i$.
\begin{subequations} \label{eqn:trapdoorpolicies}
\begin{eqnarray}
  \Src_i &=& \Src_{i-1}+\widetilde{\Src}_i \\
  \Est_i &=& \Est_{i-1}+\Out_i \label{eqn:trapdoorpolicies:dec} \\
  \State_i &=& \State_{i-1}+\widetilde{\Src}_i-\Out_{i-1} \nonumber \\
  &=& \Src_i - \Est_{i-1} \label{eqn:trapdoorpolicies:enc:1}
\end{eqnarray}
\end{subequations}
Note that the state-update equation and the decoding policy \eqref{eqn:trapdoorpolicies:dec}-\eqref{eqn:trapdoorpolicies:enc:1} are~\IRC by Example~\ref{example:inverse-reversible-compatible:b}. The output depends on the state according to the channel law $P_{\Out|\State}(\Out|\State)$ as (the inverse erasure channel) as shown in figure \ref{fig:TrapdoorchannelStatememlessChannelLaw}:
\begin{eqnarray}
\PDMC(1|\state)=\begin{cases}
             0 & \state=0 \\
             \frac{1}{2} & \state = 1 \\
             1 & \state=2 \\
           \end{cases}
\end{eqnarray}

\begin{figure}[hbtp]
\centering
\resizebox{\columnwidth}{!}{
\begin{overpic}[width=1.1\columnwidth]{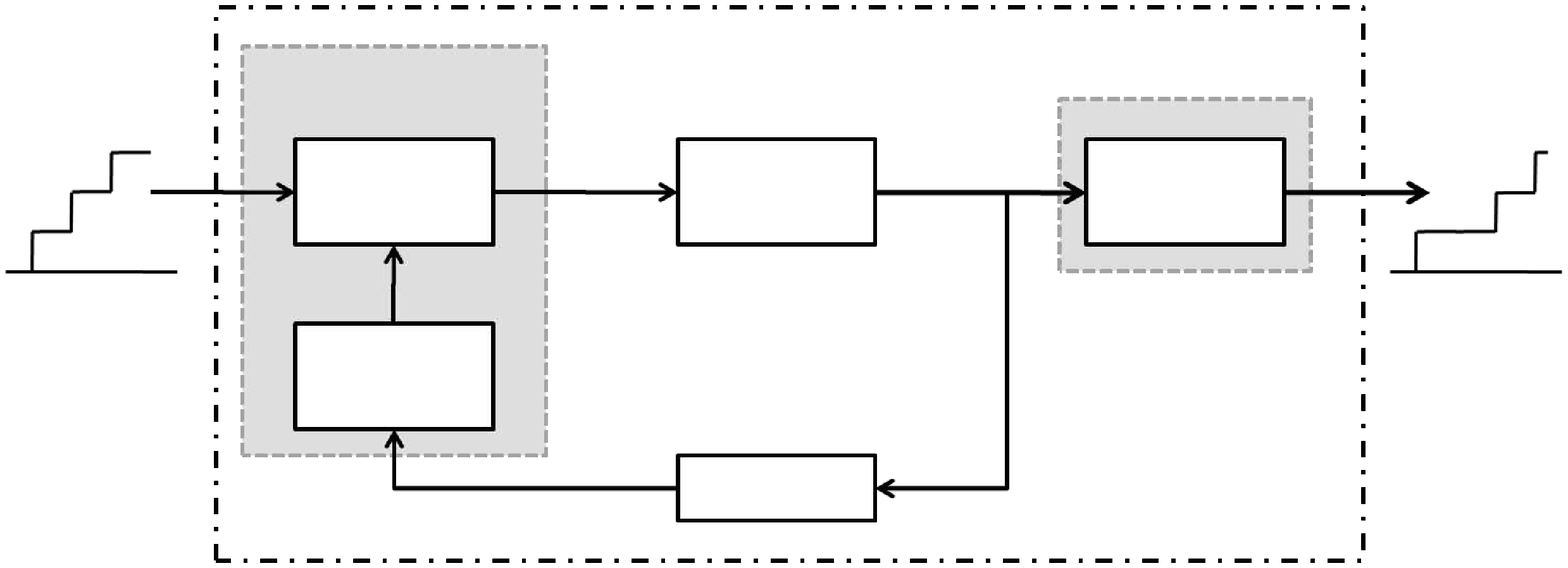}
\put(18,31){\small \it causal}
\put(18,28){\small \it encoder}
\put(23,21.5){\Large $+$}
\put(20,17){ \small $-$}
\put(27,17){$\Est_{i-1}$}
\put(19,10.5){\small accumul.}
\put(5,29){\large $\Src_i$}
\put(36.45,25){\large $\State_i$}
\put(46,22.5){$\PDMC$}
\put(36,6){\large $Y_{i-1}$}
\put(46,4){ delay}
\put(60,25){\large $Y_i$}
\put(65,31){\small \it causal decoder}
\put(70.5,23.5){\small accumul.}
\put(93,26){\large $\Est_i$}
\put(68,6){trapdoor}
\put(68,3){channel}
\end{overpic}
}
\caption{With $\Psrc$ a Markov counting process (i.i.d. $\tSrc_i$ inputs)  and an `inverted E' channel, Blackwell's trapdoor channel is inverse control optimal.}
\label{fig:DynamicsOfATrapdoorchannel}
\end{figure}

\begin{figure}[ht]
\begin{minipage}[b]{0.5\linewidth}
\centering
\begin{overpic}[width=0.7\columnwidth]{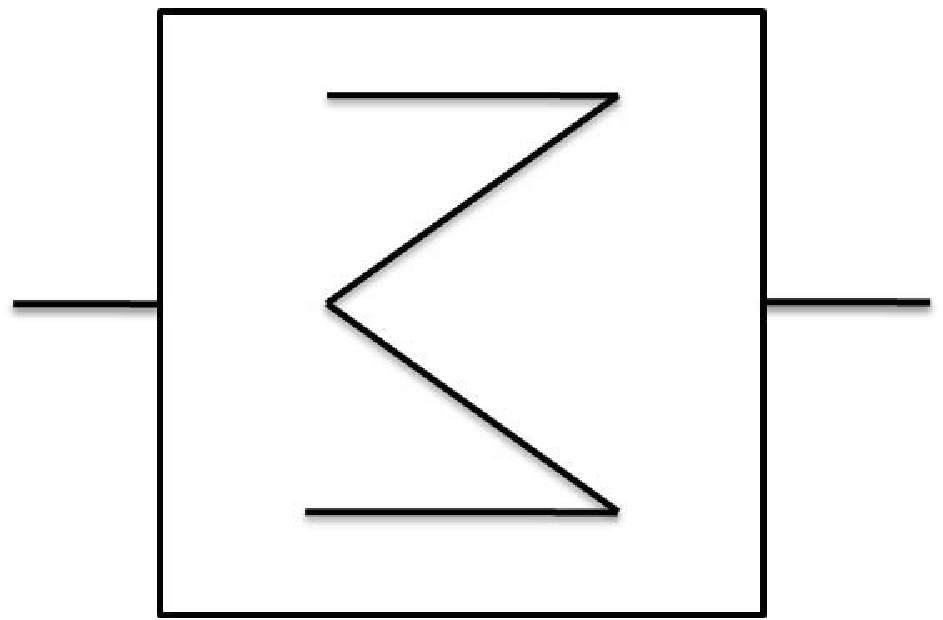}
\put(85,36){\large $\Out_i$}
\put(52,38){\small $\frac{1}{2}$}
\put(52,24){\small$\frac{1}{2}$}
\put(46,56){\small$1$}
\put(46,2.5){\small$1$}
\put(3,36){\large $\State_i$}
\put(27,50){\large $0$}
\put(27,30){\large $1$}
\put(27,10){\large $2$}
\put(68,50){\large $0$}
\put(68,10){\large $1$}
\end{overpic}
\caption{$\PDMC$ for the trapdoor channel}
\label{fig:TrapdoorchannelStatememlessChannelLaw}
\end{minipage}
\hspace{0.5cm}
\begin{minipage}[b]{0.4\linewidth}
\centering
\resizebox{0.9\columnwidth}{!}{
\begin{overpic}[width=1.1\columnwidth]{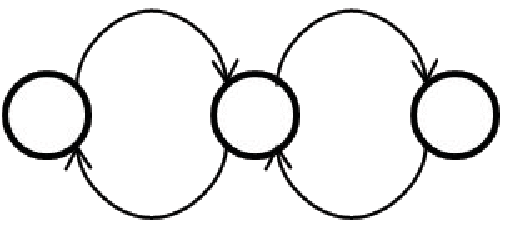}
\put(5,21){ $0$}
\put(19,48){  $1-p$}
\put(26,-5){  $\frac{1}{2}p$}
\put(47,21){  $1$}
\put(58,48){  $\frac{1}{2}(1-p)$}
\put(69,-3){  $p$}
\put(87,21){  $2$}
\end{overpic}
}
\caption{birth-death chain for $\State$ in the trapdoor channel with $\Psrc$ a Markov counting process.}
\label{fig:statetransitions}
\end{minipage}
\end{figure}
Fixing $\tSrc$ to be an i.i.d process,  with $\prob{\tSrc_i=0} = p$., and $\Est_{0}=0$.
The transition probabilities of the Markov Chain $\State_i$ are given by Fig \ref{fig:statetransitions}
\[ P = \left[ \begin{array}{ccc}
p & 1-p & 0 \\
\frac{1}{2}p & \frac{1}{2} & \frac{1}{2}(1-p) \\
0 & p & 1-p \end{array} \right] \]
 and if $\Src_{0}$ is drawn according to
\[ \prob{\Src_{0} = k} = \left\{
                              \begin{array}{ll}
                                p^2, & $k=0$; \\
                                2p(1-p), & $k=1$; \\
                                (1-p)^2, & $k=2$; \\
                                0, & \hbox{otherwise.}
                              \end{array}
                            \right.,\]
it follows that we have a birth-death chain initially in steady-state with distribution $\pi(\cdot) =  \prob{\Src_{0} = \cdot}$.
Thus from Lemma~\ref{lemma:SCmatching:Xbirthdeathchain}, we have that $\bpi$ is inverse-control optimal.
 Moreover, the from Corollary \ref{cor:SCmatching:decoderInvertible}, the trapdoor policy \eqref{eqn:trapdoorpolicies:enc} is optimal for the cost function of the form
\begin{eqnarray*}
  \dist(\src_i,\est_{i-1},\est_i) &=& \left\{
                                        \begin{array}{ll}
                                          \log p, & x_i=0,y_i=0; \\
                                          \log 2p, & x_i=1,y_i=0; \\
                                          \log 2(1-p), & x_i=1,y_i=1; \\
                                          \log (1-p), & x_i=2,y_i=1; \\
                                          +\infty, & \hbox{\mbox{otherwise}.}
                                        \end{array}
                                      \right.
\end{eqnarray*}
Note that when $p=\frac{1}{2}$, $\E\brackets{\dist(\Src,\Est_{i-1},\Est_i)} = -I(\pi,\PDMC)=-\frac{1}{2}$, which coincides with
the achievable rate coding scheme developed for the trapdoor channel in \cite{AhlswedeKaspiTrapdoor}.

\section{\bf Discussion and Conclusion}\label{sec:discussionConclusion}
In this paper, we have developed a new class of causal coding/decoding problems that can be understood
from the lens of both information theory as well as control theory.
We would like to emphasize that the primary focus of this paper is not about fundamental limits (although some new fundamental limits are presented). Rather, it is about attempting to develop a modeling framework whereby principles of traditional information theory (e.g. KL divergence, mutual information bounds, etc) and traditional concepts (e.g. dynamic programming, structural results), can be wed to elucidate things and impact the design of future real-world systems and applications.

A second aim is to demonstrate that by first formulating problems in this manner (whereby the notion of rate is not necessarily embedded directly in the problem formulation), fundamental information theoretic limits fall out as a consequence of solving the problem.  For example, in our information gain cost, we demonstrate how the posterior matching scheme
 by Shayevitz \& Feder is an optimal solution to a causal coding/decoding problem.  We did {\it not} attempt to directly
 impose the notion of achievability in the cost function, but rather constructed a cost function from the converse.  It was shown in
 \cite{shayevitz2009optimal} that in essence achievability still holds.
Moving forward, this suggests that perhaps equally as much attention should paid to the manner in which problems are formulated as what is being paid to attempt to solve already-formulated and un-solved problems.  The authors believe that a significant amount of practical and theoretical advances - including and extending beyond communication - can be made if the information theory community embraces this challenge.

In light of how the second law of thermo-dynamics appeared in Section~\ref{sec:directedInfo} and
how Markov chain time-reversibility appeared in Section~\ref{sec:timereversibility}, perhaps further work could be pursued to further understand the relationship between information theory and thermodynamics.  It has recently been suggested that such an understanding could additionally play a role in understanding brain function \cite{friston2010free,mitter2010towards}.
Recent developments in the neuroscience community have begun to posit that Bayesian decision-making could implicitly be playing a role in the processes of sensation, perception, and decision-making in the mammalian brain through  interacting neural systems sequentially handing to one another `what is missing' \cite{lee2003hierarchical}\cite{rao1999predictive}\cite{fiete2008grid,burak2009accurate}.  The sequential information gain framework perhaps could provide insight into further considering these matters.  Analogously, the inverse optimal control framework developed here, when combined with statistically inferring the coordination policies as in \cite{ng2000algorithms,abbeel2004apprenticeship,dvijothaminverse2010,abbeel2007application,kording2004loss,todorov2004optimality,baker2007goal}, could provide insight into what cost is being minimized.

\section{Acknowledgements}
The authors thank T.~Basar, T.~Bretl, B.~Hajek, N.~Kiyavash, P.~R.~Kumar, A.~Mahajan, P.~Mehta, S.~Meyn, M.~Raginsky,  R.~Srikant, and S.~Yuksel for useful discussion.

\begin{appendices}

\section{Proof of Lemma~\ref{lemma:structuralresult:fixedalphabetdynamics}}\label{appendix:proof:lemma:structuralresult:fixedalphabetdynamics}
\begin{IEEEproof}
As described in the statement of the lemma, define the state space $\cS = \cEst \times \probSimplex{\Src}$ and control space $\cU = \tilde{\alphabet{E}} \times \cEst$ with $s_i \in \cS, u_i \in \cU$ given by \eqref{eqn:defn:MDP:statesControls}:
\[ s_i = (\est_{i-1},\belief_{i|i}),  \qquad u=(\tenc_{i+1},\est_i). \]
Then
\beqa
&& \E\brackets{\dist(\Src_i,\Est_{i-1},\Est_i)  | \Est_{i-1}=\est_{i-1},\Out^i=\out^i, \Est_i=\est_i} \nonumber\\
&=&
\int_{\src_i \in \cSrc}
\dist(\src_i,\est_{i-1},\est_i) \belief_{i|i}(d\src_i) \label{appendix:proof:lemma:structuralresult:fixedalphabetdynamics:a}\\
&\deq& \bdist(s_i,z_i) \nonumber \\
&&\E\brackets{ \statecostfn(\State_{i+1}) |  \Est_{i-1}=\est_{i-1},\Out^i=\out^i, \tilde{E}_{i+1}=\tenc_{i+1}} \nonumber\\
&=& \int_{\src_{i+1} \in \cSrc}\alpha \statecostfn \parenth{\tenc_{i+1}\parenth{\src_{i+1}}} \belief_{i+1|i} (d\src_{i+1}) \nonumber\\
&=& \int_{\src_{i+1} \in \cSrc}\alpha \statecostfn \parenth{\tenc_{i+1}\parenth{\src_{i+1}}} \OSPU(\belief_{i|i})(d\src_{i+1}) \label{appendix:proof:lemma:structuralresult:fixedalphabetdynamics:c} \\
&\deq&\bstatecostfn(s_i,\tenc_{i+1}) \nonumber
\eeqa
where \eqref{appendix:proof:lemma:structuralresult:fixedalphabetdynamics:a} follows from \eqref{eqn:defn:beliefs};
\eqref{appendix:proof:lemma:structuralresult:fixedalphabetdynamics:c} follows from \eqref{eqn:defn:onestepupdaterule}.
\end{IEEEproof}

\section{Proof of Lemma~\ref{lemma:controlledMC}} \label{appendix:proofLemmaControlledMC}
\begin{IEEEproof}
Note that
\begin{subequations}
\beqa
 && P_{S_{i+1}|S^i=s^i,U^i=u^i}(ds_{i+1}) \nonumber \\
 &=& \!\!\!
  \indicatorvbl{s_{i+1,1}=u_{i,2}}
\int_{\src_{i+1} \in \cSrc}
\int_{y_{i+1} \in \cOut}
\!\!\indicatorvbl{\belief_{i+1|i+1} = \NLF(\belief_{i|i},y_{i+1},\tenc_{i+1})}\nonumber \\
&& \qquad \PDMC\parenth{dy_{i+1}|\tenc_{i+1}(\src_{i+1})} \belief_{i+1|i}(d\src_{i+1}) \label{eqn:decentralizedControl:controlledMC:a}\\
 &=&
 \indicatorvbl{s_{i+1,1}=u_{i,2}}
\int_{\src_{i+1} \in \cSrc}
\int_{y_{i+1} \in \cOut}
\!\!\!\indicatorvbl{s_{i+1,2} = \NLF(s_{i,2},y_{i+1},u_{i,1})} \nonumber \\
&& \qquad \PDMC\parenth{dy_{i+1}|u_{i,1}(\src_{i+1})} \OSPU(s_{i,2})(d\src_{i+1}) \label{eqn:decentralizedControl:controlledMC:b}\\
&=&  P_{S_{i+1}|S_i=s_i,U_i=u_i}(ds_{i+1}) \nonumber \\
&\deq& \PControlledMC(ds_{i+1}|s_i,u_i)\label{eqn:decentralizedControl:controlledMC:c}
\eeqa \label{eqn:decentralizedControl:controlledMC}
\end{subequations}
where \eqref{eqn:decentralizedControl:controlledMC:a} follows from \eqref{eqn:defn:DMC} and \eqref{eqn:defn:tSource};
\eqref{eqn:decentralizedControl:controlledMC:b} follows from \eqref{eqn:defn:MDP:statesControls}; and
\eqref{eqn:decentralizedControl:controlledMC:c} demonstrates that this is a controlled Markov chain with {\it time-invariant} statistical dynamics.
\end{IEEEproof}

\section{Proof of Lemma~\ref{lemma:problemSetup:mutualInfoSimplification}} \label{proof:lemma:problemSetup:mutualInfoSimplification}
\begin{IEEEproof}
\beqa
\!\!\!\!\!\!I(\Src^n;Y^n) \!\!&\!\!=\!\!& \!\!\sum_{i=1}^n I(\Src^n;Y_i|Y^{i-1}) \label{eqn:proof:lemma:problemSetup:mutualInfoSimplification:a} \\
              \!\!&\!\!=\!\!& \!\!\sum_{i=1}^n I(\Src^i;Y_i|Y^{i-1}) \nonumber \\
              &&\quad +I(\Src_{i+1}^n;Y_i|Y^{i-1},\Src^i)
              \label{eqn:proof:lemma:problemSetup:mutualInfoSimplification:b} \\
              \!\!&\!\!=\!\!& \!\!\sum_{i=1}^n I(\Src^i;Y_i|Y^{i-1}) \nonumber \\
              &&\quad +I(\Src_{i+1}^n;Y_i|Y^{i-1},\Src^i,\State_i)
              \label{eqn:proof:lemma:problemSetup:mutualInfoSimplification:c} \\
              \!\!&\!\!=\!\!& \!\! \sum_{i=1}^n I(\Src^i;Y_i|Y^{i-1}) \label{eqn:proof:lemma:problemSetup:mutualInfoSimplification:d}\\
                  \!\!&\!\!=\!\!& \!\!\sum_{i=1}^n I(\Src_i;Y_i|Y^{i-1}) \nonumber \\
                  &&\quad+I(\Src^{i-1};Y_i|\Src_i,Y^{i-1}) \label{eqn:proof:lemma:problemSetup:mutualInfoSimplification:g} \\
                  \!\!&\!\!=\!\!& \!\!\sum_{i=1}^n I(\Src_i;Y_i|Y^{i-1}) \nonumber \\
                  &&\quad+I(\Src^{i-1};Y_i|\Src_i,Y^{i-1},\State_i) \label{eqn:proof:lemma:problemSetup:mutualInfoSimplification:gg} \\
                  \!\!&\!\!=\!\!&\!\! \sum_{i=1}^n I(\Src_i;Y_i|Y^{i-1}) \label{eqn:proof:lemma:problemSetup:mutualInfoSimplification:h}
\eeqa
where \eqref{eqn:proof:lemma:problemSetup:mutualInfoSimplification:a} follows from \eqref{eqn:defn:chainRuleMutualInformation};
\eqref{eqn:proof:lemma:problemSetup:mutualInfoSimplification:b} follows from from \eqref{eqn:defn:chainRuleMutualInformation};
\eqref{eqn:proof:lemma:problemSetup:mutualInfoSimplification:c} follows from \eqref{eqn:defn:policy};
\eqref{eqn:proof:lemma:problemSetup:mutualInfoSimplification:d} follows follows from \eqref{eqn:defn:DMC};
\eqref{eqn:proof:lemma:problemSetup:mutualInfoSimplification:g} follows from \eqref{eqn:defn:chainRuleMutualInformation};
\eqref{eqn:proof:lemma:problemSetup:mutualInfoSimplification:gg} follows from our assumption \eqref{eqn:thm:mainThmGeneralProblem:a} that
the encoder operates on sufficient statistics; and \eqref{eqn:proof:lemma:problemSetup:mutualInfoSimplification:h} follows from
 \eqref{eqn:defn:DMC}.
\end{IEEEproof}

\section{Proof of Lemma~\ref{lemma:loglikelihoodratiocost}} \label{appendix:proof:lemma:loglikelihoodratiocost}
\begin{IEEEproof}
\eqref{eqn:lemma:loglikelihoodratiocost:b} follows directly from Lemma~\ref{lemma:structuralresult:fixedalphabetdynamics}.
Now, let us focus on \eqref{eqn:lemma:loglikelihoodratiocost:c}.  From Lemma~\ref{lemma:structuralresult:fixedalphabetdynamics}, we have that
\beqa
\tdist(s,\est) &=& \int_{\src \in \cSrc} \dist(\src,b,\est) \belief'(d\src) \nonumber \\
             &=&  \int_{\src \in \cSrc} - \log \frac{d\est}{d \OSPU(b)}(\src)\belief'(d\src) \label{eqn:proof:lemma:loglikelihoodratiocost:a}
\eeqa
where \eqref{eqn:proof:lemma:loglikelihoodratiocost:a} follows from \eqref{eqn:maximizationDirInfo:CostFn:costfn}
for any $\est$ satisfying $\est \ll \OSPU(b)$ and is infinite otherwise.  Now note that if it is not the case that
$b' \ll \est$, then there exists a set $A \in \borel{\cSrc}$ for which $\est(A)=0$ and $b'(A) > 0$ and thus it follows that
$\frac{d\est}{d \OSPU(b)}(\src)=0 \Rightarrow -\log \frac{d\est}{d \OSPU(b)}(\src)=\infty$ for all $\src \in A$.  Thus $\tdist(s,\est)=\infty$.
Now assume $b' \ll \est \ll \OSPU(b)$.  Then since if $\beta \ll \nu \ll \mu$ then $\frac{d\beta}{d\mu}= \frac{d\beta}{d\nu} \frac{d\nu}{d\mu}$, $\mu$-almost everywhere \cite[Sec 5.5]{dudley2002real}, it follows that
\beqa
\tdist(s,\est) &=&  \int_{\src \in \Src} -\log \frac{d b'}{d \OSPU(\belief)}(\src) b'(d\src) + \log \frac{d b'}{d \est}(\src) b'(d\src)   \nonumber \\
               &=& -\kldist{\belief'}{\OSPU(\belief)} + \kldist{b'}{z}
\eeqa
\end{IEEEproof}
\begin{strip}
\section{Proof of Theorem~\ref{thm:optimalDecoderMaximizationDirInfo}}\label{sec:proof:lemma:maximizationDirInfo}
\begin{IEEEproof}
In order to find the optimal cost $J_n^*$ given by $J_n^* = \E\brackets{\valuefn_0(S_0)}$, we use the standard dynamic programming approach and evaluate optimal cost-to-go functions $\{\valuefn_k:k=0,\cdots,n\}$.  Consider the final-stage problem of finding $\valuefn_n(s_n)$, where $s_n=(\est_{n-1},\belief_{n|n})$ and
describe any control $u_n$ as $u_n=(\tenc_{n+1},\est_n)$. Then the one-stage problem is
\begin{eqnarray}
\valuefn_{n}\parenth{s_n} &=&  \inf_{u_n=(\tenc_{n+1},\est_n)} \bcost_n\parenth{s_n,u_n} \nonumber \\
                                   &=& \inf_{\est_n \in \probSimplex{\cSrc}} \bdist(s_n,\est_n)  \label{eqn:maximizationDirInfo:Proof:a}  \\
                                   &=& -\kldist{\belief_{n|n}}{\OSPU(\est_{n-1})} + \inf_{\est_n \in \probSimplex{\cSrc},\;\;\belief_{n|n} \ll \est_n \ll \OSPU(\est_{n-1})} \kldist{\belief_{n|n}}{\est_n}  \label{eqn:maximizationDirInfo:Proof:b} \\
                                   &=& - \kldist{\belief_{n|n}}{\OSPU(\est_{n-1})} \label{eqn:maximizationDirInfo:Proof:d}
\end{eqnarray}
where  \eqref{eqn:maximizationDirInfo:Proof:a} follows \eqref{eqn:decentralizedControl:defn:tcost};
 \eqref{eqn:maximizationDirInfo:Proof:b} follows from \eqref{eqn:lemma:loglikelihoodratiocost:c};
  and \eqref{eqn:maximizationDirInfo:Proof:d} follows from the non-negativity of the KL divergence.  The optimal choice of $\est_n$ is the one for which the equality in \eqref{eqn:maximizationDirInfo:Proof:d} holds true and hence under an optimal policy, $z_{n}=\belief_{n|n}$. This follows the same reasoning that elicits how for in the self-information loss sequential probability assignment, the best probability assignment is the true belief \cite{merhav2002universal}.

For the second-step $k=n-1$, consider finding $\valuefn_{n-1}(s_{n-1})$, where $s_{n-1}=(\est_{n-2},\belief_{n-1|n-1})$ and
describe any control $u_{n-1}$ as $u_{n-1}=(\tenc_{n},\est_{n-1})$.  Then we have:
\begin{eqnarray}
\valuefn_{n-1}\parenth{s_{n-1}} &=&  \inf_{u_{n-1}=(\tenc_n,\est_{n-1})} \bcost_{n-1}\parenth{s_{n-1},u_{n-1}} +
   \E\brackets{\valuefn_n\parenth{z_{n-1},B_{n|n}}|S_{n-1}=s_{n-1},U_{n-1}=u_{n-1}}\\
                                   &=& \inf_{\tenc_n,\est_{n-1}} \alpha \bstatecostfn(s_{n-1},\tenc_n)+ \bdist(s_{n-1},\est_{n-1})  + \E\brackets{-\kldist{B_{n|n}}{\OSPU(\est_{n-1})}|S_{n-1}=s_{n-1},U_{n-1}=u_{n-1}}  \label{eqn:maximizationDirInfo:Proof:n-1:a}  \\
                                   &=& -\kldist{\belief_{n-1|n-1}}{\OSPU(\est_{n-2})}
                                   + \inf_{\tenc} \alpha\bstatecostfn(s_{n-1},\tenc_n)  +
                                   \inf_{\belief_{n-1|n-1} \ll \est_{n-1} \ll \OSPU(\est_{n-2})} \kldist{\belief_{n-1|n-1}}{z_{n-1}} \nonumber \\
                                   && + \E\brackets{-\kldist{B_{n|n}}{\OSPU(\est_{n-1})}|S_{n-1}=s_{n-1},\tEnc_{n}=\tenc_n}  \label{eqn:maximizationDirInfo:Proof:n-1:b}
\end{eqnarray}
where  \eqref{eqn:maximizationDirInfo:Proof:n-1:a} follows by substituting values of $\bcost_{n-1}$ and $\valuefn_n$ from \eqref{eqn:decentralizedControl:defn:tcost} and \eqref{eqn:maximizationDirInfo:Proof:d}; \eqref{eqn:maximizationDirInfo:Proof:n-1:b} follows  from \eqref{eqn:lemma:loglikelihoodratiocost:c}.

For any fixed encoder policy $\tenc_n$, the optimal choice for $\est_{n-1}$ is to pick $\est_{n-1} = \belief_{n-1|n-1}$ as shown:
\begin{eqnarray}
z^*_{n-1}(s_{n-1}) \!\!\!&=&
\arginf_{\belief_{n-1|n-1} \ll \est_{n-1} \ll \OSPU(\est_{n-2})} \kldist{\belief_{n-1|n-1}}{z_{n-1}}  - \E\brackets{\kldist{B_{n|n}}{\OSPU(\est_{n-1})}|S_{n-1}=s_{n-1},\tEnc_{n}=\tenc_n}  \label{eqn:maximizationDirInfo:Proof:n-1:f} \\
             &=&  \arginf_{\belief_{n-1|n-1} \ll \est_{n-1} \ll \OSPU(\est_{n-2})}
             \kldist{\PnextYgivenBeliefEncEqualsBWithnMinusOne}{\PnextYgivenBeliefEncEqualsZWithnMinusOne}  \nonumber \\
             &+&\!\!\!\!\underbrace{\kldist{\belief_{n-1|n-1}}{z_{n-1}}
                  -
                  \E
                  \brackets{ \kldist{\NLF\parenth{\belief_{n-1|n-1}, Y_n, \tenc_n}}{\NLF\parenth{\est_{n-1}, Y_n, \tenc_n}}|S_{n-1}=s_{n-1},\tEnc_{n}=\tenc_n}}_{\geq 0} 
                  \label{eqn:maximizationDirInfo:Proof:n-1:g} \\
             &=& \belief_{n-1|n-1}.     \label{eqn:maximizationDirInfo:Proof:i}
\end{eqnarray}
where \eqref{eqn:maximizationDirInfo:Proof:n-1:g} and the non-negativity of the difference follow because:
\beqa
z^*_{n-1}(s_{n-1}) \!\!\!\!&=&  \arginfTermNMinusOne \kldistTermAnminusOne  - \E\brackets{\kldist{\Belief_{n|n}}{\OSPU(\est_{n-1})}|S_{n-1}=s_{n-1},\tEnc_{n}=\tenc_n}\nonumber\\
             &=&  \arginfTermNMinusOne \kldistTermAnminusOne \nonumber\\
             &-& \E\brackets{ \kldist{\Belief_{n|n}}{\OSPU(\belief_{n-1|n-1})} - \E_{\Belief_{n|n}} \brackets{ \log \frac{d \OSPU(\belief_{n-1|n-1})}{d\OSPU(\est_{n-1})} \Big| \Belief_{n|n}} \Bigg| \Belief_{n-1|n-1}=\belief_{n-1|n-1}  } \label{eqn:appendix:maximizationDirInfo:Proof:a}\\
            &=&  \arginfTermNMinusOne \kldistTermAnminusOne \nonumber \\
            &-& \int_{y \in \cOut} \int_{\src \in \cSrc} \log \frac{d \OSPU(\belief_{n-1|n-1})}{d\OSPU(\est_{n-1})} (\src) \underbrace{\NLF(\belief_{n-1|n-1},y,\tenc_n)}_{\belief_{n|n}}(d\src) \PnextYEqualsyGivenBeliefEncEqualsBWithnMinusOne  \label{eqn:appendix:maximizationDirInfo:Proof:b}\\
            &=&   \arginfTermNMinusOne\kldistTermAnminusOne \nonumber \\
            &-& \!\!\int_{y \in \cOut} \int_{\src \in \cSrc} \log \parenth{\frac{d \NLF(\belief_{n-1|n-1},y,\tenc_n)}{d\NLF(\est_{n-1},y,\tenc_n)}} (\src) \NLF(\belief_{n-1|n-1},y,\tenc_n)(d\src) \PnextYEqualsyGivenBeliefEncEqualsBWithnMinusOne  \nonumber \\
            &+& \int_{y \in \cOut} \log \frac{d\PnextYgivenBeliefEncEqualsBWithnMinusOne}{d\PnextYgivenBeliefEncEqualsZWithnMinusOne}(y)
              \PnextYEqualsyGivenBeliefEncEqualsBWithnMinusOne  \label{eqn:appendix:maximizationDirInfo:Proof:c}\\
             &=&  \arginfTermNMinusOne  \kldist{\PnextYgivenBeliefEncEqualsBWithnMinusOne}{\PnextYgivenBeliefEncEqualsZWithnMinusOne} \nonumber \\ &+& \underbrace{\kldistTermAnminusOne
                  -\E_{\PnextYgivenBeliefEncEqualsBWithnMinusOne}\brackets{ \kldist{\NLF\parenth{\belief_{n-1|n-1}, Y_n, \tenc_n}}{\NLF\parenth{\est_{n-1}, Y_n, \tenc_n}}}}_{\geq 0} \label{eqn:appendix:maximizationDirInfo:Proof:d}
\eeqa
where \eqref{eqn:appendix:maximizationDirInfo:Proof:a} follows because $\Belief_{n|n} \ll \OSPU(\belief_{n-1|n-1}) \ll \OSPU(\est_{n-1})$ and so
$\frac{d\Belief_{n|n}}{d\OSPU(\est_{n-1})} = \frac{d\Belief_{n|n}}{d\OSPU(\belief_{n-1|n-1})}\frac{d\OSPU(\belief_{n-1|n-1})}{d\OSPU(\est_{n-1})}$;
\eqref{eqn:appendix:maximizationDirInfo:Proof:b} follows from the definition of the nonlinear filter \eqref{eqn:lemma:recursiveBeliefUpdate:a};
 \eqref{eqn:appendix:maximizationDirInfo:Proof:c} follows from the fact that $\belief_{n-1|n-1} \ll \est_{n-1}$ and the definition of the nonlinear filter in \eqref{eqn:defn:nonlinearfilter};
 and the difference in \eqref{eqn:appendix:maximizationDirInfo:Proof:d} being non-negative follows from mapping
 this scenario to that of the hidden Markov model and the nonlinear filter:
 \bitm
 \item Here, the latent Markov process is $\Src$ and one observation  $Y_n$ is recorded.
 \item Because in this dynamic programming problem, while in state $s_{n-1}$ and under a fixed $\tenc_n: \cSrc \to \cState$,
 the noisy channel from $\Src_n$ to $Y_n$ is the composition of the encoder map $\tenc_n$ and the input to the channel from $\State_n$ to $Y_n$:
 $P_{Y_n|\Src_n}(dy|\src_n) = P_{Y|\State}(dy|\tenc_n(\src_n))$.
 \item Two different decoders both know the statistical dynamics but have different initial beliefs about $\Src_{n-1}$. One decoder's initial belief is $\belief_{n-1|n-1} \in \probSimplex{\cSrc}$
 and the other's is $\est_{n-1} \in \probSimplex{\cSrc}$.  The initial `distance' between the beliefs is measured by the KL divergence, $\kldist{\belief_{n-1|n-1}}{\est_{n-1}}$.
\item Both decoders observe $Y_n$ and update their beliefs about $\Src_n$ according to the one-step nonlinear filter one updates its belief according to $\NLF(\belief_{n-1|n-1},Y_n,\tenc_n)$ and the other does so according to $\NLF(\est_{n-1},Y_n,\tenc_n)$.  The divergence between their beliefs after the observation is given by $\kldist{\NLF\parenth{\belief_{n-1|n-1}, Y_n, \tenc_n}}{\NLF\parenth{\est_{n-1}, Y_n, \tenc_n}}$ and on average this is smaller than the original due to Jensen's inequality and the second law of thermodynamics for hidden Markov chains.   This inequality is thus  a manifestation of how the relative entropy is a `Lyapunov function' for the stability (e.g. insensitivity to initial beliefs) of the nonlinear filter \cite[Remark 4.2]{chigansky2009intrinsic}.
 \eitm
  Hence the optimal choice for $\est_{n-1}$ is to pick $\belief_{n-1|n-1}$. Consequently,
\begin{eqnarray}
\valuefn_{n-1}\parenth{s_{n-1}} &=& -\kldist{\belief_{n-1|n-1}}{\OSPU(\est_{n-2})} + \inf_{\tenc_n} \alpha\bstatecostfn(s_{n-1},\tenc_n) + \E\brackets{\valuefn_n\parenth{\belief_{n-1|n-1},B_{n|n}}|\Belief_{n-1|n-1}=\belief_{n-1|n-1},\tEnc_{n}=\tenc_n} \nonumber\\
&=& -\kldist{\belief_{n-1|n-1}}{\OSPU(\est_{n-2})}+ \alpha\bstatecostfn(s_{n-1},\tenc^*_n[\belief_{n-1|n-1}]) \nonumber \\
&+& \E\brackets{\valuefn_n\parenth{\belief_{n-1|n-1},B_{n|n}}|\Belief_{n-1|n-1}=\belief_{n-1|n-1},\tEnc_{n}=\tenc^*_n[\belief_{n-1|n-1}]}  \label{eqn:maximizationDirInfo:Proof:n-1:e}
\end{eqnarray}

Using an inductive argument and the exact same set of arguments as above, it follows that for any $1 \leq k \leq n-1$, and any encoder policy $\tenc_{k+1}$, the optimal choice for $z_k$ is given by $\belief_{k|k}$ and that for $s_k=(\est_{k-1},\belief_{k|k})$,
  \begin{eqnarray}
  \valuefn_{k}\parenth{s_k} &=&-\kldist{\belief_{k|k}}{\OSPU(\est_{k-1})}+ \alpha\bstatecostfn(s_{k},\tenc^*_{k+1}[\belief_{k|k}])
  \nonumber \\
&+& \E\brackets{\valuefn_{k+1}\parenth{(\belief_{k|k},\Belief_{k+1|k+1})}|\Belief_{k|k}=\belief_{k|k},\tEnc_{k+1}=\tenc^*_{k+1}[\belief_{k|k}]}
  \end{eqnarray}
  For the initial step, $k=0$, by definition $\Est_0(A) = \Belief_{0|0}(A) \prob{\Src_0 \in A}$ and is known to both encoder and decoder.
  Thus the minimization is only over $\tenc_1$. For a state $s_0=(z_{-1},\belief_{0|0})$ and control $u_0 = (\tenc_1,z_0)=(\tenc_1,\belief_{0|0})$, we have:
  \begin{eqnarray}
\valuefn_{0}\parenth{s_0} &=& \inf_{\tenc_1} \alpha \bstatecostfn(s_0,\tenc_1)+ \E\brackets{\valuefn_1\parenth{\belief_{0|0},B_{1|1}}|\Belief_{0|0}=\belief_{0|0},\tEnc_{1}=\tenc_1} \label{eqn:maximizationDirInfo:Proof:0:c} \\
&=& \alpha \bstatecostfn(s_0,\tenc^*_1[\belief_{0|0}])+ \E\brackets{\valuefn_1\parenth{\belief_{0|0},B_{1|1}}|\Belief_{0|0}=\belief_{0|0},\tEnc_{1}=\tenc^*_1[\belief_{0|0}]}
\end{eqnarray}
 Next,  from \eqref{eqn:maximizationDirInfo:expansionMutualInfo} we have that
 $I(\Src_i;Y_i|Y^{i-1})= \E \brackets{\kldist{B_{i|i}}{\OSPU(B_{i-1|i-1})}}$
 and thus from \eqref{lemma:problemSetup:mutualInfoSimplification} we have:
\begin{eqnarray}
\totalcostpistar = \E\brackets{\valuefn_0(S_0)} = \min_{\enc \in \cEnc} -I(\Src^n; Y^n) + \alpha  \E_{\enc} \brackets{  \sum_{i=1}^{n}\statecostfn(\State_i)}.
\label{eqn:maximizationDirInfo:CostFn:b}
\end{eqnarray}
Lastly, it follows directly that a more `concise' sufficient statistic exists for the encoder - namely that it does not need to maintain $\est_{i-1}$ to produce $\state_{i+1}$
because under any optimal scheme, $\Est_{i-1} = \Belief_{i-1|i-1}$ and thus $\sigma(\Est_{i-1}) \subset \sigma(\Belief_{i|i})$ so the state
variable $S_i=(\Est_{i-1},\Belief_{i|i})$ can be reduced to $S_i=(\Belief_{i|i})$ with Lemma~\ref{lemma:controlledMC} still holding.
\end{IEEEproof}
\end{strip}

\section{Proof of Lemma~\ref{lemma:RlessthanC}}\label{sec:proof:lemma:RlessthanC}
\begin{IEEEproof}
Note the following standard set of inequalities:
\begin{subequations}
\begin{eqnarray}
\!\!\!R_n(\dist,P_{\Src^n},\distval)\!\!\!\!&\leq& \frac{1}{n}I(\Src^n;\Est^n) \label{eqn:RlessthanC:a} \\
&\leq&  \frac{1}{n}I(\Src^n;\Out^n) \label{eqn:RlessthanC:b}\\
&=& \frac{1}{n}\sum_{i=1}^n I(\Src_i;\Out_i|\Out^{i-1}) \label{eqn:RlessthanC:bbb}\\
\!\!\!\!&=& \!\!\!\!\frac{1}{n}\sum_{i=1}^n \kldist{P_{\Out_i|\Src_i,\Out^{i-1}}}{P_{\Out_i|\Out^{i-1}}|P_{W_i,\Out^{i-1}}} \nonumber\\
\!\!\!\!&=& \!\!\!\!\frac{1}{n}\sum_{i=1}^n \!\kldist{P_{\Out_i|\State_i}}{P_{\Out_i|\Out^{i-1}}|P_{X_i,\Out^{i-1}}} \label{eqn:RlessthanC:bc}\\
&\leq& \frac{1}{n}\sum_{i=1}^n \kldist{P_{\Out_i|\State_i}}{P_{\Out_i}|P_{X_i}} \label{eqn:RlessthanC:bd}\\
&=&  \frac{1}{n}\sum_{i=1}^n I(\State_i;Y_i) \nonumber\\
&\leq& \frac{1}{n} \sum_{i=1}^n \capacityCostFnVal{\E[\statecostfn(\State_i)]}  \label{eqn:RlessthanC:d} \\
&\leq&  \capacityCostFn   \label{eqn:RlessthanC:e}
\end{eqnarray}
\end{subequations}
where \eqref{eqn:RlessthanC:a} follows  \eqref{eqn:defn:RateDistFn};
\eqref{eqn:RlessthanC:b} follows from the data processing inequality; \eqref{eqn:RlessthanC:bbb} follows from
Lemma~\ref{lemma:problemSetup:mutualInfoSimplification}; \eqref{eqn:RlessthanC:bc} follows from the definition of conditional mutual information
\eqref{eqn:defn:conditionalMutualInformation} and the fact that $\State_i$ is a function of $\Src_i$ and $\Out^{i-1}$ under policy $\encSC$;  \eqref{eqn:RlessthanC:bd} follows from the memoryless nature of the channel \eqref{eqn:defn:DMC} and Jensen's inequality;
\eqref{eqn:RlessthanC:d} follows from \eqref{eqn:defn:capacityCostFn}; and
\eqref{eqn:RlessthanC:e} follows  from  \eqref{eqn:SCmatching:defn:a} and the concavity of the capacity-cost function \cite{mceliece2002theory}.
\end{IEEEproof}

\section{Proof of Lemma~\ref{lemma:SC:markovDecoder}} \label{appendix:proof:lemma:SC:markovDecoder}
\begin{IEEEproof} To prove \eqref{eqn:lemma:SC:markovDecoder1},
\begin{eqnarray}
 && P_{\Est_i|\Est^{i-1}=\est^{i-1},\Src^n=\src^n}(d\est_i) \nonumber \\
 &=& \int_{\cOut} \!\!\!\! P_{\Est_i|\Est^{i-1}=\est^{i-1},\Src^n=\src^b,\Out_i=\out}(d\est_i) \nonumber \\
 && \qquad \times P_{\Out_i|\Est^{i-1}=\est^{i-1},\Src^n=\src^n}(d\out) \nonumber \\
&=& \int_{\cOut} P_{\Est_i|\Est^{i-1}=\est^{i-1},\Src^n=\src^b,\Out_i=\out}(d\est_i)  \nonumber \\
&&\quad \times P_{\Out_i|\Est^{i-1}=\est^{i-1},\Src^n=\src^n,\State_i=\encSC(\src_i,\est_{i-1})}(d\out) \label{eqn:defineZplusZminusProof:1:eq1} \\
&=&  \int_{\cOut} \indicatorvbl{z_i=\decSC(\est_{i-1},\out_i)} P_{\Out|\State=\encSC(\src_i,\est_{i-1})}(d\out)  \label{eqn:defineZplusZminusProof:1:eq2} \\
   &\triangleq& Q_{\Est'|\Est,\Src'}(d\est_i|\est_{i-1},\src_i) \label{eqn:defineZplusZminusProof:1:eq3}
\end{eqnarray}
where \eqref{eqn:defineZplusZminusProof:1:eq1} follows from the stationary Markov encoder policy: $\state_i = \encSC(\src_i,\est_{i-1})$;
\eqref{eqn:defineZplusZminusProof:1:eq2} follows from defining
$\indicatorvbl{z_i=\decSC(\est_{i-1},\out_i)}$ as a Dirac measure at the point $\decSC(\est_{i-1},\out_i)$, the stationary Markov decoder policy $\est_i = \decSC(\est_{i-1},\out)$, and the
non-anticipative and memoryless nature of the channel \eqref{eqn:defn:DMC};
and \eqref{eqn:defineZplusZminusProof:1:eq3} simply denotes the time-invariant nature of the conditional distribution;

To prove \eqref{eqn:lemma:SC:markovDecoder2}, we exploit the assumption that $\{ \Out_i\}$ are i.i.d.  Because of this, we can denote
$(\Est_i: i=1,\ldots,n)$ by the following composition of independent random maps:
\[ \Est_i = \decSC(\Est_{i-1},\Out_i) \triangleq \decSC_{\Out_i}(\Est_{i-1}) = \decSC_{\Out_i} \circ \decSC_{\Out_{i-1}}\circ \cdots \circ \decSC_{\Out_2}(\Est_1)  \]
This is thus an an iterated function system (IFS) \cite{ykifer}, which is a time-homogeneous Markov chain over the state space $\cEst$.
\end{IEEEproof}

\section{Proof of Lemma~\ref{lemma:gaussianEx}}\label{appendix:gaussian}
\begin{IEEEproof}
Let $\Error_i \triangleq \Src_i - \E\brackets{\Src_i|\Out^{i-1}}$ be the error term in estimation.
We now select the statistics of $\Src_0$ such that $\State_i \sim \mathcal{N}(0,\statecostval), \forall i$.  The normalizing coefficient can be expressed as $\beta_i = \sqrt{\frac{\statecostval}{\text{Cov}(\Error_i,\Error_i)}}$, where the covariance of the error term can be recursively computed using
\begin{eqnarray}
   \text{Cov}(\Error_i,\Error_i)=\left\{
                                      \begin{array}{ll}
                                        \frac{\rho^2\sigma_n^2}{\statecostval+\sigma_n^2} \text{Cov}(\Error_{i-1},\Error_{i-1}) + \sigma_m^2, & i\geq 1; \\
                                        \text{Cov}(\Src_0,\Src_0), & i=0
                                      \end{array}
                                    \right.
      \label{eqn:gaussianEx:err1}
\end{eqnarray}
Let the steady state value of the covariance from \eqref{eqn:gaussianEx:err1} be denoted by $C$. Then,
\begin{eqnarray}
  C &\triangleq& \frac{\sigma_m^2}{1-\rho^2\frac{\sigma_n^2}{\statecostval+\sigma_n^2}} \label{eqn:gaussianEx:covErr}
\end{eqnarray}
Note that because of the choice of $\tilde{\Src}_0$ in \eqref{eqn:gaussianEx:src}, $\text{Cov}(\Error_i,\Error_i) = C$ and $\beta_i= \beta = \sqrt{\frac{\statecostval}{C}}$ for all $i \geq 0$.

Since all operations are linear and all primitive random variables $(\tSrc_i, V_i: i \geq 1)$ are i.i.d. and Gaussian, and since all other relationships are linear, all random variables are jointly Gaussian.  From standard MMSE estimation theory, $\Error_i$ thus independent of $\Out^{i-1}$.  As such, clearly $I(\State_i;\Out^{i-1})=0$.  Since the initial condition $\tSrc_0$ is chosen according to \eqref{eqn:gaussianEx:init}, $\State_i \sim \mathcal{N}(0,L)$ for all $i$. Therefore, since the variance of $V_i$'s is $\sigma_v^2$, this means that $Y$'s are i.i.d.
The policies \eqref{eqn:gaussianEx:optencdec} are thus stationary-Markov coordination strategies:
\begin{subequations} \label{eqn:gaussianEx:optencdec:ver2:app}
\begin{eqnarray}
  \State_i &=& \beta \parenth{\Src_i - \rho \Est_{i-1}} \label{eqn:gaussianEx:optencdec:ver2:app:a}\\
   \Est_i &=& \rho \Est_{i-1} + \gamma Y_i \label{eqn:gaussianEx:optencdec:ver2:app:b}
\end{eqnarray}
\end{subequations}
 where \eqref{eqn:gaussianEx:optencdec:ver2:app:a} follows  because $\E\brackets{\Src_i|\Out^{i-1}} = \E\brackets{\rho\Src_{i-1}+\tilde{\Src}_i|\Out^{i-1}} = \rho \Est_{i-1}$, and \eqref{eqn:gaussianEx:optencdec:ver2:app:b} follows by expanding $\E\brackets{\Src|\Out^{i}}$ using the innovation sequence and exploiting how  $\Out_i$ are i.i.d. The value of the parameters $\beta,\gamma$ are given by
 \begin{eqnarray}
\beta &=& \sqrt{\frac{\statecostval}{C}},\gamma = \frac{\beta C}{\statecostval+\sigma_n^2} \label{eqn:gaussianEx:alphabeta}
 \end{eqnarray}
Note that from the definition of $C$ in \eqref{eqn:gaussianEx:covErr}, $P_{\Src_i|\Est_{i-1}=\est_{i-1}} \sim \mathcal{N}(\rho \est_{i-1},C)$. Hence, using \eqref{eqn:gaussianEx:optencdec:ver2},
 \begin{eqnarray}
   Q_{\Est'|\Est,\Src'}(\cdot|\est_{i-1},\src_i) &\sim& \mathcal{N}(\rho \est_{i-1}+\beta\gamma(\src_i - \rho \est_{i-1}), \gamma^2 \NoiseVariance) \nonumber \\
   Q_{\Est'|\Est}(\cdot|\est_{i-1}) &\sim& \mathcal{N}(\rho \est_{i-1},\gamma^2 (\statecostval + \NoiseVariance)) \nonumber
 \end{eqnarray}

From Theorem \ref{thm:SCmatching}, the linear stationary Markov coordination strategy \eqref{eqn:gaussianEx:optencdec} is inverse control optimal for a $\dist$ of the form
\begin{eqnarray}
\!\!\!\!\!\!\!\!\!\!\!\!\dist(\src_i,\est_{i-1},\est_i) &\!\!\!\!\!\propto_+& -  \log
\frac{dQ_{\Est'|\Est,\Src'}(\cdot|\est_{i-1},\src_i)}{dQ_{\Est'|\Est}(\cdot|\est_{i-1})}
\parenth{\est_i} \nonumber \\
&=&  \frac{ \parenth{\est_i - \rho \est_{i-1} - \beta\gamma (\src_i  - \rho \est_{i-1})}^2}{2 \gamma^2\NoiseVariance}  \nonumber \\
 &-& \frac{\parenth{\est_i-\rho\est_{i-1}}^2}{2\gamma^2(\statecostval+\NoiseVariance)}-  \log \sqrt{\frac{\statecostval+\NoiseVariance}{\NoiseVariance}} \nonumber \\
\!\!\!\!\!\!\!\!&\propto_+& \!\!\!\!\!\parenth{\est_i - \src_i}^2 - \frac{\NoiseVariance}{\statecostval+\NoiseVariance} \parenth{\src_i - \rho \est_{i-1}}^2   \label{eqn:gaussianEx:costDerivation:final}
\end{eqnarray}
where \eqref{eqn:gaussianEx:costDerivation:final} follows from \eqref{eqn:gaussianEx:alphabeta}.
Similarly, the power-like cost for inverse control optimality is given $\statecostfn(\state) \propto_+  \kldist{P_{\Out|\State=\state}}{P_{\Out}} = \kldist{P_{V}(\cdot-\state)}{P_{\Out}(\cdot)}  \propto_+ \state^2$.
Thus we have
\begin{eqnarray}
  && \E\brackets{\sum_{i=1}^{n} \dist(\Src_i,\Est_{i-1},\Est_i)} \nonumber\\ &\propto_+& \E\brackets{\sum_{i=1}^{n} \parenth{\Src_i - \Est_i}^2 - \parenth{\frac{\NoiseVariance}{\statecostval+\NoiseVariance}} \parenth{\Src_i - \rho \Est_{i-1}}^2 } \label{eqn:gaussianEx:corollary:toMMSEerror:init}\\
  &=&
    \E\brackets{\sum_{i=1}^{n} \parenth{\Src_i - \Est_i}^2} \nonumber\\
  &-& \E\brackets {\parenth{\frac{\NoiseVariance}{\statecostval+\NoiseVariance}} \parenth{\rho \Src_{i-1} - \rho \Est_{i-1} + \tSrc_i}^2 } \label{eqn:gaussianEx:corollary:toMMSEerror:init:a}\\
  &=&\E \brackets{\sum_{i=1}^{n}  \parenth{\Src_i - \Est_i}^2 - \parenth{\frac{\NoiseVariance\rho^2}{\statecostval+\NoiseVariance}} \parenth{\Src_{i-1}  - \rho \Est_{i-1}}^2} \nonumber \\
  &-& \E \brackets{\parenth{\frac{\NoiseVariance}{\statecostval+\NoiseVariance}} \tilde{\Src}_i^2 } \label{eqn:gaussianEx:corollary:toMMSEerror:wUpdate}\\
  &=&\E\left[\sum_{i=1}^{n}  \parenth{1 - \frac{\NoiseVariance\rho^2}{\statecostval+\NoiseVariance}} \parenth{\Src_i - \Est_i}^2      - \parenth{\frac{\NoiseVariance}{\statecostval+\NoiseVariance}} \tilde{\Src}_i^2\right] \nonumber \\
 && - \frac{\NoiseVariance\rho^2}{\statecostval+\NoiseVariance} \E\left[ \Src_{0}^2 \right] + \frac{\NoiseVariance\rho^2}{\statecostval+\NoiseVariance} \E \left[ \parenth{\Est_{n} - \Src_{n}}^2 \right] 
 \nonumber \\
  &\propto_+&\E\left[\sum_{i=1}^{n} \parenth{\Src_i - \Est_i}^2 \right] + \parenth{\frac{1}{1-\frac{\NoiseVariance\rho^2}{\statecostval+\NoiseVariance}}}\E \left[ \parenth{\Est_{n} - \Src_{n}}^2 \right] 
  \nonumber 
\end{eqnarray}
where \eqref{eqn:gaussianEx:corollary:toMMSEerror:init} follows from \eqref{eqn:gaussianEx:parameterCostfn};
\eqref{eqn:gaussianEx:corollary:toMMSEerror:init:a} follows from \eqref{eqn:gaussianEx:src};
 \eqref{eqn:gaussianEx:corollary:toMMSEerror:wUpdate} follows from \eqref{eqn:gaussianEx:src:innovations}.
\end{IEEEproof}

\end{appendices}

\bibliography{GorantlaColemanSubmission}
\bibliographystyle{IEEEtran}

\end{document}